\documentclass[twocolumn]{aastex62}
\usepackage{natbib}
\usepackage[T1]{fontenc}

\let\oldAA\AA
\renewcommand{\AA}{\text{\normalfont\oldAA}}

\usepackage{hyperref}
\usepackage[capitalise]{cleveref}

\received{November 8, 2019}
\revised{October 19, 2020}
\accepted{October 29, 2020}
\submitjournal{ApJ}
\shorttitle{Improved DRW parameters for S82 QSO}
\shortauthors{Suberlak et al.}

\newcommand{\project}[1]{\textsf{#1}}

\begin{document}

\title{Improving Damped Random Walk parameters for SDSS Stripe 82 Quasars with Pan-STARRS1}

\correspondingauthor{Krzysztof Suberlak}
\email{suberlak@uw.edu}

\author[0000-0002-9589-1306]{Krzysztof L. Suberlak}
\affiliation{Department of Astronomy, University of Washington, Box 351580, Seattle, WA 98195, USA}

\author[0000-0001-5250-2633]{\v{Z}eljko Ivezi\'c}
\affiliation{Department of Astronomy, University of Washington, Box 351580, Seattle, WA 98195, USA}

\author[0000-0003-3422-2202]{Chelsea MacLeod}
\affiliation{Harvard Smithsonian Center for Astrophysics, 60 Garden Street, Cambridge, MA 02138, USA}

\begin{abstract}

We use the Panoramic Survey Telescope and Rapid Response System 1 Survey (Pan-STARRS1, PS1) data to extend  the Sloan Digital Sky Survey (SDSS) Stripe 82 quasar light curves. Combining PS1 and SDSS light curves provides a 15 yr baseline for 9248 quasars - 5 yr longer than prior studies that used only SDSS. We fit the light curves with the damped random walk (DRW) model model - a statistical description of their variability. We correlate the resulting DRW model parameters: asymptotic variability amplitude SF$_{\infty}$, and characteristic timescale $\tau$, with quasar physical properties - black hole mass, bolometric luminosity, and redshift. Using simulated light curves, we find that a longer baseline allows us to better constrain the DRW parameters. After adding PS1 data, the variability amplitude is a stronger function of the black hole mass, and has a weaker dependence on quasar luminosity. In addition, the characteristic timescale $\tau$ dependence on quasar luminosity is marginally weaker. We also make predictions for the fidelity of DRW model parameter retrieval when light curves will be further extended with Zwicky Transient Facility (ZTF) and the Rubin Observatory Legacy Survey of Space and Time (LSST) data. Finally, we show how updated DRW parameters offer an independent method of discovering changing-look quasar candidates (CLQSOs). The candidates are outliers in terms of differences in magnitude and scatter between SDSS and PS1 segments. We identify 40 objects (35 newly reported) with tenfold increase in variability timescale between SDSS and SDSS--PS1 data, which is due to a large change in brightness (over 0.5 mag) - characteristic for CLQSOs.

\end{abstract}

\section{Introduction}

Quasars are variable at rest-frame optical wavelengths at an asymptotic rms level of about 0.2 mag. These distant galaxies harbor an actively accreting supermassive black hole: an active galactic nucleus (AGN). Although it is agreed upon that the majority of optical light originates from the thermal emission of the accretion disk, the detailed origin of variability has been debated for the past 50 yr (see \citealt{sun2018} and references therein). Some favor a thermal origin of variability \citep{kelly2013} related to the propagation of inhomogeneities (`hot spots') in the disk \citep{dexter2011, cai2016}; others suggest magnetically elevated disks \citep{dexter2019} or X-ray reprocessing  \citep{kubota2018}.  Indeed, it may well be that the answer involves a combination of these; as \cite{sanchez2018} suggested, perhaps short-term variability (hours to days) is linked to the changes in X-ray flux, while long-term variability (months to years) is more intrinsic to the disk \citep{edelson2015,lira2015}. Nevertheless, quasar optical light curves have been successfully described using the damped random walk (DRW) model \citep{kelly2009, macleod2010, kozlowski2010, zu2011, kasliwal2015a}, and the DRW parameters have been linked to the physical quasar properties (\citealt{macleod2010}, hereafter M10). 

Variability is also a classification tool, allowing one to distinguish quasars from other variable sources that do not exhibit a stochastic variability pattern \citep{macleod2011}. This property is especially useful for selecting quasars in the intermediate-redshift range, which overlaps the stellar locus in color-color diagrams \citep{sesar2007, yang2017}. Variability has also been used to increase the completeness in measurements of quasar luminosity function (see \citealt{mcgreer2013, mcgreer2018, palanque2013, ross2013, alsayyad2016}). 

Power spectral density (PSD) informs us about the distribution of variability across the frequency range, from short timescales (high frequencies) to long timescales (low frequencies). Quasar -- or more broadly speaking, AGN-- variability, exhibits a broken power-law PSD, of the form 
$\log{P(f)} \propto \alpha \log{(f)}$, with $\alpha_l$ at low frequencies and $\alpha_h$ at high frequencies. For a pure DRW process,  $\alpha_{h}{=}-2$ and $\alpha_{l}{=}0$, so that

\begin{equation}
	P(f) = \frac{4\sigma^{2}\tau}{1+(2 \pi \tau f)^{2}}
\end{equation}
(with $\sigma = \mathrm{SF}_{\infty} / \sqrt{2}$, $\tau$ the characteristic timescale, and $f$ the frequency), where $P(f) \propto f^{-2}$  at high frequencies $f > (2\pi \tau)^{-1}$, and levels to a constant value at lower frequencies ~\citep{kelly2014}. 

There is a debate in the literature about the exact shape of the quasar PSD and any possible  departures from the pure DRW model. Studies using quasar data from wide-field photometric surveys (OGLE, SDSS, PS1) benefit from relatively long baselines (several years), which constrain the low-frequency part of the PSD. Overall, there is no evidence of a significant departure from DRW at these long timescales, i.e. $\alpha_{l} \approx 0$ \citep{zu2013, simm2016, kozlowski2016b, caplar2017, guo2017, sun2018}. However, these ground-based surveys suffer from a sparse sampling, which can be remedied by using a space-based telescope that can carry out near-continuous observations, like the Kepler mission ~\citep{borucki2010}. Studies using Kepler data that focused on a smaller number of well-sampled AGNs with short baselines (<100 days), found a range of power-law slopes at high frequencies -- from -1 to -3.2, which includes the DRW $\alpha_{h} \approx -2$-- but further study is needed \citep{mushotzky2011,edelson2014,aranzana2018,smith2018}. However, in this paper, the timescales probed are larger than several days; thus, we can assume that DRW is the best working description of quasar variability for available optical light curves. Furthermore, in this work, we directly compare the results of Sloan Digital Sky Survey (SDSS) light curves extended with the Panoramic Survey Telescope and Rapid Response System 1 Survey (Pan-STARRS1, PS1) to M10, who used pure DRW description (see discussion therein on a possible departure from DRW). Therefore, to allow a better comparison of our results with M10 we  use the DRW description of quasar PSD. 

Due to its stochastic nature, for an unbiased parameter retrieval of the DRW process, the light curve is required to be several times longer than the characteristic timescale (\citealt{kozlowski2010, kozlowski2017a}, hereafter K17). This means that DRW parameters recovered for short light curves (compared to the recovered timescale) may be biased, which in turn affects the correlations with physical parameters (black hole mass, Eddington ratio, absolute luminosity). 

For this reason, while some studies have restricted the probed redshift range, limiting the quasar sample to where one would expect only shorter timescales based on previous studies \citep{kelly2013, simm2016,guo2017,sun2018}, some have elected not to study timescales at all \citep{sun2018,sanchez2018} or to use the timescales recovered from short light curves primarily for classification \citep{hernitschek2016}.

By extending the available quasar light curves, we are able to better recover DRW timescales. Since almost a decade ago, when M10 published their study based on SDSS Stripe 82 (S82) data, new data sets (PS1, PTF, CRTS) have become available. They can extend the quasar light curves by almost 50\%.  For instance, \citet{li2018} combined SDSS and Dark Energy Camera Legacy Survey  (DECaLS) data to provide a 15 yr baseline. However, using all SDSS quasars, rather than only those confined to a well-observed S82  equatorial region, meant  that the majority of light curves suffered from a poor sampling (only a few epochs). Therefore, rather than directly fitting individual light curves, they had to resort to an ensemble structure function (SF) approach.

 Unlike previous studies, in this work, by combining SDSS and PS1 data for the well-observed S82, we afford an extended baseline (15 yr), a large number (9000) of quasars, and  a good cadence (N > 60 epochs), to which we fit the DRW model. The layout of this paper is as follows. We confirm in Section~\ref{sec:methods} that extending the quasar baseline is an important improvement in providing unbiased estimates of the DRW model parameters (K17);  in Section~\ref{sec:data} we describe the data sets employed and their combination into a common photometric system;  in Section~\ref{sec:simulation}, we simulate the improvement in the recovery of DRW parameters with baseline extension and realistic cadence; in Section~\ref{sec:results}, we describe the main results, analyzing correlations between physical parameters and variability; in Section~\ref{sec:discussion}, we discuss the physical meaning of relevant timescales; and in Section~\ref{sec:conclusions}, we summarize the main conclusions. In this work, we adopt a $\Lambda $CDM cosmology with $h_{0} = 0.7$ and $\Omega_{m} = 0.3 $. 
\newline
\section{Methods}
\label{sec:methods}
\subsection{DRW as a GP}
The DRW (Ornstein--Uhlenbeck process, \citealt{rasmussen2006}) can be modeled as a member of a class of Gaussian processes (GPs). Each GP is described by a mean and a kernel: a covariance function that contains a measure of correlation between two points, $x_{n}$ and $x_{m}$, separated by $\Delta t_{nm}$ (autocorrelation). For the  DRW process, the covariance  between two observations spaced by  $\Delta t_{nm}$ is

\begin{eqnarray}
k(\Delta t_{nm}) &=& \sigma^{2}\exp{(-\Delta t_{nm} / \tau)}  \nonumber \\
                 &=& \sigma^{2} \mathrm{ACF}(\Delta t_{nm}) 
\label{eq:covariance}
\end{eqnarray} 

Here $\sigma^{2}$ is an amplitude of correlation decay as a function of $\Delta t_{nm}$,  while $\tau$ is the characteristic timescale over which correlation drops by $1/e$. For a DRW,  the correlation function $k(\Delta t_{nm})$ is also related to the autocorrelation function.

Not explicitly used in this paper, but of direct relevance to the DRW modeling, is the SF. It can be found from the data as the rms scatter of  magnitude differences $\Delta m$  calculated as a function of temporal separation $\Delta t$ (we drop the subscripts $n$ and $m$ for brevity). The SF is directly related to a DRW kernel $k(\Delta t)$:

\begin{equation}
\mathrm{SF}(\Delta t) = \mathrm{SF}_{\infty} (1-\exp{(-|\Delta t|/\tau)})^{1/2}
\end{equation}

For quasars, SF approximately follows  a power law, $\mathrm{SF} \propto \Delta t^{\beta}$,  and for large time separation $\Delta t$, as epochs in the light curve cease to be correlated, it levels out to a constant value $\mathrm{SF}_{\infty}$: the asymptotic SF.  Note that $\sigma$ in Equation~\ref{eq:covariance} is related to the asymptotic SF as $\mathrm{SF}_\infty = \sqrt{2} \sigma$ (also see \citet{bauer2009, macleod2012, graham2015a} for an overview).

\subsection{Fitting}
We evaluate the likelihood of the DRW model with a particular set of $\tau,\sigma$ given the data with \project{celerite}, a fast GP solver \citep{foreman2017}. The underlying matrix algebra is similar to that used by \cite{rybicki1992}, \cite{kozlowski2010}, and M10. Also, as in previous work, we use a prior on the DRW parameters that is  uniform  in log space:  $1 / (\sigma \tau)$. The main difference in our approach is that rather than adopting the maximum a posteriori (MAP) as the `best-fit' value for the DRW parameters (as in \citealt{kozlowski2010}, K17; \citealt{kozlowski2016b}, M10; \citealt{macleod2011}),  we find the expectation value of the marginalized posterior. This is advantageous because of the non-Gaussian shape of the posterior; otherwise, if the posterior was a 2D normal distribution, the expectation value would coincide with the maximum of the posterior (MAP solution). 

\subsection{The Impact of Light Curve Baseline}\label{sec:baseline}

It was reported by K17 that one cannot trust any results of DRW fitting unless the light curve is at least 10 times longer than the characteristic timescale. In this section, we revisit the relationship between recovered and input timescales as a function of light curve baseline by following the K17 setup. We confirm that the bias in retrieved DRW timescale depends on how many times the light curve is longer than the timescale. However, we find that the baseline does not have to be as many as 10 times longer to provide meaningful, rather than unconstrained, results. Assuming a fixed baseline of $\Delta T = 8$ yr, we simulate 10,000 light curves, exploring 100 values of input timescales, but identical $\mathrm{SF}_{\infty}=0.2$ mag with either SDSS ($N=60$), or OGLE-like ($N=445$) cadence. Defining $\rho$ as the ratio of input timescale to baseline, we probe a range of $\rho$ between 0.01 and 15, uniform in a logarithmic grid.

\begin{figure*}
	\plottwo{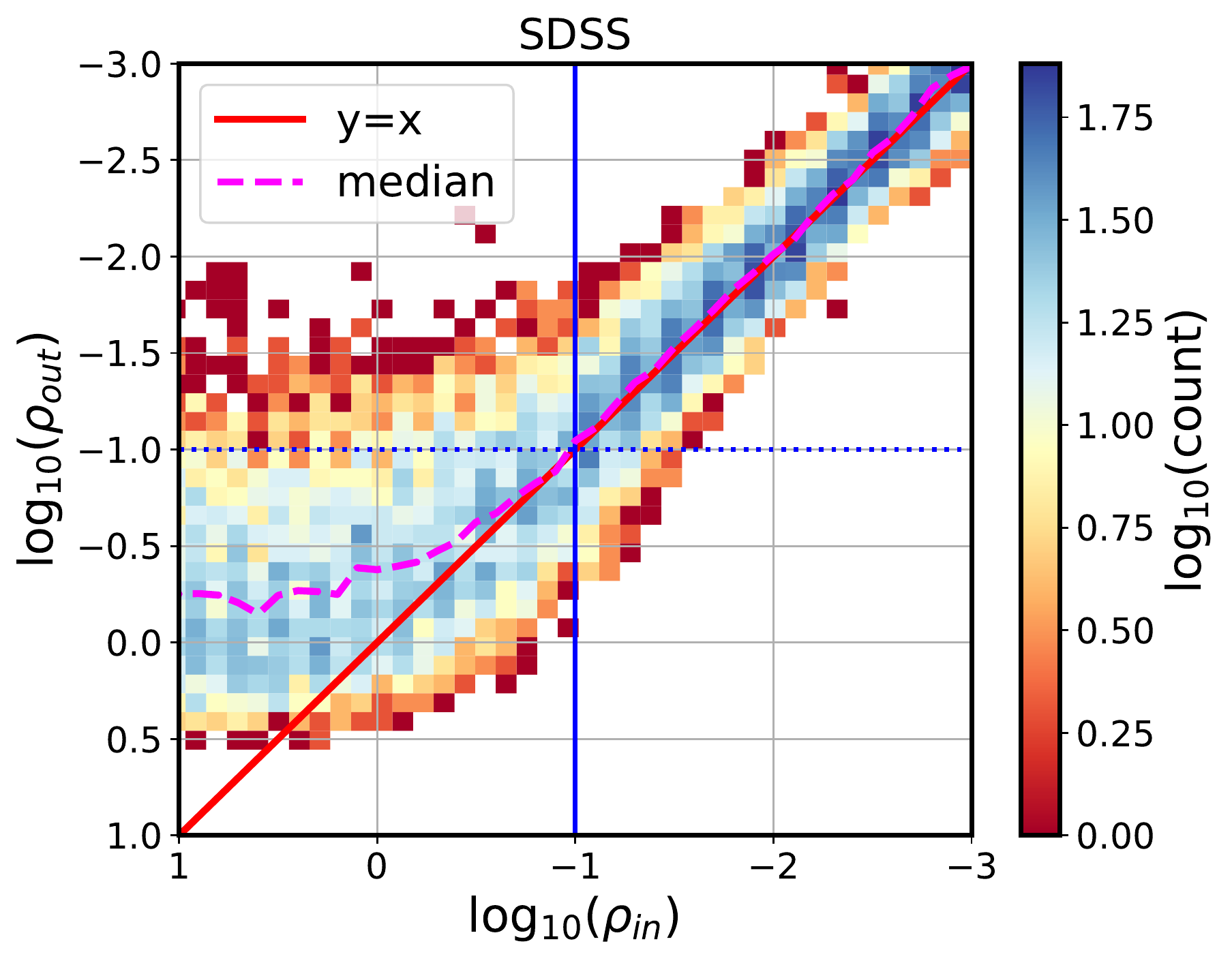}{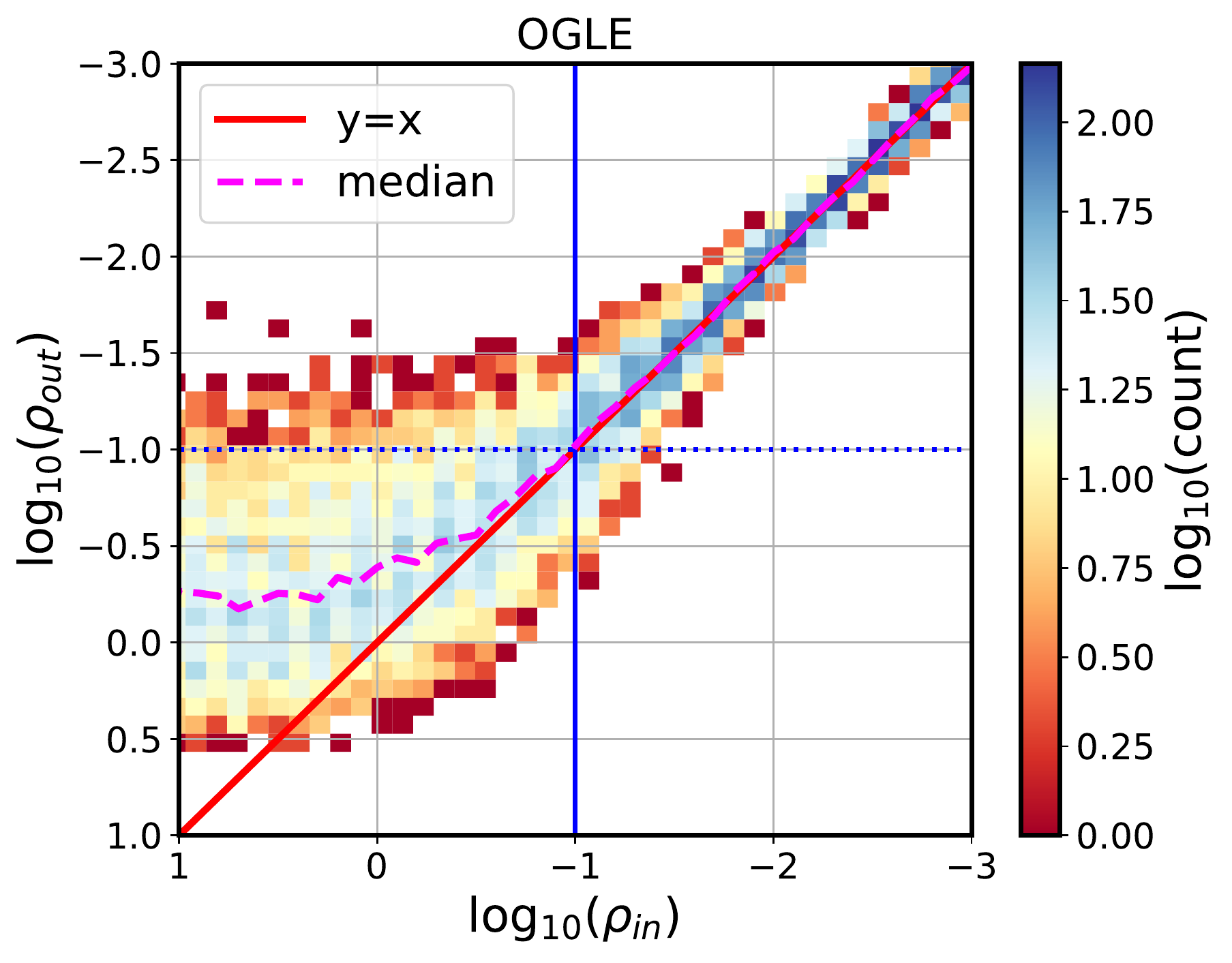}
	\caption{Recovery of the input DRW timescale with baseline fixed to $\Delta T = 8$ yr. We explore 100 logarithmically spaced values of $\rho \equiv \tau / \Delta T$, simulating 100 realizations of the DRW process at each $\rho$. The impact of photometric uncertainties and cadence is small in this case: the left panel (SDSS; $N=60$ epochs) does not significantly differ from the right panel (OGLE; $N=445$ epochs). The dotted horizontal and solid vertical lines mark $\rho = 0.1$, i.e. the baseline being 10 times longer than the timescale. The solid diagonal line corresponds to a perfect recovery of DRW parameters (where $\rho_{in} =\rho_{out} $). For any quasar, extending its light curve  moves it toward the top right (well-constrained) portion of the diagram, since for a fixed $\tau_{in}$, increasing $\Delta T$ decreases $\rho$. For baselines shorter than $\tau$, best-fit $\tau$ is underestimated and becomes biased to $\Delta T /2 $. } 
	\label{fig:rho_space}
\end{figure*}

For each light curve, we simulate the underlying DRW signal $s(t)$ by iterating over the array of time steps $t$.  At each step, we draw a point from a Gaussian distribution, for which the mean and standard deviation are recalculated at each time step (see Equations A4 and A5 in \citealt{kelly2009}; Sec. 2.2 in M10; K17). Initially, at $t_{0}$, the signal is equal to the mean magnitude, $s_{0} = \langle m \rangle$. After a time step $\Delta t_{i} = t_{i+1} - t_{i}$, the signal $s_{i+1}$ is drawn from  a normal distribution $\mathcal{N}(loc, stdev)$, with 

\begin{equation}
loc = s_{i} e ^ { - r  }  + \langle m \rangle \left( 1 - e ^{ - r }\right)
\end{equation}

and 

\begin{equation}
stdev^{2} =  0.5  \, \mathrm{SF}_{\infty}^{2} \left( 1 - e ^{  - 2 r  }  \right)
\end{equation}

where  $r = \Delta t_{i} / \tau$, and $\tau$ is the damping timescale.

Like K17, we add to the true underlying signal with zero mean $s(t)$ and a mean magnitude  ($\mathrm{r_{SDSS}}=17$ mag and $\mathrm{I_{OGLE}}=18$ mag) and calculate a magnitude-dependent estimate of photometric uncertainty:

\begin{eqnarray}
\sigma_{\mathrm{SDSS}}^{2} &=& 0.013^{2} + \exp{[2 (\mathrm{r_{SDSS}}-23.36)]} \\
\sigma_{\mathrm{OGLE}}^{2} &=& 0.004^{2} + \exp{[1.63 (\mathrm{I_{OGLE}} - 22.55)]}
\end{eqnarray}

To simulate observational conditions, we add the Gaussian noise $n(t) = \mathcal{N}(0,\sigma(t))$:

\begin{equation}
y(t) = s(t) + n(t) 
\end{equation}

The resulting distribution of fitted timescales  as a function of input timescales scaled by the  8 yr baseline, $\rho_{out}$, versus $\rho_{in}$, is shown in Figure~\ref{fig:rho_space}. We confirm the findings of K17: for short light curves, the best-fit $\tau$ becomes ${\sim}1/5$ of the light curve length (where $\log_{10}{(\rho_{out})} \approx -0.7$, the `unconstrained' region; bottom left in each panel). However, as long as the light curve is several times longer than the timescale ($1/\rho \gtrapprox 3, i.e. \log_{10}{(\rho)} \lessapprox 0.5$), we can recover the timescale without substantial bias (the dashed line approaches the solid diagonal line in both panels). In summary, in this section, we showed that when using a DRW description for quasar light curves, extension of the light curve baseline moves the fit results from the biased region (Figure~\ref{fig:rho_space}, bottom left of each panel) to the unbiased regime (Figure~\ref{fig:rho_space}, top right of each panel). This is the basis for this study,  in which we extend the baselines of quasar light curves from SDSS only (10 yr) to combined SDSS--PS1 (15 yr).

\section{Data}
\label{sec:data}

\begin{figure*} 
	\plotone{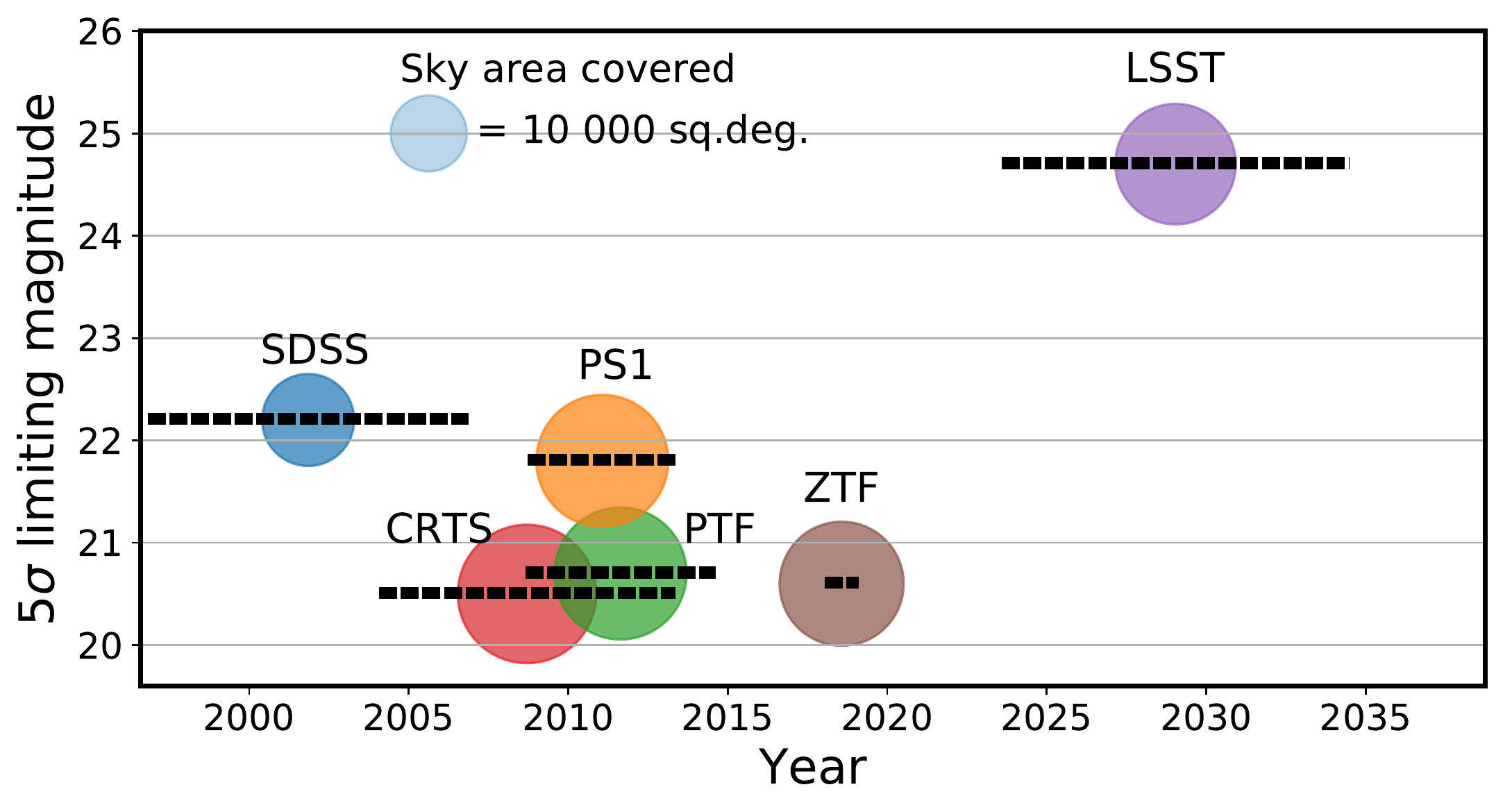}
	\caption{Illustration of survey baseline, sky area covered, and depth. The \textbf{length} of each thick dashed line corresponds to the extent of the real or simulated light curves for S82 quasars for each survey. This includes SDSS DR7, CRTS DR2, PS1 DR2, PTF DR2, ZTF DR1, and, for LSST, the full 10 yr survey. The vertical location of each line corresponds to the $5\sigma$ limiting magnitude (SDSS $r$, PS1 $r$, PTF $R$, ZTF $r$, LSST $r$, CRTS $V$). The size of each circle represents the total survey area (for SDSS, up to DR15).  Note how PS1 and PTF extend the baseline of SDSS by approximately $50\%$ and that inclusion of LSST roughly triples the SDSS baseline. For reference, the area covered by LSST is $20,000$ deg${^2}$.}
	\label{fig:lc_extent}
\end{figure*}

We focus on the data pertaining to a 290 deg$^{2}$ region of the southern sky known as S82, repeatedly observed by the SDSS between 1998 and 2008. Originally aimed at supernova discovery, objects in this area were reobserved 60 times, on average (see \citealt{macleod2012}, Section 2, for overview and \citealt{annis2014} for details). Availability of well-calibrated \citep{ivezic2007}, long-baseline light curves spurred variability research \citep{sesar2007}. The DR9 catalog \citep{schneider2008} contains 9258 spectroscopically confirmed quasars within S82. Using a $0.5''$ matching radius, we find corresponding data for 9248 quasars from  the PS1 DR2 \citep{chambers2016,flewelling2018,flewelling2020}, 7737 from the Catalina Real-Time Transient Survey (CRTS; \citealt{drake2009}),  6455 from the Palomar Transient Factory (PTF; \citealt{rau2009}), and 8001 from the Zwicky Transient Facility (ZTF) DR1 \citep{bellm2019,masci2019}. In this section, we first consider the possibility of utilizing data from all surveys and the challenges involved in combining the data from varying filter sets. We conclude that, with its shallower photometry (Figure~\ref{fig:lc_extent}) and baseline overlapping with SDSS--PS1, the added data from the CRTS would be compromised by photometric uncertainty (Figure~\ref{fig:lc_errors}), and the necessity of converting from a broad CRTS $V$-band filter (white light) to SDSS $r$. Similarly, the ZTF and PTF  were deemed too shallow for optimal data combination (median 0.1 mag uncertainty, as opposed to 0.02 mag for SDSS and 0.03 mag for PS1; see Figure~\ref{fig:lc_errors}). Finally, utilizing photometric offsets, although considered and calculated (Figure~\ref{fig:offsetsPS1}), would complicate the investigation into quasar variability by adding an additional  layer of uncertainty, and for this reason, we use only SDSS and PS1 $r$-band data, as these bandpasses are sufficiently similar that no offset is required (see Figure~\ref{fig:offsetPS1mag}).

Figure~\ref{fig:lc_extent} illustrates the improvement in baseline coverage when combining various surveys. The length of each thick dashed line corresponds to the duration of each survey (survey baseline), and the size of each circle corresponds to the area covered by each survey. The vertical location of each dashed line marks the $5\sigma$ depth in the $r$-band (or equivalent). The Rubin Observatory Legacy Survey of Space and Time (LSST) stands out in that it will provide the best extension of SDSS baseline and depth.

Combining data from different photometric standards requires applying color transformation or photometric offsets. We first seek to combine PS1 $gri$,  PTF $gR$, and CRTS $V$ into a common SDSS $r$ band (best photometry). To this end, we calculate  color terms using the SDSS standard star catalog~\citep{ivezic2007}. Focusing on a 100,000 randomly chosen stars, we find their CRTS, PS1, and PTF matches.~\footnote{CRTS from B.Sesar, priv.comm., PS1 from MAST (\url{http://panstarrs.stsci.edu}), and PTF from IRSA PTF Object Catalog (\url{https://irsa.ipac.caltech.edu/})}

The difference  between the target (SDSS) and source (e.g.,PS1) photometry can be written as a function of the mean SDSS $g-i$ color: 

\begin{equation}
m_{\mathrm{PS1}} - m_{\mathrm{SDSS}} = f(g-i)
\end{equation}

\begin{figure*} 
	\plotone{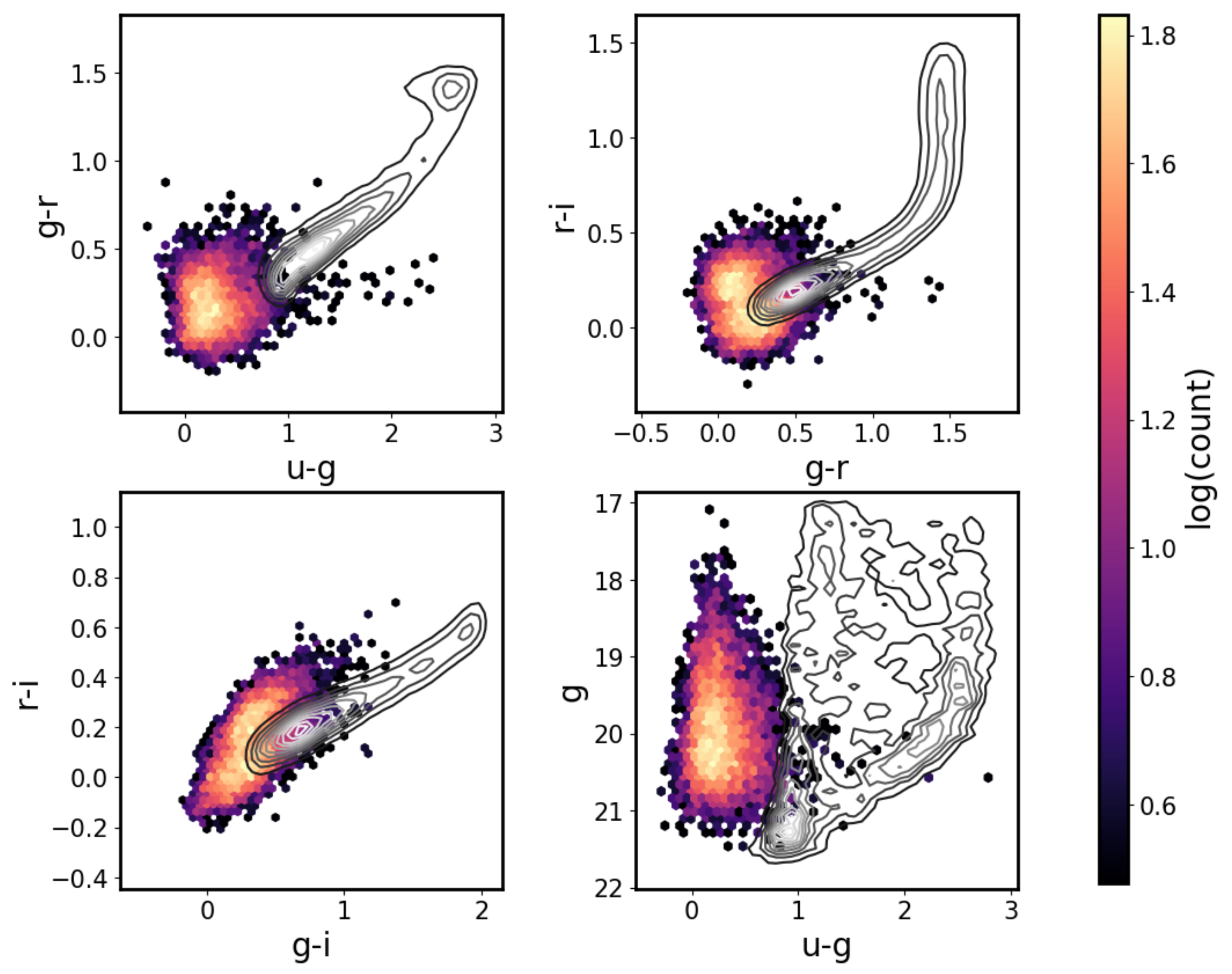}
	\caption{Regions of color-color (top left, top right, bottom left) and color-magnitude (bottom right)  space occupied by SDSS S82 quasars (colors) and stars (contours). We use quasar median photometry from \citet{schneider2010} and the standard star catalog of \citet{ivezic2007}, showing a random subset of 10,000 stars. As seen in the bottom left panel, quasars occupy a particular range of SDSS $g-i$ color. Therefore, in fitting the linear color transformations, we limit the color range to $-0.35<(g-i)<0.75$ (vertical dashed lines in Figure~\ref{fig:offsetsPS1}). Quasars also overlap other variable sources (e.g., RR Lyrae) not shown here \citep{sesar2007}. }
	\label{fig:quasarColors}
\end{figure*}

\begin{figure*}
	\epsscale{1.2}
	\plotone{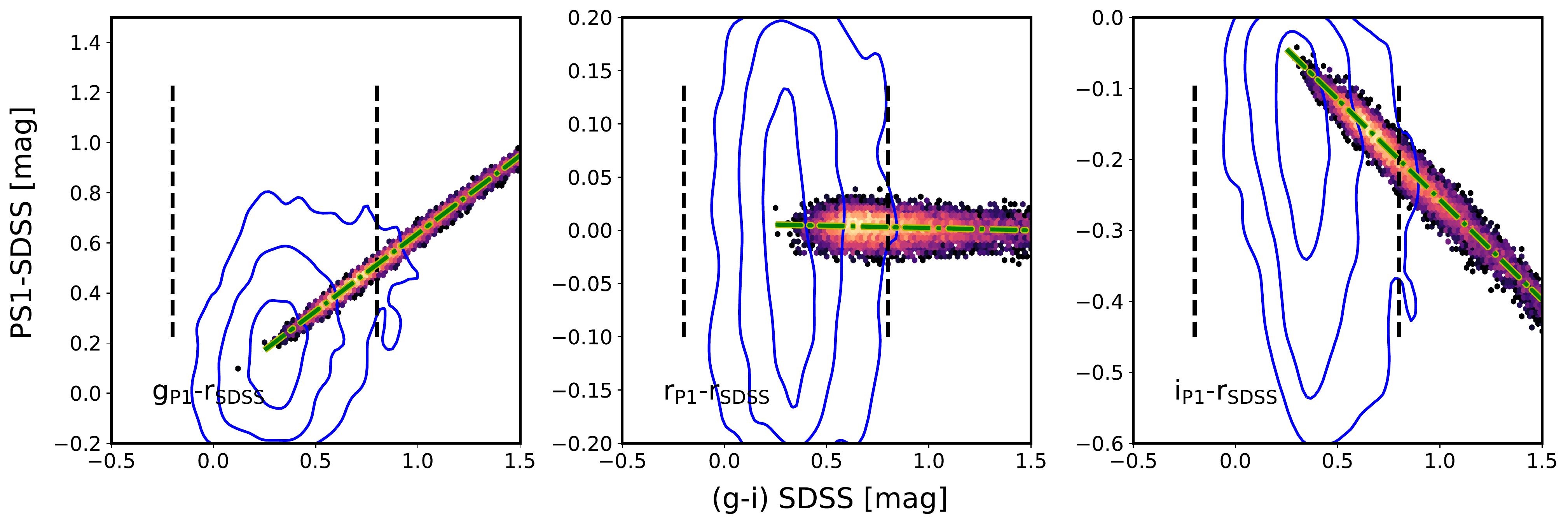}
	\caption{The SDSS--PS1 offsets, derived with the SDSS standard stars~\citep{ivezic2007}. From a randomly chosen subset of 100,000 SDSS stars, 95,000 have PS1 DR2 data. To minimize scatter due to larger errors, we select 40,000 stars with  $r < 19$ mag. Vertical dashed lines mark the region in the SDSS color space occupied by quasars, indicated by the contours enclosing  90\%, 70\%, and 30\% quasar data (also see Figure~\ref{fig:quasarColors}). Stars with $-0.2<g-i<0.8$ were used to fit the stellar locus with a first-order polynomial, marked by the dotted-dashed line. The best-fit slopes of 0.619, -0.04, and -0.283 for PS1 $g$, $r$, and $i$, respectively, are listed as $B_{1}$ in Table~\ref{tab:offsets}.}
	\label{fig:offsetsPS1}
\end{figure*}

\begin{figure}
	\epsscale{1.2}
	\plotone{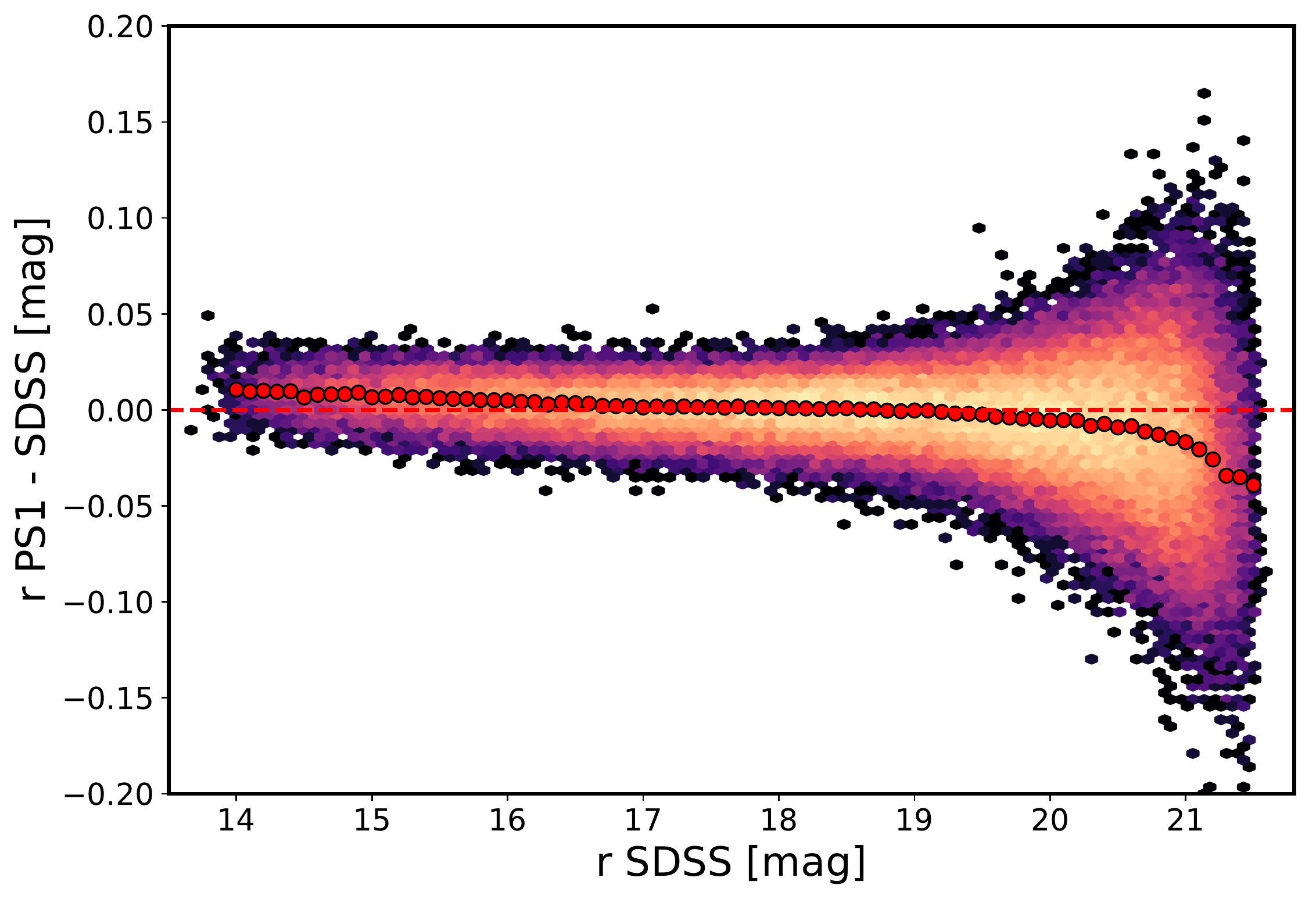}
	\caption{The PS1 $r$ versus SDSS $r$ as a function of SDSS $r$ for 100,000 randomly selected standard stars from the \citet{ivezic2007} catalog. Almost 95\% of SDSS stars have PS1 DR2 photometry. The filled circles represent the median offset, a slight slope at the 1\% (0.01 mag) level, up to $r<20.5$ mag.}
	\label{fig:offsetPS1mag}
\end{figure}

\begin{figure}
	\epsscale{1.25}
	\plotone{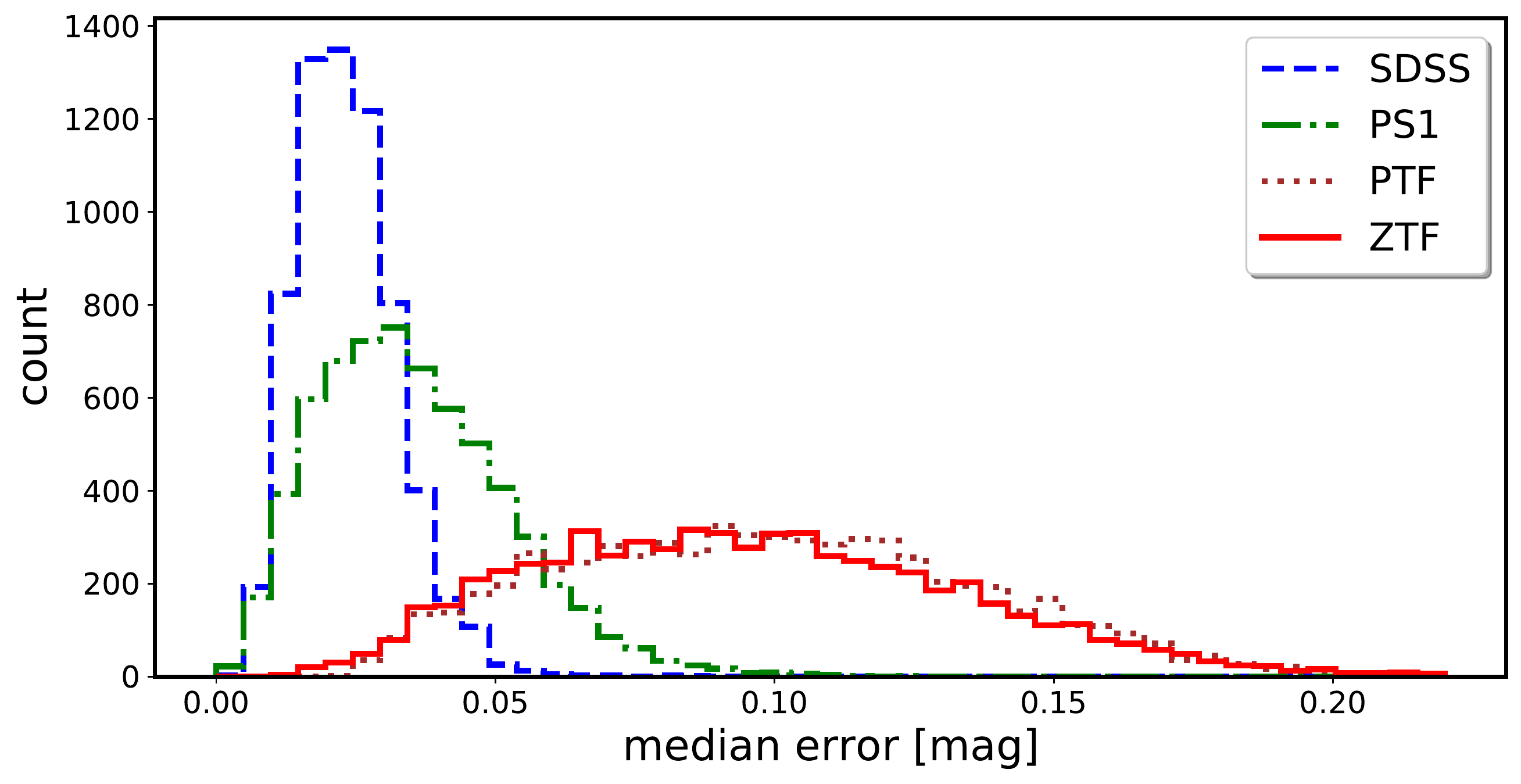}
	\caption{Distribution of median photometric uncertainties (`errors') in $r$-band real light curves. The PTF and ZTF surveys are shallower than SDSS and PS1 (see Figure~\ref{fig:lc_extent}), and thus for faint objects, such as quasars, they have much larger median errors. The CRTS errors (not shown) are, on average, 50\% larger than PTF.}
	\label{fig:lc_errors}
\end{figure}

Some authors (e.g., \citealt{li2018}) allow the transformation to be a higher-order polynomial, but as Figure~\ref{fig:quasarColors} shows, quasars occupy a relatively narrow region of $g-i$ color space, and we find that the linear fit is sufficient. The derived linear coefficients for photometric transformations between SDSS $r$ and PS1 $gri$, PTF $gR$, and CRTS $V$ as a function of SDSS $g-i$ color are listed in Table~\ref{tab:offsets}. We illustrate the process, showing, in Figure~\ref{fig:offsetsPS1}, the SDSS--PS1 standard star data used to calculate the offsets. Note that the PS1 $r$ (middle panel) is very close to the SDSS $r$, within  $0.01$ mag (1\% level) across the $g-i$ color range. We focus on the SDSS $r$--PS1 $r$ offset as a function of magnitude in Figure~\ref{fig:offsetPS1mag}; the near-equivalence of bandpass coverage is valid at the 1\% level up to $r < 20.5$.  

In selecting the most beneficial datasets to complement SDSS $r$, we also consider the associated photometric uncertainties (aka `errors'). As shown in Figure~\ref{fig:lc_extent},  PTF and CRTS are shallower than SDSS or PS1 ($\sim20.5$ mag vs $~22$ mag). Therefore, for faint objects like quasars (for the S82 sample, the population median is SDSS $r{\sim}20$ mag), PTF and CRTS have larger photometric uncertainties than SDSS or PS1. Indeed, as Figure~\ref{fig:lc_errors} shows, the distribution of median errors for PTF, CRTS, and ZTF quasar data (median $\sim0.1-0.15$ mag) is wider than the corresponding SDSS and PS1 data (median ${\sim}0.02-0.04$ mag). As simulations show (Sec.~\ref{sec:simulation}), although PTF and CRTS data do extend the SDSS baseline, their error properties decrease their utility in complementing the SDSS data set. After all, the SDSS baseline extension afforded with PTF and CRTS is comparable to that achieved with PS1 data alone (Figure~\ref{fig:lc_extent}). 

Furthermore, to mitigate problems that could arise when applying photometric transformations (such as spurious variability due to incorrect offsets or color-dependent variability), we choose to combine SDSS $r$ with only PS1 $r$, since, as Figures~\ref{fig:offsetsPS1} and ~\ref{fig:offsetPS1mag} show, SDSS $r$ and PS1 $r$ are sufficiently similar (at a 1\% level up to 20.5 mag) that no photometric transformation is required. 

Finally, we clean the combined SDSS $r$--PS1 $r$ quasar light curves using standard procedures of  $\sigma$-clipping in magnitude and error space and error-weighted day averaging to mitigate the impact of bad photometry and average out the intra-night variability (as in \citealt{charisi2016,suberlak2017}). Of 9248 SDSS--PS1 quasars, 8516 have the PS1 $r$ data with 662,092 epochs.  We remove points that have errors departing from the median SDSS(PS1) light curve segment by more than $7 \sigma$, and we visually inspect all photometry with magnitudes departing by more than  $7\sigma$ from the median magnitude. Of 585 flagged light curves, 253 required removal of individual epochs containing bad photometry. To avoid unphysically small errors,  we add in quadrature $0.01$ mag if the combined nightly error is $<0.02$ mag. In the final sample, there are 580,321 epochs.

\begin{deluxetable}{c|cc}
	\tablecaption{Color terms (photometric offsets) between CRTS, PTF, and PS1 Passbands and SDSS Using the SDSS Mean $g-i$ Color to Spread the Stellar Locus.\label{tab:offsets}}  

	\tablehead{
	\colhead{Band (x)} & 
	\colhead{\hspace{1.15cm}$B_{0}$}\hspace{1.05cm} & 
	\colhead{\hspace{1.15cm}$B_{1}$}\hspace{1.05cm}
	}
	\startdata
	CRTS V & -0.0464  & -0.0128 \\
	PTF g &  -0.0294  &  0.6404 \\
	PTF R &  0.0058   & -0.1019 \\
	PS1 g &  0.0174   &  0.6194 \\
	PS1 r &  0.0065   & -0.0044 \\
	PS1 i &  0.0260   & -0.2830 \\
	\enddata

	\tablecomments{The SDSS $r$ synthetic magnitude, $r_{s}$, can be found as $r_{s} = x-B_{0}-B_{1}(g-i)$. This linear trend is illustrated in Figure~\ref{fig:offsetsPS1}, where we plot $(x-r_{\mathrm{SDSS}})$ as a function of $(g-i)_{\mathrm{SDSS}}$ for $x=g_{\mathrm{P1}}, r_{\mathrm{P1}},i_{\mathrm{P1}}$. To derive the color terms, we used a subset of 100,000 stars randomly chosen from the SDSS standard star catalog \citep{ivezic2007}. To minimize scatter, we selected bright stars with $r<19$ mag.}
\end{deluxetable}

In conclusion, we use SDSS S82 data spanning 10 yr in the observed frame extended by the PS1 DR2 data that add, on average, 5 yr of data. Although we considered utilizing ZTF, PTF, and CRTS data, their larger photometric uncertainties offset any gain due to the extended baseline. Furthermore, only the SDSS and PS1 $r$ bands are sufficiently similar (at a ${<} 0.01$ mag level; Figure~\ref{fig:offsetPS1mag}) that no photometric offsets are required, which would unnecessarily complicate merging light curves from different surveys.

\section{Simulations : lessons learned}\label{sec:simulation}

\begin{figure}
\epsscale{1.2}
	\plotone{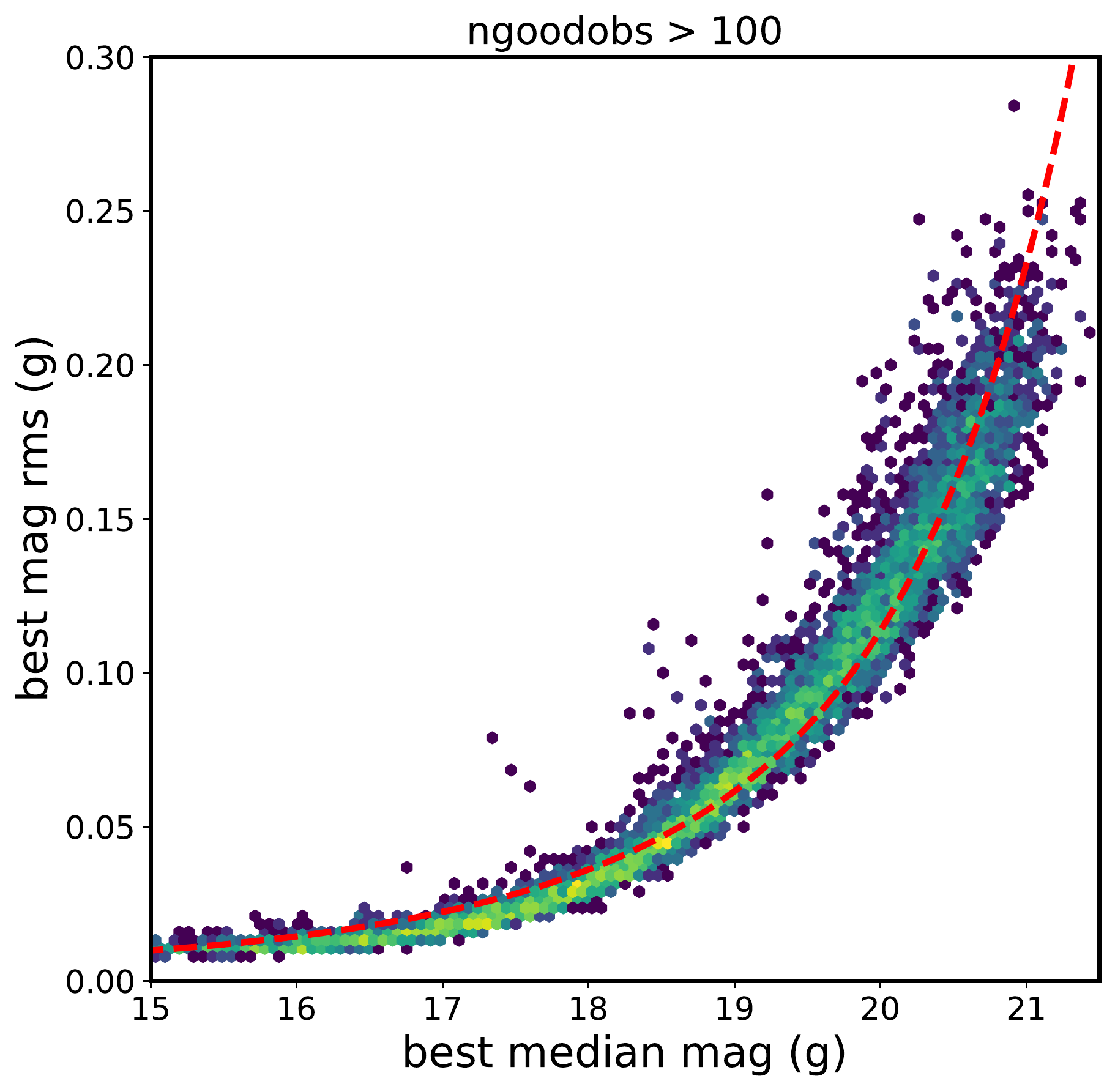}
	\caption{Best mag rms plotted as a function of magnitude for ZTF nonvariable stars with over 100 observations. We overplot the adopted error model, with $\gamma = 0.05$, $\sigma_{sys} = 0.005 $, and $m_{5} = 20.8$ (see Equation~\ref{eq:errorModel}). The properties of the ZTF photometric uncertainties are largely similar to the PTF uncertainties.}
	\label{fig:ztf_errors}
\end{figure}

\begin{figure*}
	\plotone{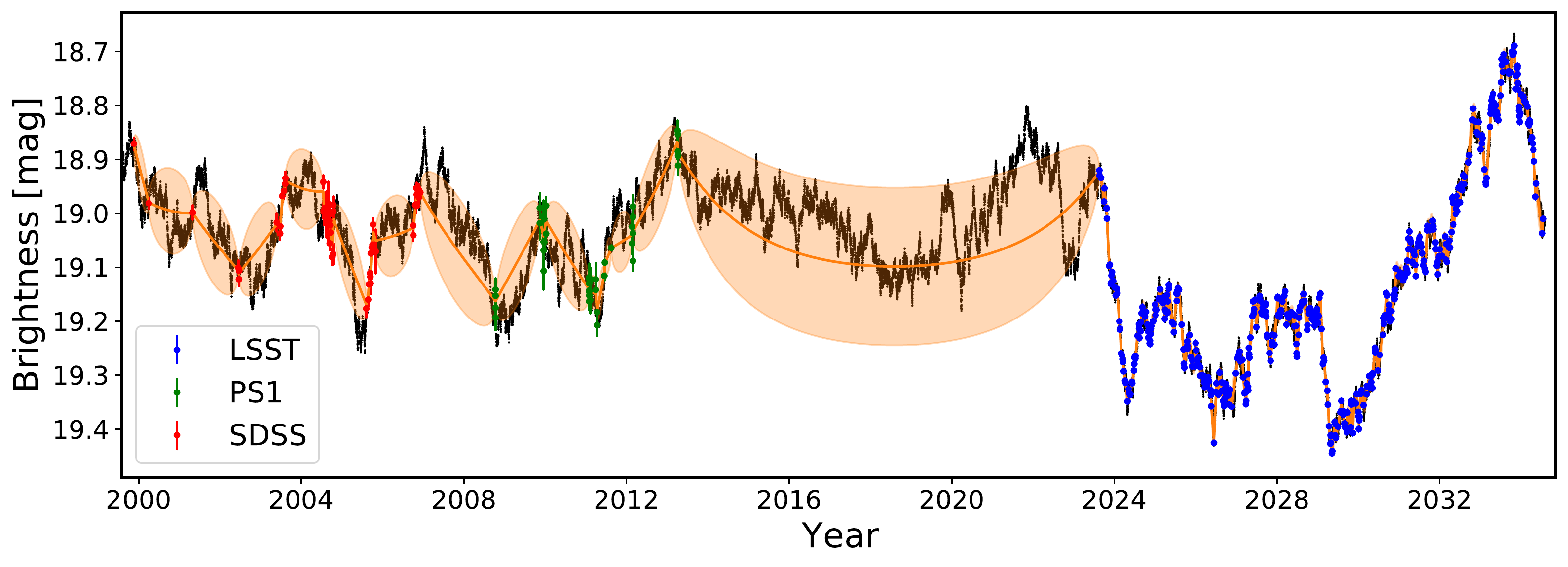}
	\caption{Simulated well-sampled underlying DRW process, one of `master' light curves ($\tau=575$ days, SF$_{\infty} = 0.2$ mag, 4 points day$^{-1}$) shown with  small black dots. To simulate observations, the cadence is degraded (subsampled) to match the ground-based cadence corresponding to real quasar data from SDSS (red), PS1 (green) segments, and simulated LSST (blue) epochs (here we use SDSS--PS1 epochs for quasar dbID=3537034). The orange `error snake' is an envelope marking the standard deviation of the fit to the data using a GP with a DRW kernel (Sec.~\ref{sec:simulation}).}
	\label{fig:lc_simulated}
\end{figure*}

We simulate the theoretical improvement of the DRW parameter retrieval in extended light curves. We generate long and well-sampled  `master' light curves, all with input $\tau = 575 $ days, SF$_{\infty} = 0.2$ mag (the median of S82 quasar distribution in M10), with zero mean.  We subsample at real observed epochs for SDSS and PS1 and at predicted cadences for ZTF and LSST  (see Figure~\ref{fig:lc_simulated}). To each simulated light curve, we add a magnitude offset corresponding to the mean of the combined SDSS--PS1 light curve. That way, the magnitude distribution of simulated light curves is similar to that of the observed SDSS--PS1 data. For the LSST 10 yr segment (finishing in 2031), we assumed 50 randomly distributed  epochs per year, with the following error model:

\begin{eqnarray}
\label{eq:errorModel}
\sigma_{LSST}(m)^{2} &=& \sigma_{sys}^{2} + \sigma_{rand}^{2} \,\, \mathrm{(mag)}^{2} \\
\sigma_{rand}^{2} &=& (0.04-\gamma)x + \gamma x^{2} \nonumber \\
x &=& 10^{0.4(m-m_{5})} \nonumber
\end{eqnarray}

with  $\sigma_{sys} = 0.005$, $\gamma=0.039$, and $m_{5} = 24.7$ (see \citealt{ivezic2019}, Sec. 3.2). For the ZTF 1 yr segment (spring 2019 ZTF DR1, including the data from 2018), we assumed 120 observations (every 3 nights) in $g_{\mathrm{ZTF}}$ and $r_{\mathrm{ZTF}}$, deriving the magnitude-dependent error model by plotting the best mag rms as a function of the best median magnitude for ZTF matches to S82 standard stars in Figure~\ref{fig:ztf_errors}. We find that the LSST error model (Equation~\ref{eq:errorModel}) with $\gamma = 0.05$, $\sigma_{sys} = 0.005 $, and $m_{5} = 20.8$ adequately describes the ZTF photometric uncertainty.

\begin{figure*} 
	\plotone{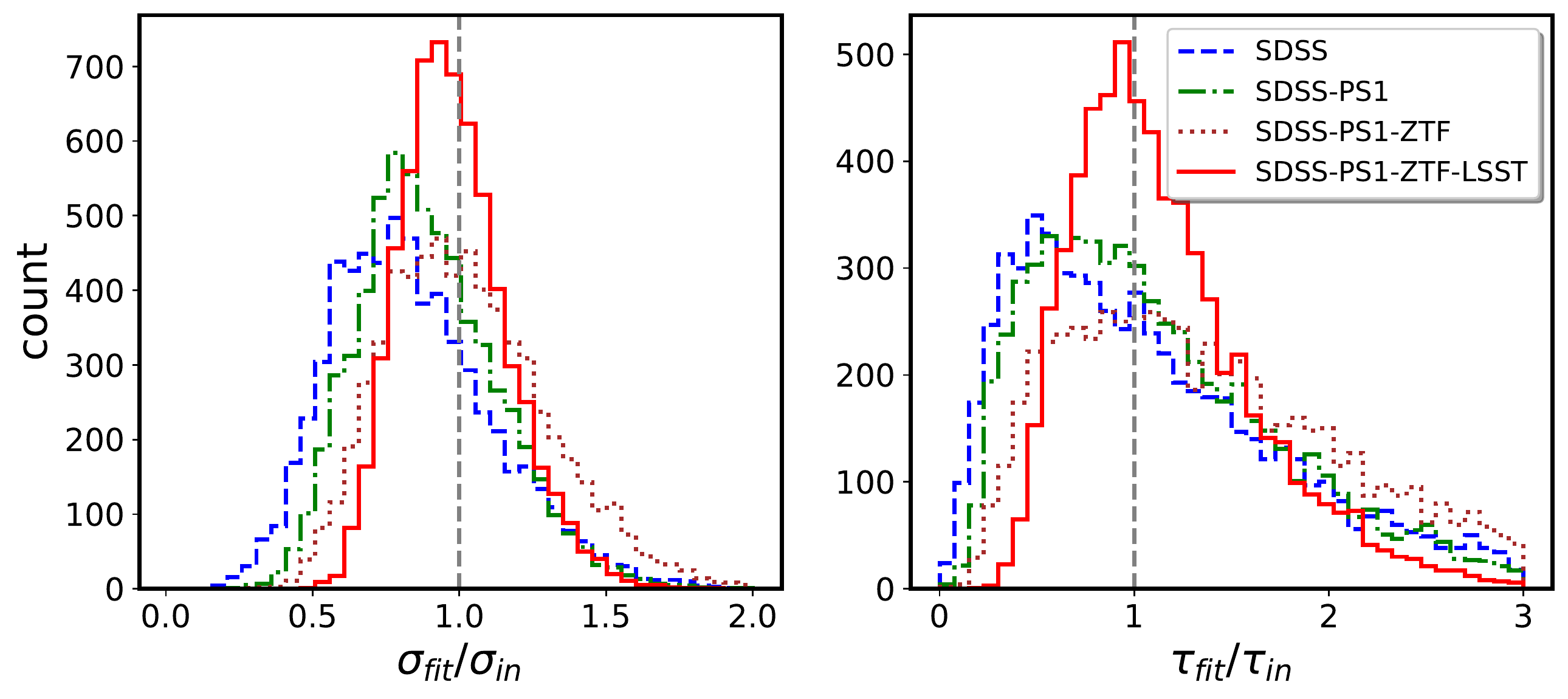}
	\caption{Ratio of DRW parameters fitted with \project{celerite}: $\tau$ and $\sigma$, to the input $\tau_{in} = 575 $d, $\sigma_{in} = 0.2 / \sqrt{2} {\sim} 0.14$  (SF$_{\infty}=0.2$ mag). We simulated 9258 `master' light curves, subsampled at real SDSS or PS1 $r$-band cadence and photometric uncertainties, and simulated ZTF and LSST cadence. To simulate observing conditions, the underlying DRW signal was convolved with a Gaussian noise corresponding to epochal errors. For each light curve, we start with the SDSS segment only, and as we add more segments (PS1, ZTF, LSST), we refit for DRW model parameters with \project{celerite}. Thus,	each distribution corresponds to a different segment of the simulated  combined SDSS--PS1--ZTF--LSST light curves. Extending the baseline shifts the distribution of the recovered DRW parameters toward the unbiased regime; the vertical dashed line marks the input matching the output. This corresponds to the top right (well-constrained) portion of Figure~\ref{fig:rho_space}.}
	\label{fig:lc_simulated_results}
\end{figure*}

To mirror observational conditions, we add a Gaussian noise to the true underlying DRW signal with variance defined by photometric uncertainties for corresponding surveys. Figure~\ref{fig:lc_simulated} illustrates the simulated `master' light curve (black dots, 4 day$^{-1}$), subsampled at SDSS (red), PS1 (green), and LSST (blue) cadence. While PS1 provides a 50\% improvement of the SDSS baseline, LSST will nearly triple it. Figure~\ref{fig:lc_simulated_results} shows how the simulated distribution of DRW parameters $\sigma$, $\tau$ changes as the SDSS quasar light curves are extended with PS1, ZTF, and LSST data. We quantify the improvement in recovery of DRW parameters with an rms error calculated for $\sigma_{fit}/\sigma_{in}$ and $\tau_{fit} / \tau_{in}$. For ratio $r$,  rms$(r)^{2} = \mathrm{bias}(r)^{2} + \sigma_{G}(r)^{2}$. For $r=\sigma_{fit}/\sigma_{in}$, the rms changes from 0.322 (SDSS only), to 0.273  (SDSS--PS1), to 0.305 (SDSS--PS1--ZTF), to 0.182 (SDSS--PS1--ZTF--LSST). In the future (after more data have been assembled and recalibrated), ZTF will help, but not as dramatically as LSST (which provides a factor of 1.8 improvement in rms errors in $\sigma_{fit}/\sigma_{in}$ and a factor of 1.5 for $\tau_{fit} / \tau_{in}$). Note that, due to larger errors (Figure~\ref{fig:lc_errors}), including ZTF causes a widening of the recovered $\tau$ distribution - the rms increases from 0.715 with SDSS to 0.890 with SDSS--PS1--ZTF (see Figure~\ref{fig:lc_simulated_results}). Using PS1 data with their excellent  deep photometry (as compared to ZTF or PTF) is the best improvement over existing SDSS results (factor of 1.2 decrease of rms errors for $\sigma$, and a smaller increase for rms errors for $\tau$; factor of 1.25 comparing SDSS--PS1--ZTF, versus factor of 1.07 for SDSS--PS1). For this reason, we use only the SDSS--PS1 portion of quasar light curves as the best trade-off between adding more baseline vs introducing more uncertainty with noisy data.

In summary, to gauge the improvement in the recovery of DRW parameters due to extending quasar light curves, we simulate $8516$ well-sampled light curves, subsampled at real cadence for SDSS and PS1 segments, and predicted cadences for ZTF and LSST segments. The mean brightness and the photometric uncertainties for each simulated light curve  mimic those in the real SDSS--PS1 data. Importantly, since the true underlying distribution of $\tau$, $\sigma$ is not known, we assume that the DRW parameters for all light curves correspond to the mean of the results of the M10 SDSS-based study: the same input timescale $\tau=575$ days, and amplitude $\sigma_{in}=0.14$ mag.  We find that combining SDSS and PS1 provides a factor of 1.2 decrease in rms errors for $\sigma$, balanced by a modest increase in rms errors for $\tau$.

\section{Results: variability parameters for S82 Quasars}\label{sec:results}

We extend the S82 quasar light curves by combining the SDSS $r$-band data with  the PS1 $r$-band data without any photometric offsets. For each quasar, we fit the SDSS and SDSS--PS1 segments  with the DRW model. This yields two sets of DRW parameters per quasar: $(\tau_{SDSS}, \sigma_{SDSS})$, and $(\tau_{SDSS-PS1},\sigma_{SDSS-PS1})$. Because variability is inherent to the quasar, for the remaining analysis, we shift all fitted timescales to the quasar rest frame and implicitly assume that the DRW timescales are considered in the rest frame: $\tau_{\mathrm{RF}} = \tau_{\mathrm{OBS}} / (1+z)$.

\begin{figure} 
\epsscale{1.2}
	\plotone{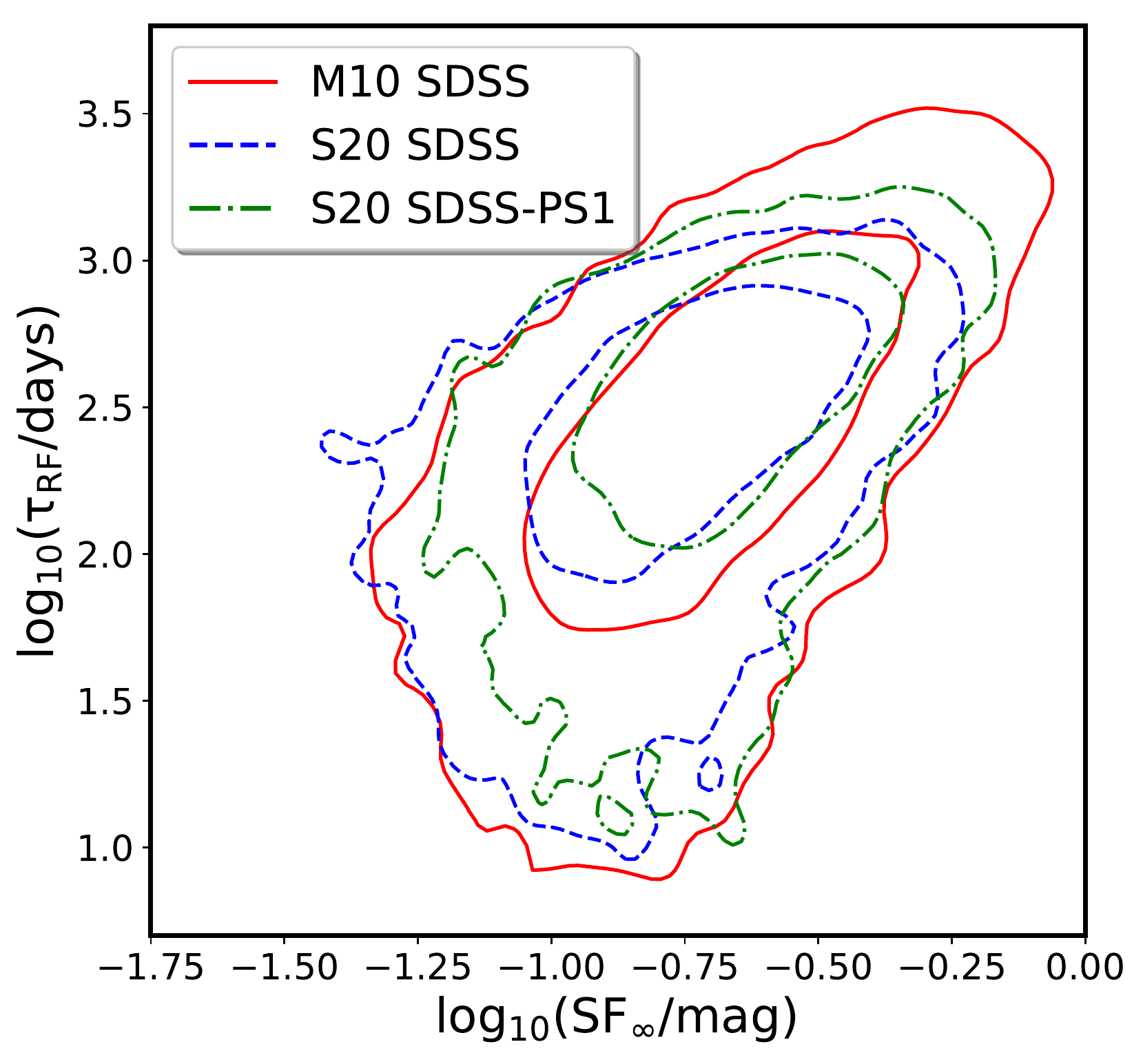}
	\caption{Comparison of distributions of the rest-frame variability timescale $\tau_{RF}$ against the  asymptotic variability amplitude SF$_{\infty}$, for M10 SDSS $r$-band (red solid)  and \project{celerite} fits using  SDSS (blue dashed) or SDSS--PS1 (green dotted-dashed) segments of combined S82 quasar light curves. Contours show the 1$\sigma$ and 2$\sigma$ levels (enclosing $68.3\%$ and $95.5\%$ of the data). The timescales and SF$_{\infty}$ from M10 and this work overlap, as we recover the same underlying distributions. }
	\label{fig:tau_sf_dist}
\end{figure}

In this section, we first correct the fitted $\tau$, $\sigma$ for wavelength dependence. Then we show consistency with the M10 results and consider the trends between DRW parameters and physical quasar properties: black hole mass $M_{\mathrm{BH}}$, absolute $i$-band  magnitude $M_{i}$, or redshift $z$.

\subsection{Comparison to M10}
\begin{figure*}
	\epsscale{1.1}
	\plottwo{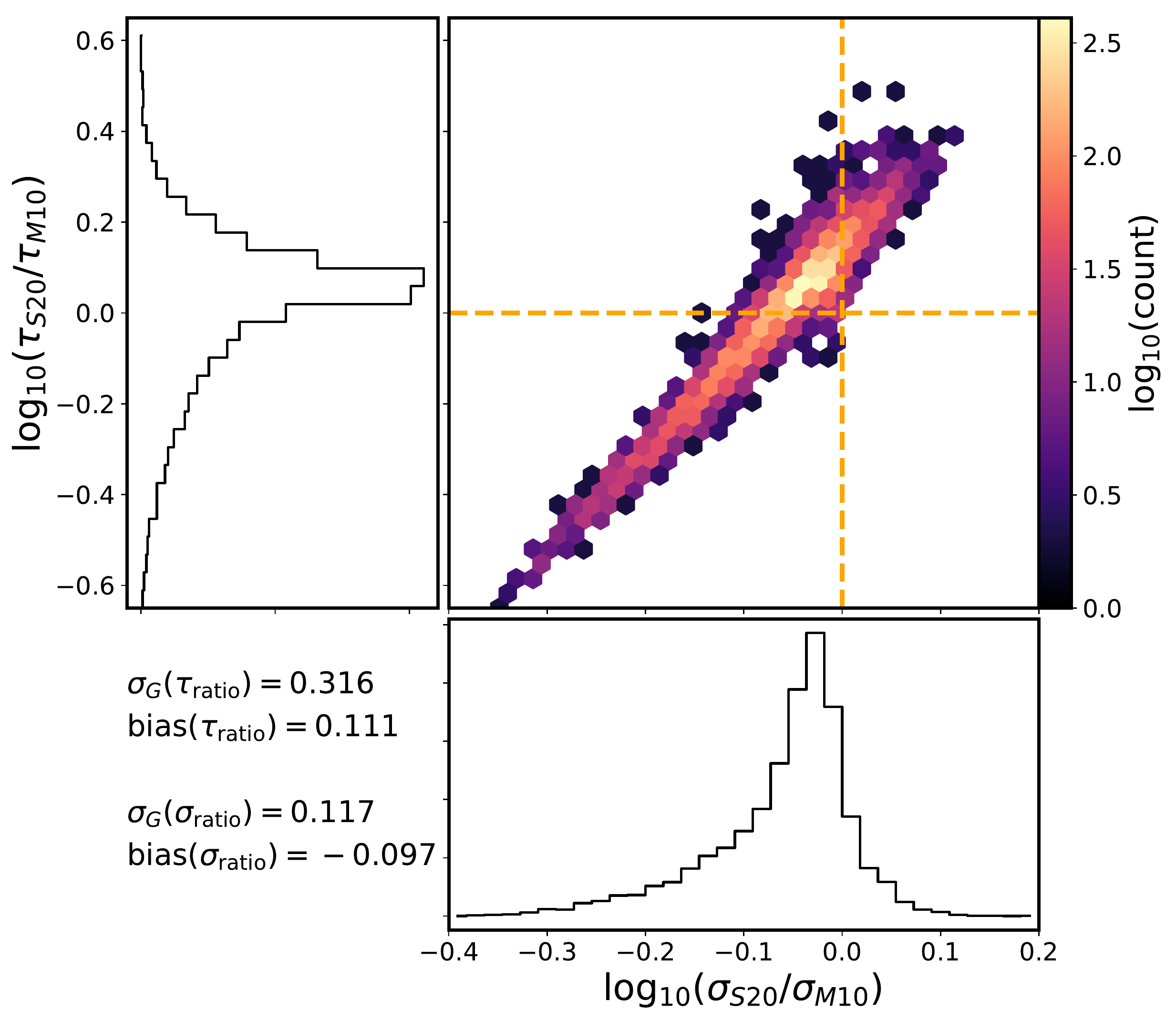}{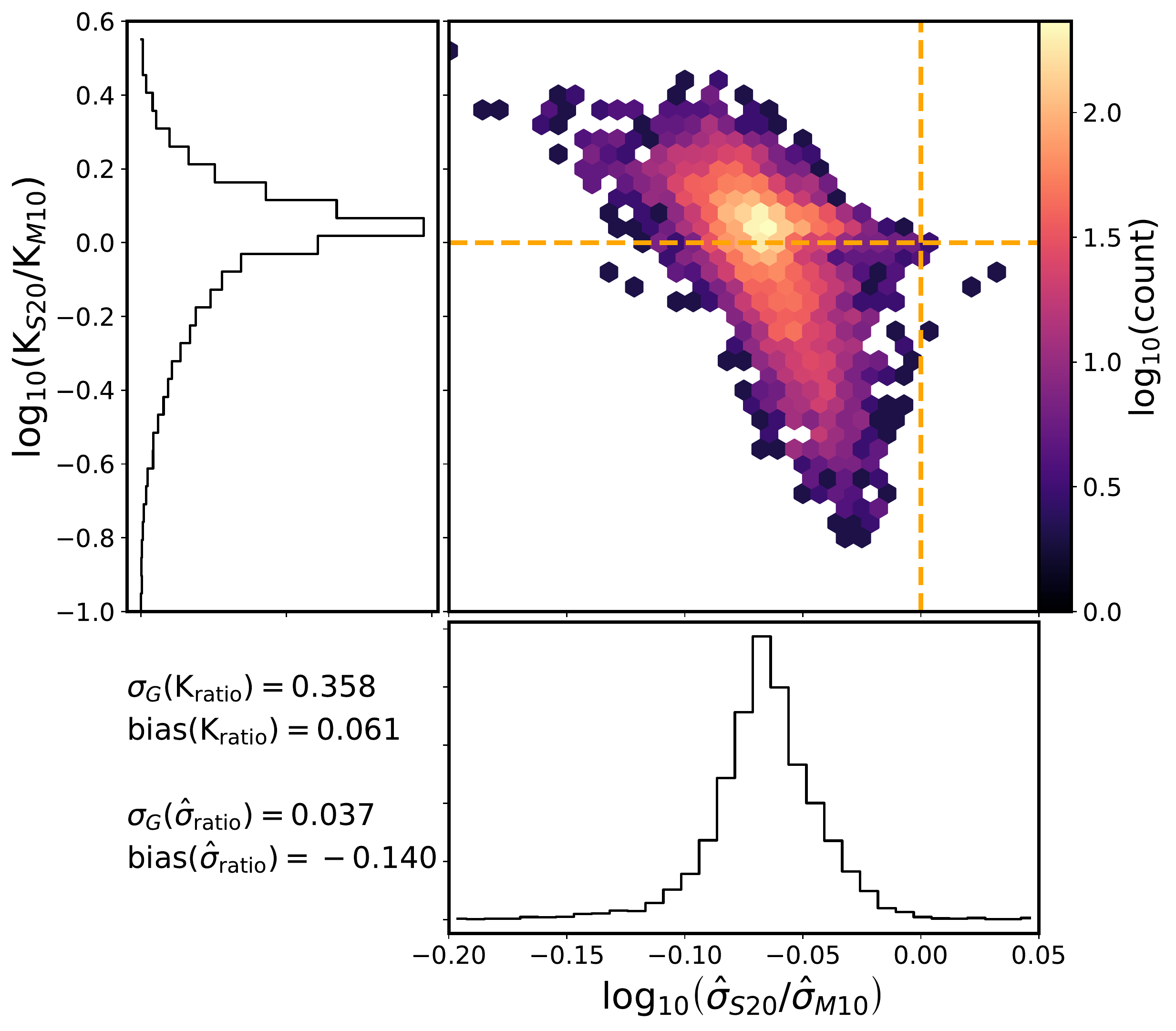}
	\caption{Comparison of \project{celerite} fits using only the  SDSS $r$-band segments of S82 quasars ($\sigma_{\mathrm{SDSS}}, \tau_{\mathrm{SDSS}}$) against M10 results for the SDSS $r$ band ($\sigma_{M10}, \tau_{M10}$) object by object. Note that for each ratio, the median-based bias (bottom left corner) is calculated before taking the logarithm. The small offset ($<0.05 $ dex) can be attributed to data cleaning and software differences. See Figure~\ref{fig:tau_sf_dist} for a comparison of rest-frame $\tau$ and SF$_{\infty}$ distributions. This is similar to Figure 3 in M10, except we plot only the $r$-band SDSS results. The right panel shows the comparison in an orthogonal $K-\hat{\sigma}$  space, where $K$ is the direction along the diagonal in the left panel and $\hat{\sigma}$ is perpendicular to the diagonal. For this reason, the right panel has a 10 times smaller scatter along $\hat{\sigma}$ (0.037) than $K$ (0.358).} 
	\label{fig:sigma_tau_ratios_M10}
\end{figure*}

The DRW parameters recovered with \project{celerite} are  broadly consistent with M10; Figure~\ref{fig:tau_sf_dist} shows the rest-frame  $\tau$ and SF$_{\infty}$ distributions for our results for the SDSS segment (blue dashed contours),  SDSS--PS1 combined light curves (green dotted-dashed contours), and  M10 SDSS for $r$-band only (red solid contours). When using exactly the same data as M10 (SDSS), our results agree. The offset of 0.05 dex  between our results and those of M10 results for SDSS, seen in the left panel of Figure~\ref{fig:sigma_tau_ratios_M10}, can be attributed to data cleaning and software differences. The right panel of Figure~\ref{fig:sigma_tau_ratios_M10} shows the same distribution in terms of $K-\hat{\sigma}$ space, orthogonal to $\tau-\sigma$, where $\hat{\sigma} = \sigma\sqrt{2 / \tau}$, and $K = \tau \sqrt{\sigma} 2^{1/4} $.

\begin{figure*} 
	\epsscale{1.1}
	\plottwo{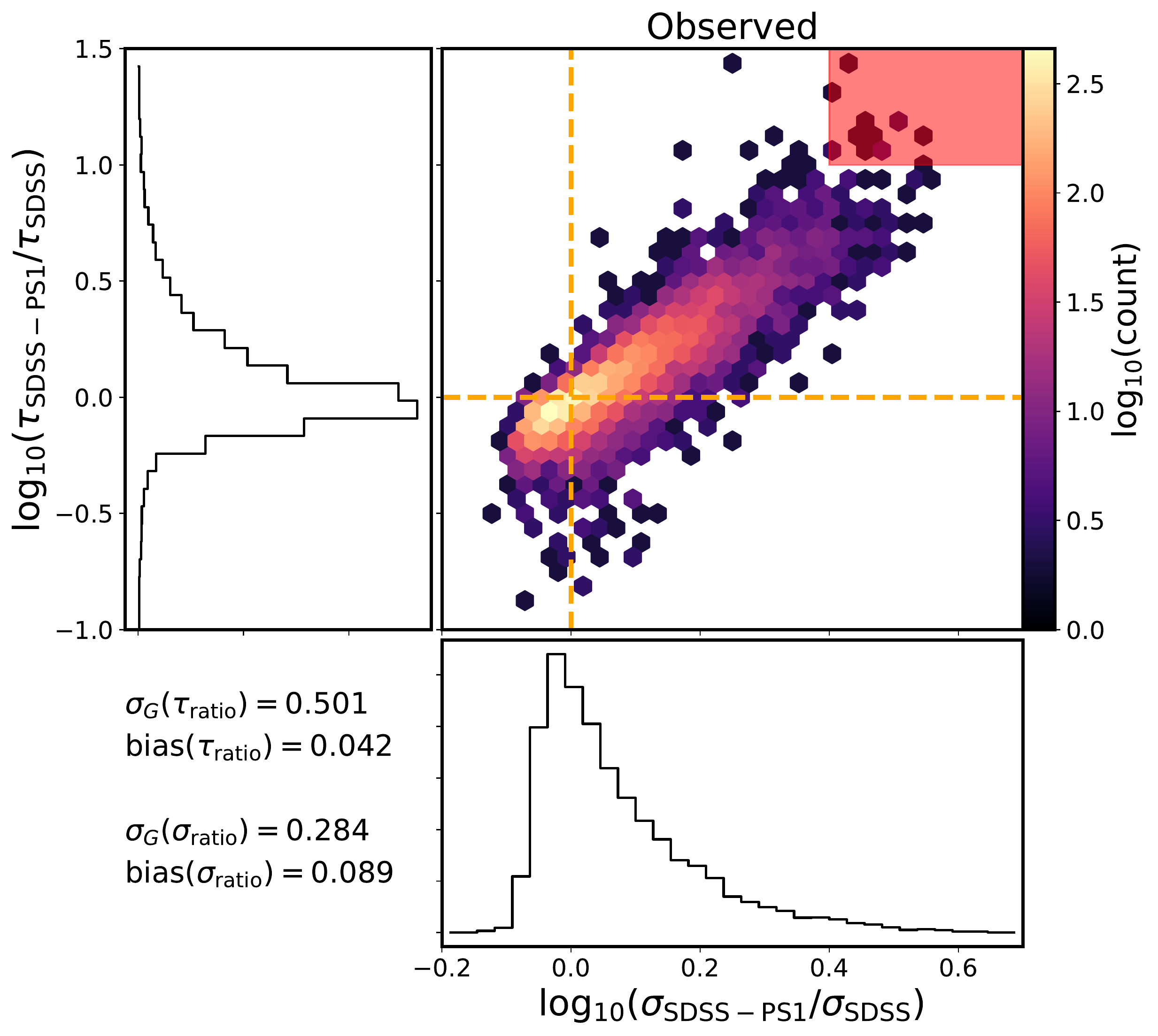}{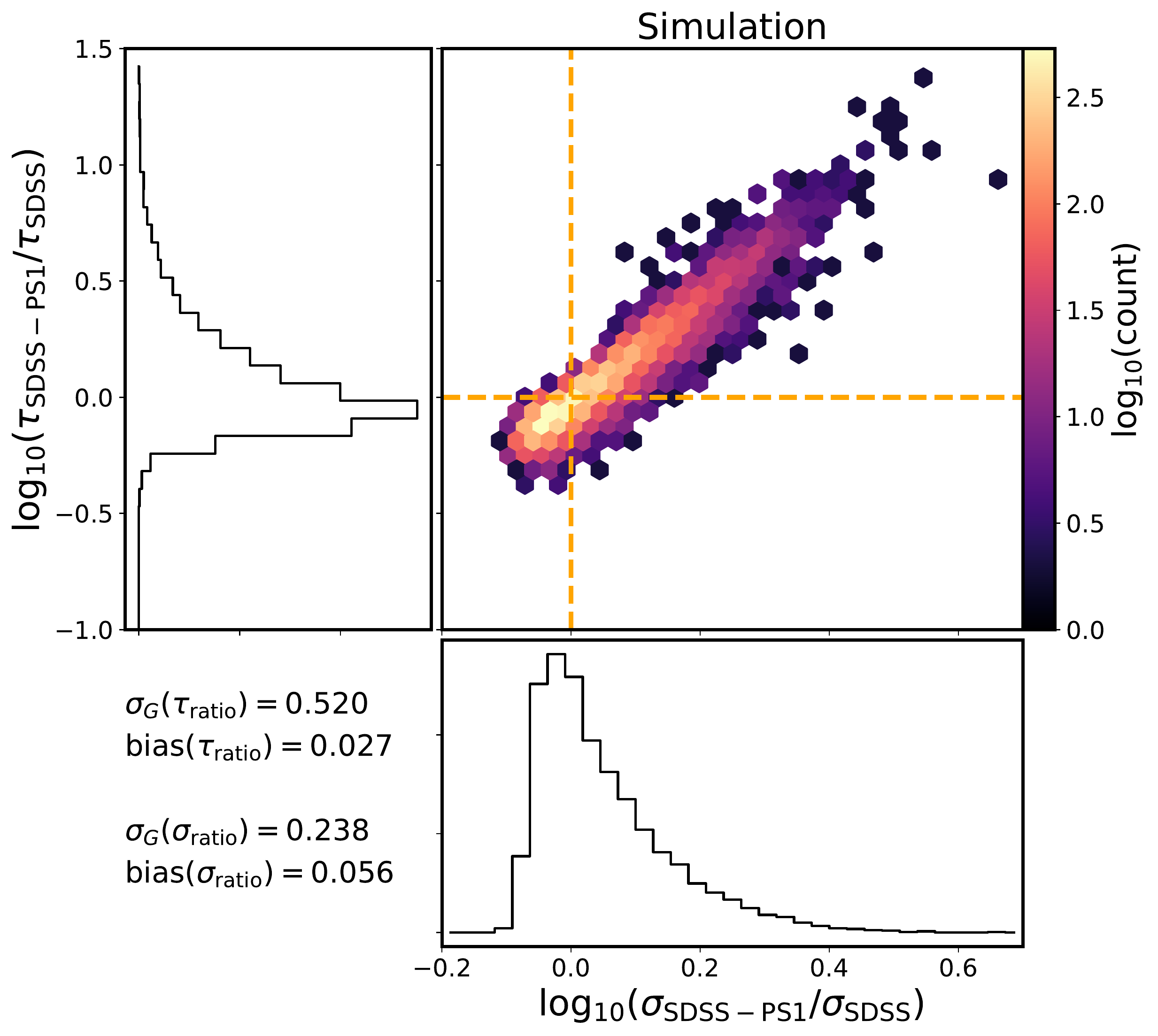}
	\caption{Ratios of fitted DRW parameters  ($\tau$, $\sigma$), comparing the value of the parameter recovered using the combined light curve length (SDSS--PS1) to the shorter, SDSS-only light curve (labeled S20 in Figure~\ref{fig:sigma_tau_ratios_M10}). The left panel shows the results for S82 quasars using real data, whereas the right panel shows the simulated quasars with realistic cadence, with  $\tau_{in}=575$ days and  SF$_{\infty}=0.2$. The general trend when using the real data (despite having a range of underlying timescales and amplitudes) is similar to that when using simulated data: the diagonal scatter is along the lines of constant $\hat{\sigma}$, and there is much less scatter in the perpendicular direction of $K$ (see Figure~\ref{fig:K_sigma_ratios}). There is no major change in the shape of the distribution as a function of mean quasar magnitude. The red rectangle marks the outliers belonging to the tail end of the population, with $\log{(\tau_{\mathrm{SDSS-PS1}} /  \tau_{\mathrm{SDSS}})} > 1$   and   $\log{(\sigma_{\mathrm{SDSS-PS1}} / \sigma_{\mathrm{SDSS}})}   > 0.4  )$, discussed in Sec.~\ref{sec:outliers}.}
	\label{fig:sigma_tau_ratios}
\end{figure*}

\begin{figure*}
	\epsscale{1.1}
	\plottwo{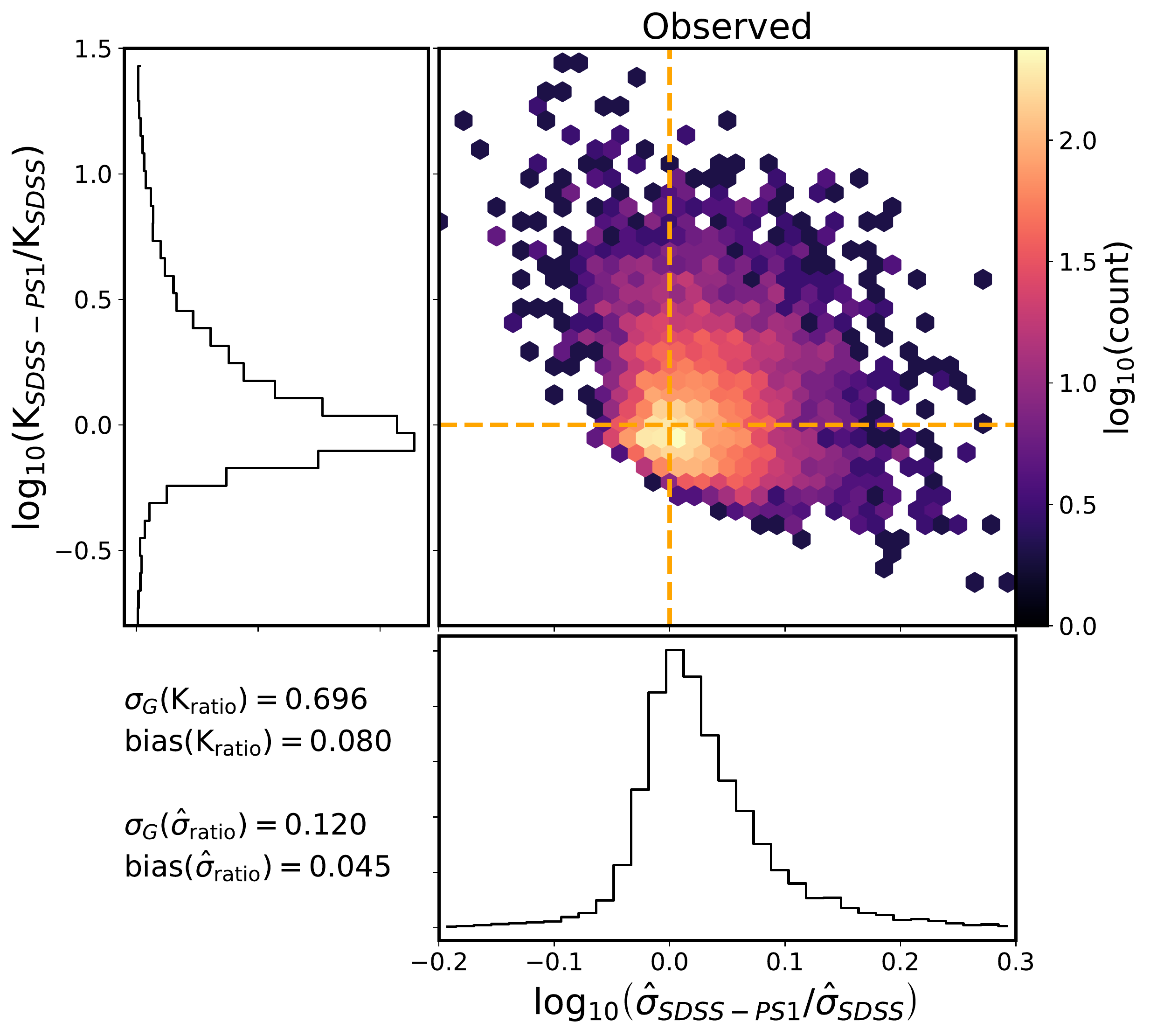}{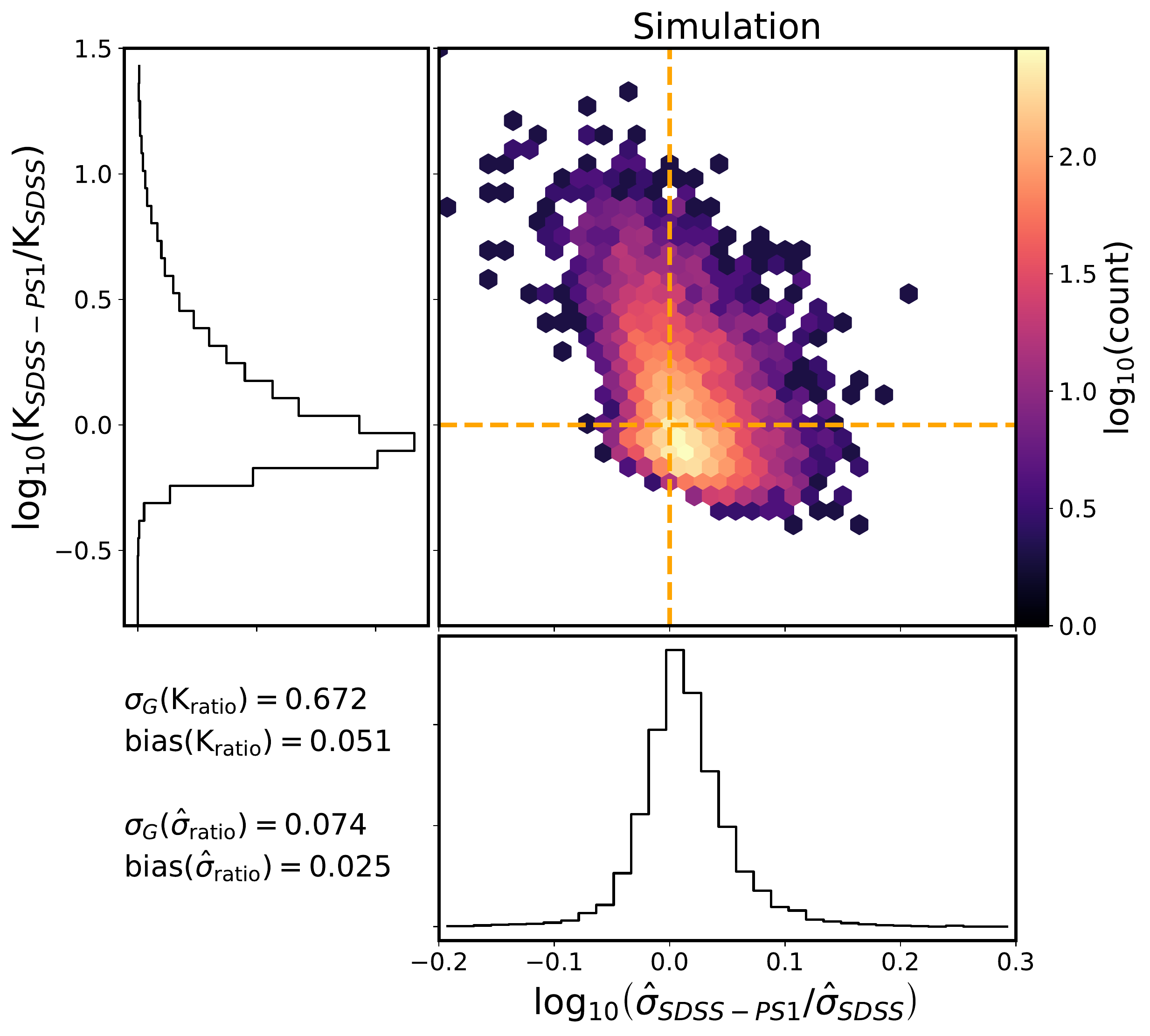}
	\caption{Same as Figure~\ref{fig:sigma_tau_ratios}, but in $K-\hat{\sigma}$ space, which is orthogonal to the $\tau-\sigma$ space, since $K = \tau \sqrt{\mathrm{SF}_{\infty}} = \tau \sqrt{\sigma} 2^{1/4}$ and $\hat{\sigma} = \mathrm{SF}_{\infty} / \sqrt{\tau} = \sigma \sqrt{2/\tau}$.
	}
	\label{fig:K_sigma_ratios}
\end{figure*}

\subsection{Outliers: possible CLQSO candidates}
\label{sec:outliers}
Figures~\ref{fig:sigma_tau_ratios} and ~\ref{fig:K_sigma_ratios} show the change in recovered DRW parameters between SDSS and combined SDSS--PS1 light curves. In Figure~\ref{fig:sigma_tau_ratios} the distribution of $f_{\sigma} \equiv \log_{10}{(\sigma_{\mathrm{SDSS-PS1}} / \sigma_{\mathrm{SDSS}})}$ and $f_{\tau} \equiv \log_{10}{(\tau_{\mathrm{SDSS-PS1}} / \tau_{\mathrm{SDSS}} )}$ for real light curves (left) matches the predicted distribution for simulated light curves (right). Analogously, on Figure~\ref{fig:K_sigma_ratios} these ratios converted to a $K-\hat{\sigma}$ space also show close correspondence between the observed (left) and simulated (right) light curves. Studies show that about 0.1\% of quasars will exhibit large variability (in excess of 0.5 mag rms; see Figure 18 in \citealt{macleod2012}). Visual inspection of light curves in the top right region of the left panel of Figure~\ref{fig:sigma_tau_ratios} (marked by the red rectangle) reveals large changes in brightness, similar to those seen in changing-look quasars (CLQSO,see \citealt{elitzur2014,lamassa2015,schawinski2015,guo2016,ruan2016,runnoe2016,blanchard2017,gezari2017, stern2017, sheng2017,lawrence2018,ross2018,stern2018,yang2018,frederick2019,macleod2019,ruan2019,trakhtenbrot2019,shen2019,sheng2020}). The rectangle in Figure~\ref{fig:sigma_tau_ratios} marks the region with  $f_{\tau} > 1$, $f_{\sigma} > 0.4$. There are 28 objects in the simulated sample and 48 objects in the observed sample with that property, which reflects the fact that we do expect quasars to exhibit changes in their brightness as a function of time, with larger variance possible over longer timescales. We further select 40 of these that are brighter than  $20.5$ median PS1 $r$-band magnitude, since PS1 observations correspond to more recent epochs and provide a better indication of the possibility of follow-up. The light curves and properties of these CLQSO candidates are further discussed in Appendix~\ref{app:clqso_cands}.

\begin{figure} 
	\epsscale{1.2}
	\plotone{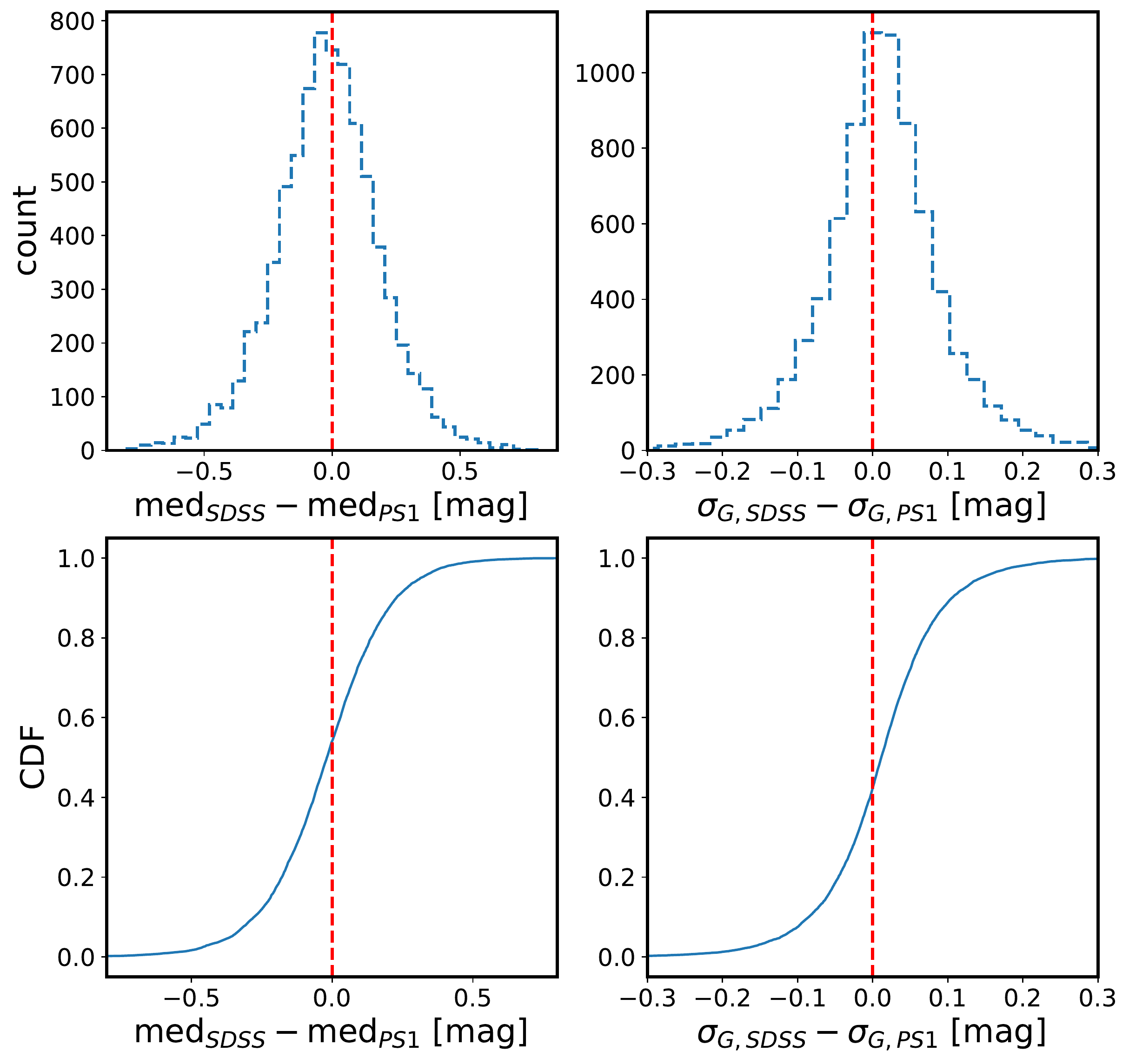}
	\caption{Differences between SDSS and PS1 segments of combined quasar $r$-band light curves. First, we show the difference between the median SDSS and median PS1 portion, plotted as a histogram (top left panel) and cumulative distribution function (bottom left panel). Then, we show difference between the $\sigma_{G}$ calculated for each portion of the light curve ($\sigma_{G}$ is a robust estimate of the standard deviation and is related to the difference between the 75th and 25th percentiles: $\sigma_{G} = 0.7413(Q_{75} - Q_{25})$). The outliers in the median offset space are also outliers in the DRW parameter space (e.g., objects with $\log{(\tau_{\mathrm{SDSS-PS1}} /  \tau_{\mathrm{SDSS}})} > 1$ and $\log{(\sigma_{\mathrm{SDSS-PS1}} / \sigma_{\mathrm{SDSS}})}   > 0.4  )$, and $r > 20.5$ have $\Delta(\mathrm{median}) > 0.1$).}
	\label{fig:median_offsets}
\end{figure}

Such large differences in timescale and amplitude of variability can also be inferred directly from the light curves. Consider the difference in magnitude and scatter between the SDSS portion of the light curve (spanning approximately 10 yr between 1998 and 2008) and the PS1 portion (spanning ${\sim}5$ yr between 2009 and 2014; see Figure~\ref{fig:lc_extent}). We measure the median magnitudes offset as $\Delta(\mathrm{median}) =  \mathrm{median}(SDSS) - \mathrm{median}(PS1)$, and the scatter difference as $\Delta(\sigma_{G}) = \sigma_{G}(SDSS)-\sigma_{G}(PS1)$. The resulting distributions of  $\Delta(\mathrm{median}) $ and $\Delta(\sigma_{G})$ for S82 quasars are shown in Figure~\ref{fig:median_offsets}. Indeed, when plotting $\Delta(\mathrm{median})$ as a function of $f_{\tau}$ and $f_{\sigma}$ there is a gradient indicating that the  CLQSO candidates, outliers in $(f_{\tau}, f_{\sigma})$ space, are also outliers in $\Delta(\mathrm{median})$-- $\Delta(\sigma_{G})$ space. Thus, the by-product of extending light curves to recalculate the DRW parameters with increased fidelity is an independent method to discover the CLQSO. 

\subsection{Rest-frame Wavelength Correction}

\begin{figure}
	\epsscale{1.2}
	\plotone{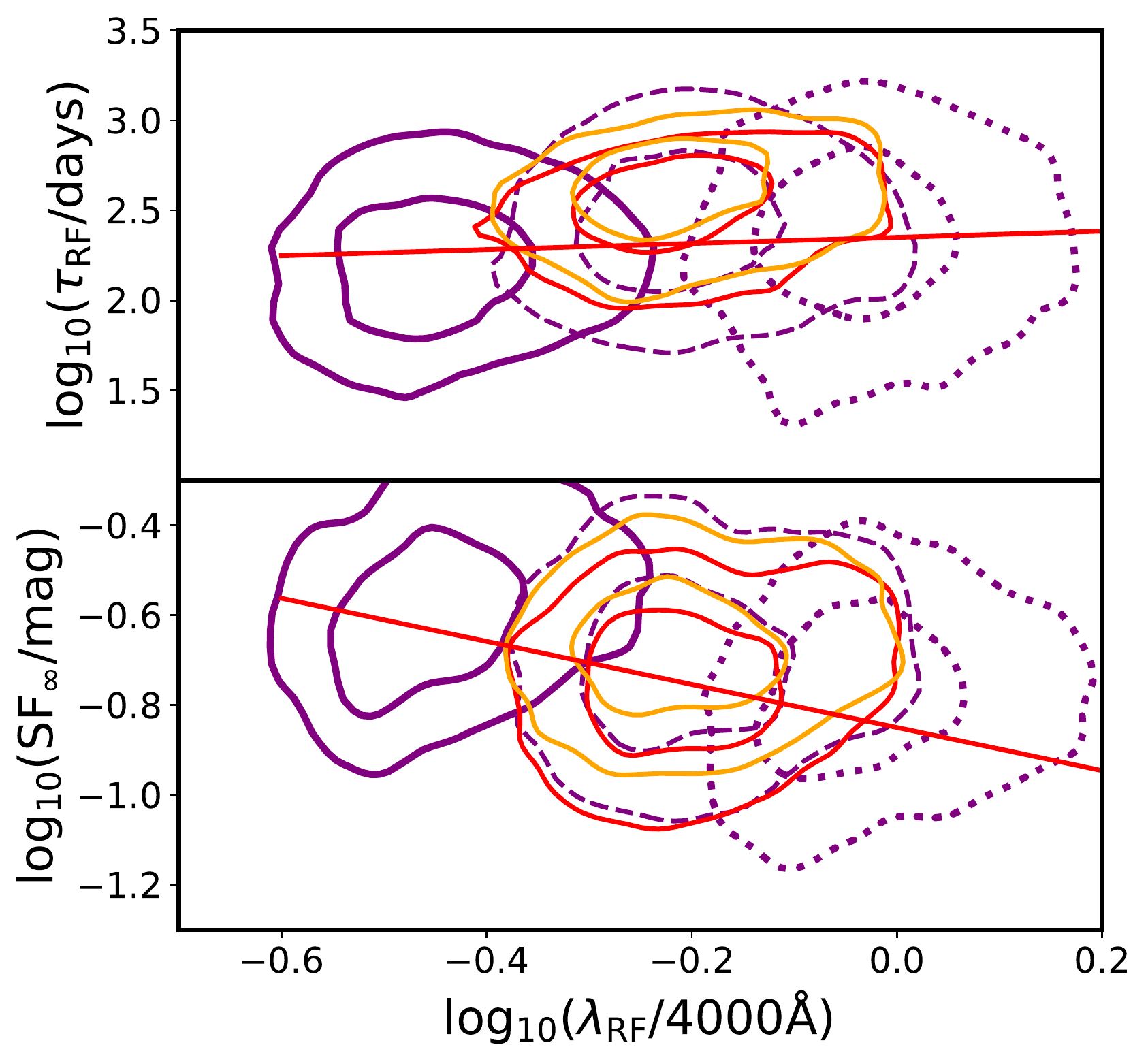}
	\caption{Rest-frame timescale $\tau$ (top panel) and asymptotic SF: SF$_{\infty}$ (bottom panel) as a function of rest-frame wavelength $\lambda_{RF}$. The background contours show the 30\% and 70\% levels for  M10 SDSS $ urz $  data, and the foreground contours  denote our results using  SDSS (red) and SDSS--PS1 (orange) segments. The red line indicates the best-fit power law to M10 data, with $B=0.17$ and $-0.479$ for $\tau_{RF}$, and SF$_{\infty}$, respectively. As M10 showed, this means that the timescale is almost independent from the bandpass, while the variability amplitude decreases toward redder rest-frame wavelengths.  We take the center of each bandpass to approximate the  observed wavelength; that is, for SDSS $urz$ bandpasses,  $\lambda_{obs} = 3520$, $6250$, and $9110$ $\mbox{\AA}$, respectively, and given the redshift of each quasar, we find $\lambda_{RF}=\lambda_{obs} / (1+z)$.}
	\label{fig:lambda_dependence}
\end{figure}

Objects at cosmological distances are embedded in the Hubble flow due to the expansion of the universe \citep{riess2019}. Therefore, light observed from a distant quasar would have been emitted at shorter wavelengths in the quasar's rest frame, $\lambda_{RF} = \lambda_{obs} / (1+z)$, where $z$ is the cosmological redshift. Quasars at different redshifts probe different regions of rest-frame spectra (see Figure 7 in \citealt{shen2019}). Thus, before correlating the DRW parameters with quasar properties, we correct $\sigma, \tau$ for the $\lambda_{RF}$ dependence studied by M10 with SDSS $ugriz$ light curves. We plot the DRW parameters in Figure~\ref{fig:lambda_dependence}: SF$_{\infty}$ and $\tau$ as a function of $\lambda_{RF}$. A solid line marks the M10 best-fit power-law trend,
\begin{equation}
\label{eq:lambda}
f \propto \left( \frac{\lambda_{RF}}{4000 \mbox{\AA}} \right)^{B}
\end{equation}

with  $B=-0.479$ and $0.17$ for SF$_{\infty}$ and $\tau$, respectively.

\subsection{Trends with Black Hole Mass, Absolute Luminosity}

In the era of large synoptic surveys such as ZTF or LSST, the large increase in the number of discovered quasars means that due to limited observational resources, we can afford a spectroscopic follow-up for only a few percent of AGNs with optical time series~\citep{ivezic2019}. Therefore, a relationship between the quasar variability parameters ($\tau, \sigma$) and physical properties $M_{BH}$, $M_{i}$ could provide an estimate of the latter for millions of quasars. We inspect correlations between $\tau, \sigma$  and $M_{BH}$, $M_{i}$, using the \cite{shen2011} catalog, based on single-epoch SDSS spectra. Here $M_{i}$ is $k$-corrected to  $z=2$, corresponding to the peak of quasar activity~\citep{richards2006a}. For details, see Appendix~\ref{app:measureBHmass}.

In Figure~\ref{fig:quasar_properties}, we examine the distribution of $M_{BH}$, $M_{i}$, as a function of $z$ for S82 quasars. The upward gradient in the top two panels reflects the selection effect that higher-redshift quasars have to be brighter to be included in the magnitude-limited sample (luminosity--redshift degeneracy; see Section 5, Figure 12 in M10, and \citealt{dong2018}). Higher-redshift quasars are also more active  and have higher black hole masses due to cosmological downsizing (see \citealt{mclure2004,babic2007,labita2009}). The distribution in the bottom left panel of Figure~\ref{fig:quasar_properties} is peaked at $z=2,$ which corresponds to the peak of quasar activity.

\begin{figure*} 
	\plotone{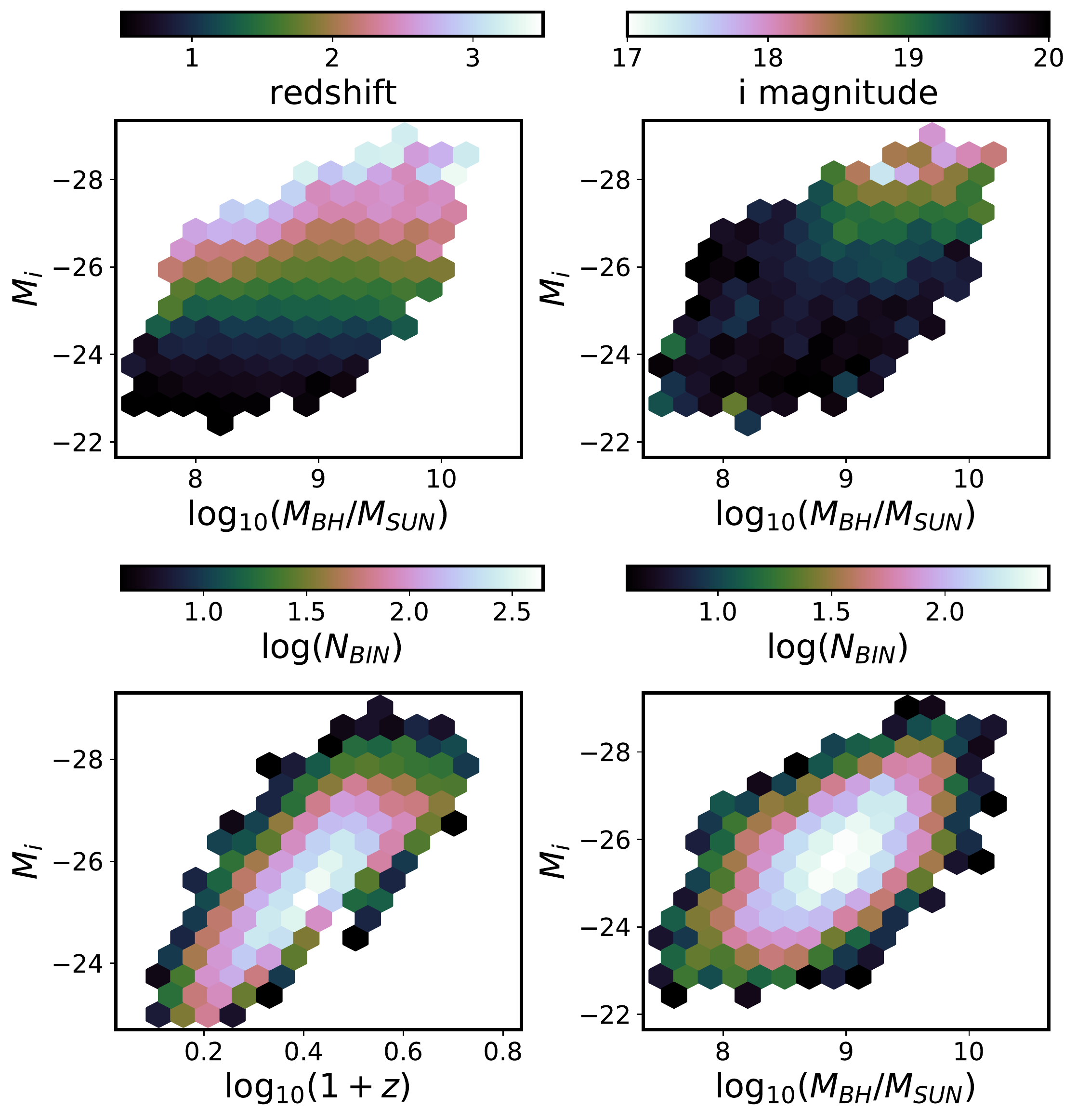}
	\caption{Distribution of quasars as a function of  redshift, observed $i$-band magnitude, absolute $i$-band magnitude ($k$-corrected to z=2), and virial black hole mass. All data are from \cite{shen2011}.}
	\label{fig:quasar_properties}
\end{figure*} 

\begin{figure*}   
	\plotone{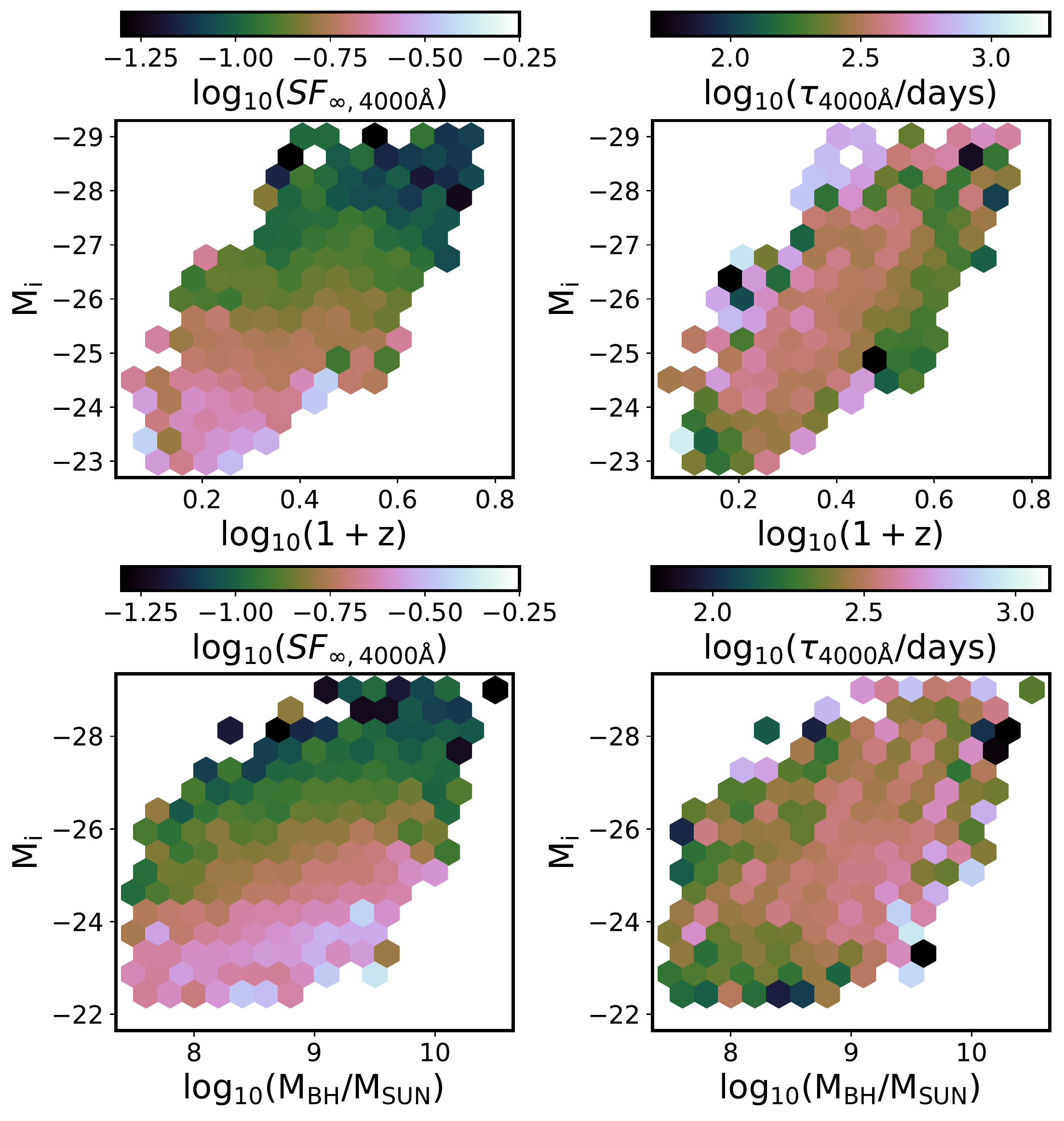}
	\caption{Absolute $i$-band magnitude $M_{i}$  ($k$-corrected to $z=2$) as a function of the virial black hole mass $M_{BH}$ (bottom panels) and redshift $z$ (top panels), colored by the  long-term variability amplitude SF$_{\infty}$ (left panels) or characteristic timescale $\tau$ (right panels). Here $M_{i}$  is a proxy for quasar bolometric luminosity, and the DRW variability parameters are for combined SDSS--PS1 $r$-band data.}
	\label{fig:quasar_trends}
\end{figure*}

Figure~\ref{fig:quasar_trends} shows the DRW parameters for S82 quasars, $\tau$ and SF$_{\infty}$, plotted as a function of quasar physical properties $M_{BH}$, $M_{i}$, and $z$. The left  panels in Figure~\ref{fig:quasar_trends} contain a gradient of SF$_{\infty}$  with 
$M_{i}$,$z$: brighter quasars have lower variability amplitude, largely independent of black hole mass.

We investigate these relations in more detail by fitting $f$ ( $\tau$ or SF$_{\infty}$) as a power-law function of $M_{BH}$, $M_{i}$, and $z$,

\begin{eqnarray}
\label{eq:powlawmodel}
\log_{10}{f} = &A& + B \log_{10}\left( \lambda_{RF} / 4000 \mbox{\AA} \right) + C (M_{i} + 23) \nonumber \\
&+& D \log_{10}{\left( M_{BH} / 10^{9} M_{\odot}  \right)} 
\end{eqnarray} 

using a Bayesian linear regression method that incorporates measurement uncertainties in all latent variables~\citep{kelly2007b}. This ansatz is identical to that used by M10, and very similar to the relation used by \cite{scaringi2015}, since, for black holes, their Equation (1) becomes $\log{t_{b}} = A' \log{M_{BH}} + B' \log{L_{Bol}}+C'$, where $t_{b}$ is the PSD break timescale. 

\begin{figure*}  
	\epsscale{1.2}
	\plotone{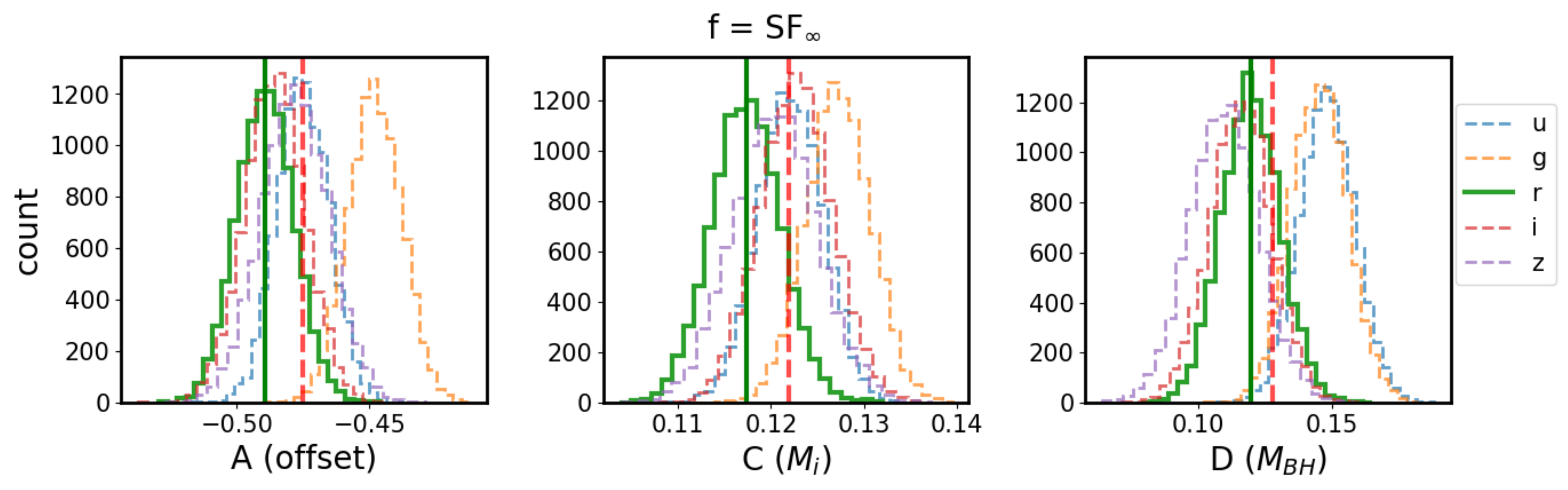}
	\caption{Posterior MCMC draws for fitting Equation~\ref{eq:powlawmodel} with M10 variability amplitude SF$_{\infty}$ against $M_{BH}$, $M_{i}$, and $z$ \citep{shen2011}. Since M10 treated the near-simultaneous SDSS $ugriz$ data for 9258 quasars independently for each band, this resulted in  DRW fit parameters for 7014 $u$-,  7408 $g$-,  6871 $r$-, 6814 $i$-, and 5111 $z$-band SDSS quasar light curves that fulfilled the M10 quality of DRW fit selection criteria. The M10 values for SF$_{\infty}$ are corrected to 4000 $\mbox{\AA}$ using Equation~\ref{eq:lambda}, with the power-law coefficient $B=-0.479$. Each distribution corresponds to a different SDSS band. We compare the results of fitting the SDSS--PS1 $r$ band directly against the M10 results for the SDSS $r$ band (green solid). Note that Table 1 in M10 reported band-averaged values for the A, C, and D coefficients (red vertical dashed  line), while we cite in Table~\ref{tab:coefficients} the mean for the $r$ band (green vertical solid line).}
	\label{fig:ugriz_drw_M10}
\end{figure*}

\begin{figure*} 
	\plotone{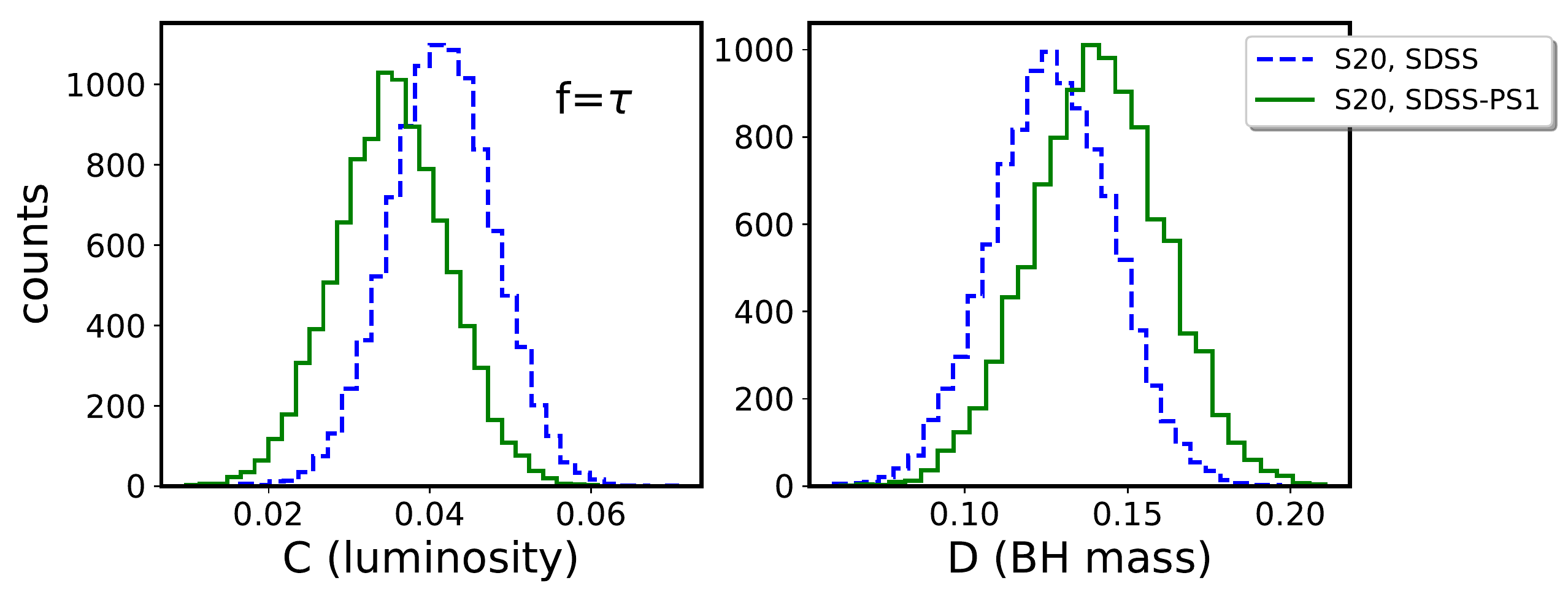}
	\caption{Distribution of MCMC posterior draws fitting Equation~\ref{eq:powlawmodel} for the characteristic timescale ($f=\tau$) based on SDSS $r$-band results (dashed blue line) and new SDSS--PS1 combined $r$-band results (solid green line). These are considered simultaneously as a function of quasar absolute magnitude $M_{i}$ (left panel) and black hole mass $M_{BH}$ (right hand side panel). Of 9258 spectroscopically confirmed quasars in S82, we employed 8516 that had PS1 matches, of which 6371  fulfill the M10 selection criteria (see M10, Sec 2.2). The results from the SDSS--PS1 light curves are consistent with M10 for the SDSS $r$ band. }
	\label{fig:drw_tau_posterior}
\end{figure*}

\begin{figure*}
	\plotone{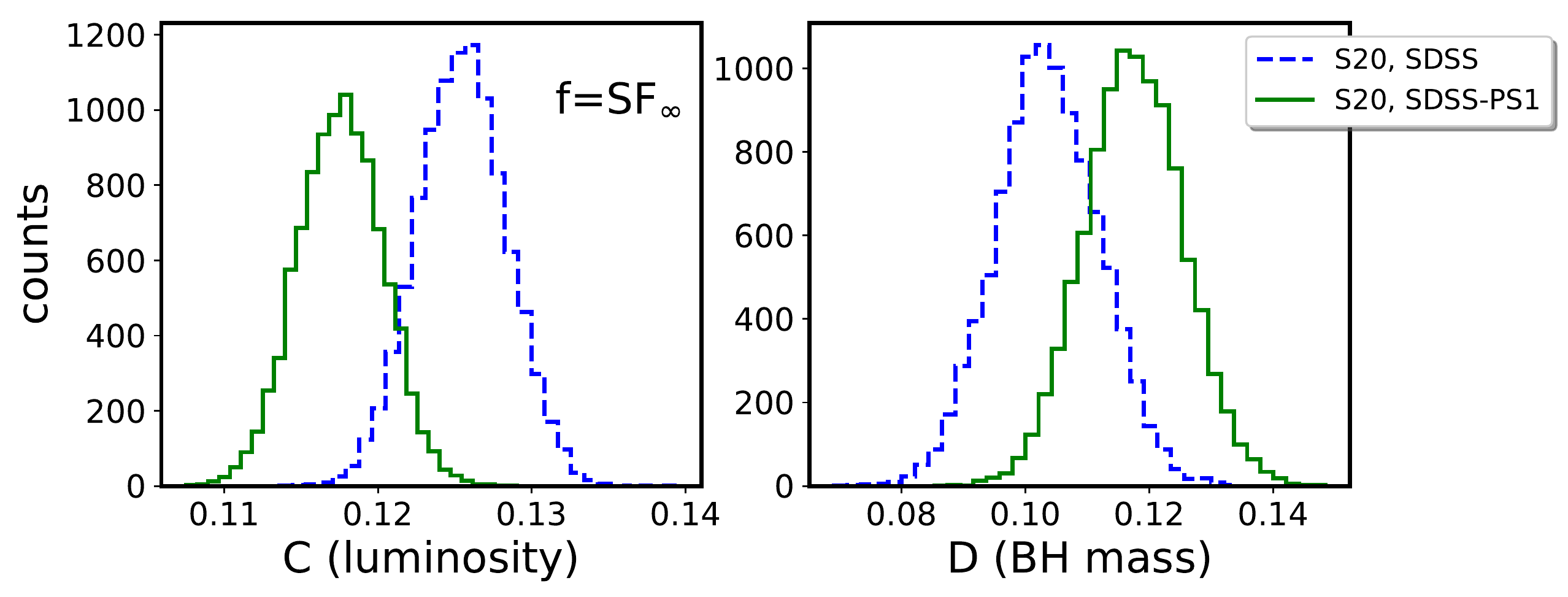}
	\caption{Same as Figure~\ref{fig:drw_tau_posterior}, but fitting the DRW asymptotic amplitude ($f = \mathrm{SF}_{\infty}$ in Equation~\ref{eq:powlawmodel})  as a function of absolute magnitude $M_{i}$, and black hole mass $M_{BH}$. New data from PS1 are consistent with the earlier results of M10 on luminosity dependence but support a slightly weaker dependence of $\mathrm{SF}_{\infty}$ on $M_{BH}$ (by 0.06 dex).}
	\label{fig:drw_sf_posterior}
\end{figure*}

We compare the change in retrieved fit coefficients caused by adding PS1 data to SDSS against the M10 SDSS-only study. Note that M10 fitted a DRW model treating each of the five SDSS bands as a separate light curve, resulting in over 30,000 values of $\tau, \mathrm{SF}_{\infty}$ for 9000 S82 quasars. Grouping the fitted quasar parameters by band, they were correlated to the quasar physical parameters with Equation~\ref{eq:powlawmodel}. Figure~\ref{fig:ugriz_drw_M10} shows the posterior samples for fitting  Equation~\ref{eq:powlawmodel} to $f=\mathrm{SF}_{\infty}$ for quasar data separately for each SDSS bandpass.  Each band yields a slightly different fit coefficient. As the fit result, M10 reported the band mean (red vertical dashed line). Since this study uses only $r$-band data, we compare the fit coefficients to M10 SDSS $r$ data (green vertical solid line in Figure~\ref{fig:ugriz_drw_M10}). We show the results of fitting Equation~\ref{eq:powlawmodel} to new SDSS and SDSS--PS1 parameters in Figures~\ref{fig:drw_tau_posterior} and ~\ref{fig:drw_sf_posterior}. First, with $f=\tau$ in Equation~\ref{eq:powlawmodel} (Figure~\ref{fig:drw_tau_posterior}), the SDSS--PS1 data confirm M10 for luminosity dependence (the posterior Markov Chain Monte Carlo (MCMC) samples overlap), but the dependence of $\tau$  on $M_{BH}$  is marginally weaker (by 0.007 dex). Second, in Figure~\ref{fig:drw_sf_posterior} with  $f=\mathrm{SF}_{\infty}$, $\mathrm{SF}_{\infty}$ has a slightly weaker dependence on $M_{BH}$ (by 0.04 dex compared to M10). The difference between the \project{celerite} SDSS-only results and M10 can be attributed to data cleaning that was not performed by M10, and software differences. Each distribution from Figures~\ref{fig:drw_tau_posterior} and ~\ref{fig:drw_sf_posterior} is summarized in Table~\ref{tab:coefficients}, with the uncertainty in the A,C, and D fit coefficients estimated from  the standard deviation of the posterior samples. 

We also searched for a signal of Mg\,{\sc ii} variability (see \citealt{cackett2015} for a review). Like H$\alpha$  and  H$\beta$ , Mg\,{\sc ii} is a permitted low-ionization line \citep{yang2020}, but on average, it is being emitted by gas further away from the ionizing source than the  H$\beta$, possibly at the edge of the broad-line region (BLR; \citealt{guo2020}). \cite{ivezic2004} and \cite{macleod2012} studied the SDSS--POSS sample of quasars and reported the detection of a decrement in the difference between the data and the best-fit model  (residuals) around 2800 $\mbox{\AA}$ when plotting the residuals as a function of $\Delta t$ and $\lambda_{RF}$. We investigate the residuals for $f=\mathrm{SF}_{\infty}$ in Equation~\ref{eq:powlawmodel}. We find that, using the SDSS data, the decrement in the median($\tau_{\mathrm{RF}}$) around 2800 $\mbox{\AA}$ is visible at ${\sim}3 \sigma$  relative to the smooth model, but by adding the PS1 data, the significance rises to ${\sim}5 \sigma$. We do not see much difference with regard to whether using the subset of 6371 quasars for which M10 had reliable results (listed in Table~\ref{tab:coefficients}) versus  the full set of 8516 quasars fitted with \project{celerite} for DRW parameters. The effect is interesting but does not produce a very significant signal; see Appendix~\ref{app:mgii_variability} for more details. 

\begin{deluxetable*}{cc|CCCC}

	\tablecaption{Comparison of best-fit coefficients for Equation~\ref{eq:powlawmodel} using M10 Results and This Work (S20). \label{tab:coefficients} }

	\tablehead{\colhead{$f$} & \colhead{Source} & \colhead{$A$(offset)} & \colhead{$B(\lambda_{RF})$} & \colhead{$C (M_{i})$} & \colhead{$D (M_{\mathrm{BH}})$} }
	\startdata
	$\tau$ & M10, SDSS & $2.5\pm0.027$ & $0.17\pm0.02$ & $0.03\pm0.009$ & $0.178\pm0.027$ \\
	 & S20, SDSS & $2.515\pm0.019$ & $0.17\pm0.02$ & $0.042\pm0.007$ & $0.127\pm0.019$ \\
	 \tableline
	& S20, SDSS--PS1 & $2.597\pm0.02$ & $0.17\pm0.02$ & $0.035\pm0.007$ & $0.141\pm0.02$ \\
	  \tableline
	SF$_{\infty}$ & M10, SDSS & $-0.486\pm0.012$ & $-0.479\pm0.005$ & $0.119\pm0.004$ & $0.121\pm0.012$ \\
	 & S20, SDSS & $-0.543\pm0.009$ & $-0.479\pm0.005$ & $0.125\pm0.003$ & $0.104\pm0.008$ \\
	  \tableline
	 & S20, SDSS--PS1 & $-0.476\pm0.008$ & $-0.479\pm0.005$ & $0.118\pm0.003$ & $0.118\pm0.008$ \\
	 \enddata

	 \tablecomments{Here $B$ is fixed to $0.17$ or $-0.479$ from fitting a power law between $\lambda_{RF}$ and $\tau$, SF$_{\infty}$ (see Figure~\ref{fig:lambda_dependence}). Of 8516 quasars with SDSS--PS1 data, for consistency we use here an unbiased subset of 6371 quasars for which M10 had reliable results. For $f=\tau$, $C$ is almost the same between M10 and this work for SDSS--PS1 (rows 1 and 3). However, $D$ based on SDSS--PS1 data is larger than M10 by 0.01 dex (row 3). For $f = \mathrm{SF}_{\infty}$, SDSS--PS1 based C is within 0.01 dex from M10 (rows 4,6), and $D$ based on SDSS--PS1 data is almost identical to M10. When using \project{celerite} $\tau$ ($\sigma$) results for all 8516 quasars, the luminosity dependence is unchanged to within 0.01 dex, and the dependence on the black hole mass is stronger by 0.05 dex (0.02 dex), respectively. As shown on Figure~\ref{fig:sigma_tau_ratios_M10}, there is a small  offset between $\log_{10}{(\tau_{\mathrm{S20,SDSS}}/ \tau_{\mathrm{M10,SDSS}})}$  and $\log_{10}{(\sigma_{\mathrm{S20,SDSS}}/ \sigma_{\mathrm{M10,SDSS}})}$, attributed to data cleaning procedures and software differences, which contributes to a shift between $C$ and $D$ parameters for $\tau$ and SF$_{\infty}$ between M10 and S20. }

\end{deluxetable*}

\subsection{Comparison to other studies: Eddington ratio}

\begin{figure*}
	\epsscale{1.2}
	\plotone{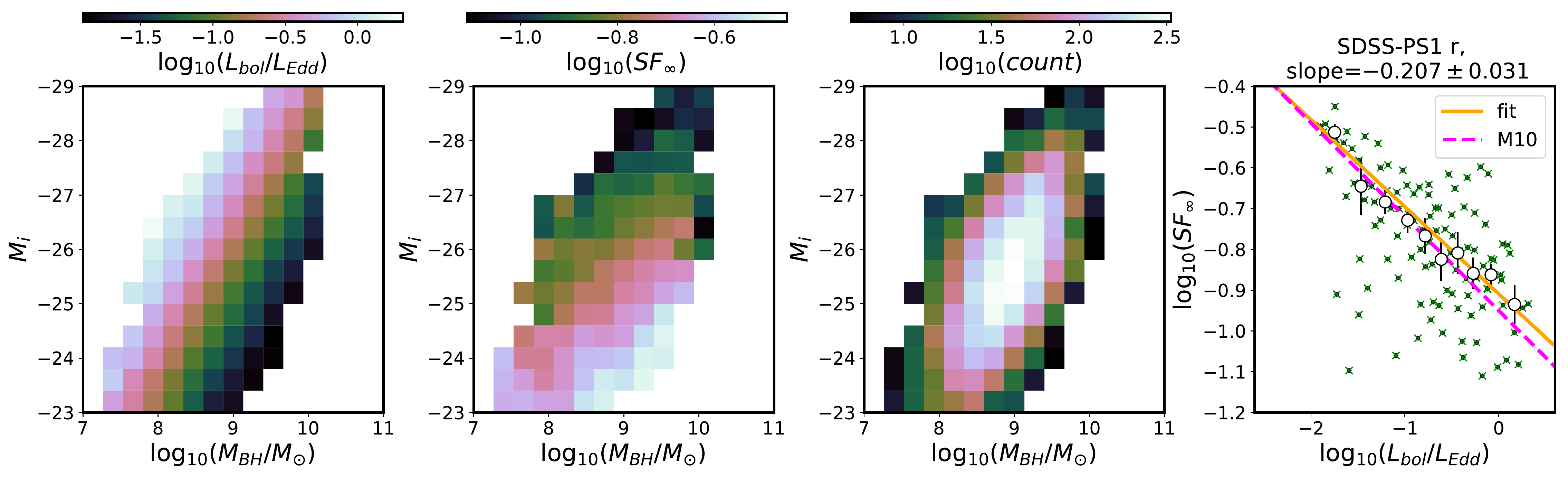}
	\caption{Absolute $i$-band magnitude $M_{i}$ as a function of black hole mass $M_{BH}$, where color encodes the Eddington ratio $f_{Edd} = L/L_{Edd}$ (first panel), variability amplitude SF$_{\infty}$  (second panel), or quasar count (third panel). 
 We only plot bins with more than five quasars. The fourth panel shows the median SF$_{\infty}$ as a function of the median $f_{Edd}$ (green crosses),  averaged in bins of $f_{Edd}$ (open circles). The bin width $w$ is found to ensure an equal number of points (crosses) per bin ($N$). The error bars are $\sigma_{y} = 1.25  \sigma_{G}(\mathrm{bin}) / N$, where $\sigma_{G}$ is the robust estimate  of the standard deviation ($\sigma_{G} = 0.7413 (Q_{75}-Q_{25})$). We assume the uncertainty along $f_{Edd}$ as $\sigma_{x} = w/\sqrt{12} $ (see \citealt{ivezic2014}). The solid orange line is the best-fit slope,  $-0.207 \pm 0.031$, with the slope uncertainty estimated from the standard deviation of the posterior samples. The best-fit slope agrees with the M10 results ($-0.23 \pm 0.03$), plotted as a dashed magenta line.}
	\label{fig:eddington_ratio}
\end{figure*}

The Eddington ratio $(f_{Edd} {=} L_{Bol}/L_{\mathrm{Edd}})$ encodes accretion strength: the proximity of quasar bolometric luminosity to the theoretical Eddington limit, where $L_{\mathrm{Edd}} {=} 1.26 {\times} 10^{38} (M_{\mathrm{BH}} / M_{\odot})$ erg s$^{-1}$ \citep{shen2011}. Since $\tau$ and SF$_{\infty}$ depend on $M_{i}$ and $M_{BH}$, we investigate the possibility of the Eddington ratio being the driver of these observed trends. In Figure~\ref{fig:eddington_ratio}, we show $f_{Edd} $ as a function of $M_{i}$, $M_{BH}$, and SF$_{\infty}$. The first two panels depict  $f_{Edd} $ and SF$_{\infty}$ binned as a function of $M_{i}$ and $M_{BH}$. The third panel shows the quasar counts, and the fourth panel shows the bin means (black dots). The means are further binned along  $f_{Edd} $ (as in M10). Combined SDSS--PS1 data support SF$_{\infty}$  being inversely related to  $f_{Edd} $, with a power-law slope of $-0.207 \pm 0.03$, consistent with  $-0.23 \pm 0.03$ reported by M10. Observations are generally consistent with basic predictions from Table~\ref{tab:theory}: $A$ increases with increasing  $M_{\mathrm{BH}}$, and  $f_{Edd} $ decreases as $L_{Bol}$ increases, while $\tau$ increases with  $L_{Bol}$. No model from  Table~\ref{tab:theory} is rejected. 

\begin{deluxetable*}{c|c}
	\tablecaption{Theoretical Predictions of Various Models Concerning Amplitude $A$ and/or Timescale $\tau$ of Variability, Eddington Ratio $f_{Edd}$, Bolometric Luminosity $L_{Bol}$, and Accretion Rate $\dot{m}$. \label{tab:theory}}  

	\tablehead{ \colhead{Theory} & \colhead{Prediction}}
	\startdata
	Standard thin disk \citep{shakura1973,netzer2013} &  $ A \nearrow$ as $f_{Edd} \searrow$ , $\tau \propto L_{Bol}^{1/2}$ (in \citealt{caplar2017}) \\
	Strongly inhomogeneous disk \citep{dexter2011} / local fluctuations \citep{cai2018} &  $ A \nearrow$ as $f_{Edd} \searrow$ \\
	Variations in global accretion rate \citep{hawkins2007,li2008,zuo2012} &  $ A \nearrow$ as $M_{BH} \nearrow$,  $ A \searrow$ as  $L_{Bol} \nearrow$ \\
	Eddington ratio reflecting AGN age \citep{martini2003,hopkins2005} & $A \nearrow$ as $f_{Edd} \searrow$ \\
	X-ray reprocessing \citep{kubota2018} & $A \nearrow$ as $f_{Edd} \searrow $, and  $A \nearrow$ as $\dot{m} \searrow$ \\  
	\enddata
    
    \tablecomments{Observations cannot reject any model.}
\end{deluxetable*}

\begin{deluxetable*}{c|cccccccc}
	\tablecaption{Comparison of Published Results on Correlating Observed Quasar Light-curve Properties: A Measure of Variability Amplitude A, and Timescale of Variability $\tau$ against the Physical Quasar Properties Black Hole mass $M_{\mathrm{BH}}$, Eddington Ratio $f_{\mathrm{Edd}}$, and Bolometric Luminosity $L_{Bol}$.  \label{tab:comparison}}  

	\tablehead{
	\colhead{Publication} & 
	\colhead{Measure of Amplitude / Timescale} & 
	\colhead{} &  \colhead{A vs.} & \colhead{} &  \colhead{} &  \colhead{$\tau$ vs.} &  \colhead{} & \\
	\colhead{} & \colhead{} & 
	\colhead{$M_{\mathrm{BH}}$} & \colhead{$f_{\mathrm{Edd}}$} &  \colhead{$L_{\mathrm{Bol}}$} & 
	\colhead{$M_{\mathrm{BH}}$} & \colhead{$f_{\mathrm{Edd}}$} &  \colhead{$L_{\mathrm{Bol}}$} 
	}
	\startdata
	\cite{wilhite2008} & SF (ensemble study) & + & - & - & x &x & x \\
	\cite{kelly2009} & DRW: $\tau$-decorrelation timescale, $\sigma$   & -- & 0 & !-- & + & x &  $\sim$+ \\
	\cite{macleod2010} & DRW: $\tau$,$\sigma$ & + & !-- & !-- & + & x & $\sim$+ \\ 
	\cite{morganson2014}\tablenotemark{a} & SF: A, $\gamma$ &  x & x & - & x & x & + \\
	\cite{kozlowski2016a} & DRW: $\tau$,$\sigma$  & x & x & !-- & + & x & $\sim$+ \\
	\cite{simm2016}\tablenotemark{b}& EV, and PSD (break timescale) & 0 & -- & -- &  0 & 0 & 0 \\ 
	\cite{caplar2017} & SF: $\tau$,$\sigma$ & $\sim$ & x & !-- & x & x & + \\
	\cite{rakshit2017}\tablenotemark{c} & DRW: $\tau$,$\sigma$ & + & !-  & $\sim$+ & x & x& x \\ 
	\cite{sun2018}\tablenotemark{d} & SF: $\tau$,$\sigma$ & x & x& !-- & x & + & !+ \\ 
	\cite{li2018}\tablenotemark{e} & SF: A, $\gamma$& $\sim$+  & --  & -- & + & x & + \\
	\cite{sanchez2018}\tablenotemark{f} & SF: A & 0 & -- & x & x & x & x \\
	\hline
	This work & DRW: $\tau$-decorrelation timescale, $\sigma$ & + & !-- & + & + & x & $\sim$+ \\
	\enddata

	\tablecomments{We list the correlations as positive (+), negative (--, i.e. anti-correlation), not found (0), or not studied (x). We further note if the correlation is strong (!), or weak ($\sim$).}
	\tablenotetext{a}{The 105,783 quasars with SDSS--PS1 sparse data,  SF parametrized as $V(\Delta t | A,\gamma) = A (\Delta t/  1 \mathrm{yr})^{\gamma}$; $\gamma$ is the increase of SF with time lag $\Delta t$}
	\tablenotetext{b}{The 90 X-ray-selected AGN, PS1 optical data. Variability characterized by normalized excess variance (EV). PSD characterized by break timescale. }
	\tablenotetext{c}{Narrow- and broad-line Seyfert 1 AGNs, $z<0.8$, CRTS data ($>50$ epochs, $5--9$ years baseline), DRW fitted with JAVELIN~\citep{zu2011}, no timescale correlations considered due to short baseline.  }
	\tablenotetext{d}{The 1004 SDSS quasars with  $0.5 \le z \le 0.89$}
	\tablenotetext{e}{The 119,305 sparse quasar light curves from DECaLS and SDSS; SF as in \cite{morganson2014} above.}
	\tablenotetext{f}{The 1348 QUEST-La Silla quasars, only amplitude of variability and excess variance.   }
\end{deluxetable*}

In summary, following the standard procedure of shifting the DRW parameter $\tau$ to the quasar rest-frame, and correcting for wavelength dependence, we compare our results to M10. We show in  Figure~\ref{fig:tau_sf_dist} that the new rest-frame distributions of $\tau$,$\sigma$ with SDSS, SDSS--PS1 are similar to those found by M10. There is a small ($<0.05$ dex) offset between the parameters fitted for SDSS light curves between this study and M10, that we attribute to data cleaning and software differences (Figure~\ref{fig:sigma_tau_ratios_M10}). When comparing SDSS and SDSS--PS1 combined data, we expect that there will be a long tail of objects that have a larger timescale and amplitude of variability with an extended baseline, corresponding to a true timescale that is longer than the initially probed time range with SDSS (Figure~\ref{fig:sigma_tau_ratios}). Changing-look quasars belong to the tail end of quasar variability (Sec.~\ref{sec:outliers}), and we suggest 40 candidates in Appendix ~\ref{app:clqso_cands}, as well as a quick diagnostic plot of differences in mean magnitudes and rms scatter between PS1 and SDSS light-curve portions that could also be used to hunt for CLQSOs independently of DRW fitting (Figure~\ref{fig:median_offsets}). We explore the correlation between the DRW parameters obtained for SDSS and SDSS--PS1 light curves and quasar physical properties. We find that the general trends are in agreement with M10 at the $2\sigma$ level, with the largest deviations caused by different pipeline procedures, as outlined in Table~\ref{tab:coefficients}.

\section{Discussion}
\label{sec:discussion}
\subsection{Trends with Eddington ratio}
Anticorrelation of  variability amplitude with Eddington ratio  has a variety of possible theoretical explanations. In the thin disk theory \citep{shakura1973, frank2002, netzer2013}, the radius of the emission region at a given wavelength increases with Eddington ratio and is inversely proportional to temperature \citep{rakshit2017}. Thus, a hotter disk means that the emission observed in a given bandpass is emitted from a larger radius. From causality, a smaller region can be more variable than a larger one. Therefore, a  hotter disk would be less variable at a given wavelength than a colder one, and  the variability amplitude as studied in a particular bandpass (here SDSS $r$ band) would be anticorrelated with Eddington ratio \citep{fausnaugh2016,edelson2015}. 

On the other hand, in the strongly inhomogeneous disk model, independent temperature fluctuations in $N$ zones drive the variability \citep{dexter2011}. In that framework, the inverse trend of variability amplitude against $L/L_{Edd}$  and $L_{Bol}$  can be understood qualitatively if more luminous quasars also have a higher mass accretion rate and thus a greater number of disk inhomogeneities, resulting in smaller flux variability \citep{simm2016}. The inhomogeneous disk model was consistent with mean SDSS spectral analysis in \citet{ruan2014} but was not a preferred explanation for \citet{kokubo2015}. 

Both \cite{rumbaugh2018} (with the Dark Energy Survey SF study) and  \citet{sun2018}  (with a low-$z$ subsample of S82 SDSS quasars) confirmed the anticorrelation between quasar variability and luminosity. However, \citet{graham2020} did not find support for this trend with the sample of extremely variable quasars (EVQs) in the CRTS dataset,  but when selecting for lower-luminosity sources ($M_{V} < -23$), the anticorrelation is recovered. This agrees with an interpretation that a dwindling fuel supply may correspond to higher variability. Furthermore, \citet{sanchez2018} combined the SDSS spectra with the 5 yr light curves of 2345 quasars obtained with the Quasar Equatorial Survey Team (QUEST)-La Silla AGN Variability Survey, and  using the Bayesian parametrization of SF \citep{schmidt2010}, they also found that the amplitude of variability $A$ is anticorrelated with rest-frame emission wavelength  and Eddington ratio (also see \citealt{simm2016}, \citealt{rakshit2017}). See Table ~\ref{tab:comparison} for a comparison of published studies correlating the statistical measures of observed quasar light curve variability (amplitude/timescale),  and the physical quasar properties (black hole mass, bolometric luminosity). 

Indeed, $f_{Edd}$ is a proxy for the strength of accretion, which, together with orientation, may be the key to explaining the quasar main sequence (QMS; \citealt{shen2014, marziani2018}). The QMS, defined by so-called Eigenvector-1, is the anticorrelation between the broad-line Fe{\sc ii} emission and the strength of the narrow O{\sc iii} ($5007$ $\mbox{\AA}$) line \citep{wang1996}. An analysis of quasar clustering by \citet{shen2014}, later confirmed by \citet{sun2015} with measurements of  black hole mass from the quasar host galaxy stellar dispersion \citep{ferrarese2000, kormendy2013}, showed that the entire diversity of quasars in the QMS can be explained by the variation in accretion (affecting $R_{\mathrm{Fe  II}}$, the ratio of the  Fe{\sc ii} equivalent width between $4435$ and $4685$ $\mbox{\AA}$ and H$\beta$) or orientation effects (affecting the FWHM of the H$\beta$). However, \citet{panda2019a, panda2019b} found that these are insufficient, and variations in metallicity, as well as a range of cloud densities and turbulences, are required.  \cite{jiang2016} also found that metallicity, and in particular the iron opacity bump, may have a strong influence on the stability of an accretion disk, and thus linking metallicity to AGN variability. This is also consistent with the findings of \cite{sun2018}: quasars with high  Fe{\sc ii} strength have higher metallicity and have more stable disks.

\subsection{Variability Timescales}

In the era of changing-look active galaxies (including initially distinct classes of changing-look quasars \citep{lamassa2015, macleod2019}, changing-look AGNs (CLAGN, see \citealt{bianchi2009, risaliti2009,marchese2012}), and changing-look LINERS \citep{frederick2019}, to name a few), there is a revived interest in possibly linking the behavior of stellar-sized accreting systems (e.g., black hole binaries) to that of galactic-scale systems (e.g., AGNs, QSOs, LINERS; \citealt{noda2018, ruan2019}). 

Several relevant timescales are involved, and there are various interlinked mechanisms that could drive the variability. A standard optically thick, geometrically thin, $\alpha$-disk model has a hierarchy of timescales: dynamical, thermal, front, and viscous, with   $t_{\mathrm{dyn}} < t_{\mathrm{th}} < t_{\mathrm{front}}  < t_{\mathrm{visc}} $ \citep{netzer2013, frank2002}. We proceed to briefly describe  each timescale, concluding with our interpretation of the mechanism that could drive the variability observed from the data. 

The dynamical, or gas orbital, timescale is simply  an inverse of the Keplerian orbital angular frequency $ \Omega$  at radius $R$: 

\begin{equation}
t_{\mathrm{dyn}} {\sim}  1 / \Omega = \left( \frac{GM}{R^{3}}\right)^{-1/2}
\end{equation}

The main parameter  describing the accretion disk is $\alpha$, the ratio of the (vertically averaged) total stress to thermal (vertically averaged) pressure: 

\begin{equation}
\alpha= \frac{\langle \tau_{r\varphi}  \rangle_{z} }{\langle P \rangle _{z}} 
\end{equation}

After \cite{lasota2016},  the hydrodynamical stress tensor (corresponding to  kinematic viscosity $\nu$) is

\begin{equation}
\tau_{r\varphi } = \rho \nu \frac{\partial v_{\varphi}}{\partial R} = \rho \nu \frac{d \Omega}{d \ln{R}} = \frac{3 \rho \nu \Omega}{2}  
\end{equation}

so  with  $c_{s}$ (local sound speed) at radius $R$ (isothermal sound speed is $c_{s} = \sqrt{P/\rho}$),

\begin{equation}
\alpha  =   \frac{3 \rho \nu \Omega}{2 P} =  \frac{3 \Omega \nu}{2 c_{s}^{2}}
\end{equation}

This means that a smaller $\alpha$ corresponds to less viscous disks.

The thermal timescale, related to the time needed for readjustment to the thermal equilibrium (derived in detail in \citealt{frank2002}), is the ratio of heat content per unit disk area to dissipation rate per unit disk area: $(dE / A) / (dE/dt /  A) = dt $.  The heat content per unit volume is ${\sim} \rho k T / \mu m_{p} {\sim} \rho c_{s}^{2}$, and the heat content per unit area is  ${\sim} \rho c_{s}^{2} / h {\sim} \Sigma c_{s}^{2}$. Meanwhile, the dissipation rate per unit area, $D(R)$, is 

\begin{equation}
D(R) = \frac{9}{8} \nu \Sigma R^{-3} G M
\end{equation}

(eq. 4.30 in \citealt{frank2002}), so :

\begin{equation}
t_{\mathrm{th}} {\sim} \frac{c_{s}^{2}R^{3}}{G M \nu } = \frac{c_{s}^{2}}{\nu \Omega} = \frac{t_{\mathrm{dyn}}}{\alpha}
\end{equation}

Thus, if the disk is inviscid ($\nu \rightarrow 0$), then $t_{\mathrm{th}}\rightarrow\infty$; i.e., there is no contact with adjacent disk elements. 

The cooling and heating fronts propagate through the disk at  $\alpha c_{s} $ \citep{hameury2009}; in that description,  with no viscosity, there is no communication between neighboring disk annuli and thus no front propagation \citep{balbus1998, balbus2003}. Following \cite{stern2018}, if we define the disk aspect ratio as $h/R$ with a disk height $h = c_{s} / \Omega$, the characteristic time for front propagation is

\begin{equation}
t_{\mathrm{front}} {\sim} (h/R) ^ {-1} t_{\mathrm{th}}
\end{equation}

The viscous timescale is the characteristic time it would take for a parcel of material to undergo a radial transport due to the viscous torques from the radius $R$ to the black hole \citep{czerny2006}. Note that while viscosity probably has a magnetic origin \citep{eardley1975, grzedzielski2017}, in this simplistic order-of-magnitude estimate, we use a hydrodynamical description of accretion flow.  With $\nu = \eta / \rho$ (kinematic viscosity being the ratio of dynamical viscosity to density), \cite{frank2002} showed (chapter 5.2) that 

\begin{equation}
t_{\mathrm{visc}} {\sim} R^{2} / \nu {\sim}  R / v_{R} = (h/R)^{-2} t_{\mathrm{th}}
\end{equation}

 We can parametrize each timescale for a black hole mass $M_{\mathrm{BH}} = 10^{8} M_{\odot}$, at $R {\sim} 150 r_{g}$, with the gravitational radius $r_{g} = GM_{\mathrm{BH}} / c^{2} {\sim} 4 \, \mathrm{au}$, using Equations (5)-(8) in \cite{stern2018}:

 \begin{equation}
 t_{\mathrm{dyn}} {\sim} 10  \mathrm{days} \left(\frac{M_{\mathrm{BH}}}{10^{8} M_{\odot}} \right) 
 \left( \frac{R}{150 r_{g}}\right) ^{3/2} 
 \end{equation}

 \begin{equation}
 t_{\mathrm{th}}   {\sim} 1 \,\mathrm{year} \left( \frac{\alpha}{0.03}\right)^{-1}  
 \left( \frac{M_{\mathrm{BH}}}{10^{8} M_{\odot}}\right) \left( \frac{R}{150 r_{g}}\right)^{3/2} 
 \end{equation}

  \begin{eqnarray}
  t_{\mathrm{front}} {\sim} 20 \,\mathrm{years} \left( \frac{h/R}{0.05}\right)^{-1}   \left( \frac{\alpha}{0.03}\right)^{-1}  \nonumber  \\ 
  \left( \frac{M_{\mathrm{BH}}}{10^{8} M_{\odot}}\right)     \left( \frac{R}{150 r_{g}}\right) ^{3/2} 
 \end{eqnarray}

  \begin{eqnarray}
  t_{\mathrm{visc}}  {\sim} 400 \, \mathrm{years} \left( \frac{h/R}{0.05}\right)^{-2}   \left( \frac{\alpha}{0.03}\right)^{-1} \nonumber  \\  
  \left(\frac{M_{\mathrm{BH}}}{10^{8} M_{\odot}} \right)     \left( \frac{R}{150 r_{g}}\right) ^{3/2}  
 \end{eqnarray}

In summary,  of the  considered timescales, only the thermal and dynamical are short enough to be related to  the observed short-term stochastic variability. It may be that the variability on the scale of days is driven by local changes and that on the longer scale (perhaps hundreds of days) by a different mechanism \citep{kokubo2015}. The other time scales may be more related to the dramatic changes in brightness of the continuum as observed in changing-look AGNs. Indeed, \citet{noda2018} favored a change in mass accretion rate, followed by a propagation of the cooling front \citep{simm2016,lawrence2018}. \citet{noda2018} also suggested that perhaps some short-term variability could be related to the amount of the disk swept by the thermal front propagation due to hydrogen ionization instability, similar to white dwarf systems (see also \citealt{ross2018, ruan2019, sniegowska2019}).

The variability on a several-year timescale could also be explained by the X-ray reprocessing model \citep{kokubo2015, kubota2018}, assuming that the AGN UV--optical variability is a result of reprocessing of X-ray or far-UV emission \citep{krolik1991}.  The idea of X-ray reprocessing  over time has gained more and more support, with evidence from simultaneous X-ray--UV--optical AGN time series \citep{edelson2014, mchardy2018,  zhu2018}. In particular, the accretion disk blackbody emission is insufficient to explain the broadband AGN spectrum. The total spectral energy distribution with a soft X-ray excess and a hard X-ray tail requires additional sources of emission. A recent model by \citet{kubota2018} divides the flow into blackbody emission, a warm Comptonization region (the disk), and a hard X-ray hot Comptonization component (the corona, or a hot material filling the region close to the black hole below the truncation radius).  Since the soft X-rays are correlated with the hard X-rays, at least part of the picture consists of reflection or reprocessing of hard X-rays by the disk \citep{lawrence2018}. This model predicts an increase of variability amplitude (SF$_{\infty}$) with $M_{\mathrm{BH}}$, and adds the insight that the observed slope is due to changes in accretion rate $\dot{m}$, explaining that smaller $\dot{m}$ corresponds to the highest variability. This qualitatively agrees with the picture that a dwindling fuel supply makes the flow more variable. Previous worries about X-ray reprocessing concerned the seemingly insufficient solid angle subtended by the source of the hard X-rays to cause the observed soft X-ray and optical response. This is addressed by realizing that reprocessing could be taking place in the extended region \citep{gardner2017}, such as an inflated inner disk (corresponding to a warm Comptonizing region in \citealt{kubota2018}, or even the BLR region, serving as an additional `complex reprocessor'\citep{mchardy2018}. Also, for \citet{panda2019a}, a warm corona helps decrease the dependence of $R_{\mathrm{Fe  II}}$ on $f_{\mathrm{Edd}}$. 

Thus, while CLAGN may be related to the state change to Advection-Dominated Accretion Flow \citep{sniegowska2019}, similar to that of X-ray binaries \citep{noda2018,ruan2019}, with cooling and heating fronts \citep{ross2018}, the short-timescale variability requires approximately three distinct emission regions \citep{kubota2018} with an extended reprocessor (such as a diffuse, hot, puffed-up inner disk and BLR; \citealt{mchardy2018}) that reverberates the rapid hard X-ray variability in soft X-rays to the optical via UV \citep{fausnaugh2018}. Some emission (especially soft X-rays) seems to require the warm Comptonizing corona \citep{kubota2018}. The warm corona, coupled with metallicity changes and variation in turbulence level and cloud density, also helps explain the QMS in the optical \citep{panda2019a,panda2019b}. Finally, the \citet{kubota2018} model, apart from being consistent with other mechanisms \citep{lawrence2018, mchardy2018, panda2019a, ross2018, ruan2019, sniegowska2019}, explains the observed correlation of variability amplitude with black hole mass as corresponding to variations in mass accretion rate.

\section{Summary and Conclusions}
\label{sec:conclusions}
We model the optical variability of ${\sim} 9000$ S82 quasars as the DRW \citep{kelly2009}. The DRW is a GP, described by two parameters: characteristic timescale $\tau$ (representing the decorrelation timescale, or light-curve smoothness) and the asymptotic amplitude SF$_{\infty}$ (which relates to the amplitude of variability). We fit observed and simulated light curves with \project{celerite} - a fast GP solver \citep{foreman2017}. By simulating and fitting DRW light curves, we explore the impact of the ratio of input timescale and the light-curve baseline. We find that the light-curve length needs to be several times larger than the input timescale to allow unbiased timescale retrieval, confirming K17. Motivated by this result, we consider extending SDSS with PS1, PTF, CRTS, and ZTF data.  We calculate appropriate photometric offsets (color terms) to relate PS1 $gri$, PTF $gR$,  CRTS $V$, and ZTF $r$ to the SDSS $r$ band. However, due to larger photometric uncertainties of PTF, ZTF, and CRTS at the faint magnitudes of SDSS quasars, we decided to use only PS1 $r$-band data. Furthermore, the SDSS and PS1 $r$ bands are sufficiently similar that no photometric transformation is required.  Thus, by extending the SDSS $r$-band light curves with PS1 DR2 $r$-band data, we improve upon the fidelity of recovered DRW parameters (e.g., in Figure~\ref{fig:lc_simulated_results}, showing a simulated population of $\tau=575$ days, the rms of $\log{(\tau_{fit}/ \mathrm{baseline})}$ decreases from 1.75 dex with SDSS to 1.5 dex with PS1, and in the future, with the inclusion of ZTF and LSST data, it will decrease to $\sim 1$ dex).  

We identify 40 objects that exhibit a tenfold increase in variability timescale when using the SDSS--PS1 data set, as compared to the timescale inferred from SDSS alone. Their light curves show characteristics of changing-look quasars (magnitude difference larger than 0.5 mag; \citealt{macleod2016}). Of these, five are confirmed in the literature ~\citep{macleod2019, lamassa2015}. We recommend spectroscopic follow-up and further monitoring of the brightest targets (see Appendix~\ref{app:clqso_cands}).

We investigate the correlation of quasar physical properties, such as black hole mass $M_{\mathrm{BH}}$ and absolute $i$-band magnitude $M_{i}$,  with DRW model parameters. The SDSS--PS1 data, coupled with the \citet{shen2011} quasar catalog, imply that the damping timescale $\tau$  is correlated with $M_{\mathrm{BH}}$ with a power-law index of $0.141 \pm 0.019$ and almost independent of quasar bolometric luminosity as in M10, \citet{wilhite2008}, and \citet{vandenberk2004}. The asymptotic variability amplitude SF$_{\infty}$ is \textbf{correlated} with $M_{i}$ (i.e. anticorrelated with luminosity) with a power-law index of $0.118 \pm 0.003$, and correlated with $M_{\mathrm{BH}}$ with a slope of $0.118\pm 0.008$. This can be explained if the driving variable was the Eddington ratio, $f_{Edd}$ \citep{wilhite2008}. Indeed, there is an anticorrelation of SF$_{\infty}$ and $f_{\mathrm{Edd}}$, with a power-law slope of $-0.207 \pm 0.031$ (similar to M10).  As suggested by \citet{kubota2018}, this gradient of SF$_{\infty}$ in the plane of $M_{\mathrm{BH}}$ vs $M_{i}$ could be explained if the lower mass accretion rate corresponds to higher variability, so that when the supply of fuel decreases, the flow becomes less stable and more clumpy and inhomogeneous ~\citep{rakshit2017, kokubo2015, dexter2011}.  This is also consistent with the X-ray reprocessing model, whereby the hard X-ray variability of the inner disk is reflected/reprocessed by the extended warm Comptonization region (inflated disk) and perhaps a complex reprocessor, including the clouds of the BLR~\citep{kubota2018, panda2019b}. Changes on recovered timescales are too fast to be driven by changes in disk viscosity or thermal front propagation alone; a thermal or dynamical timescale of response to the changes in X-ray emission seems most consistent with our results ~\citep{stern2018}.

More data extending the light curves would help improve the DRW fit coefficients, potentially decreasing the scatter in the observed correlations - for instance, for simulated $\tau$ recovery, an improvement of the rms by a factor of 1.8 (from 0.322 for SDSS only to 0.182 with combined SDSS--PS1--ZTF--LSST). Moreover, given that the uncertainty in black hole mass is one of the biggest sources of error, better measurements of quasar properties would be of high utility~\citep{shen2011}.  This will be possible with the upcoming AGN reverberation mapping campaigns (e.g., SDSS-V black hole mapper), providing better calibration for line width--based methods of estimating black hole masses~\citep{kollmeier2017}. All quasars in this study were spectroscopically confirmed, but some spectra had low signal-to-noise ratios, resulting in a higher likelihood of incorrect redshift measurement. Better spectroscopy and follow-up of S82 quasars, afforded by SDSS-V panoptic spectroscopy, would not only help improve on the spectrum-based  properties (redshift, absolute magnitude, black hole masses) but also allow the study of spectral changes and further new CLAGN discoveries~\citep{macleod2019}. 

If this study were to be expanded onto a sample of quasars with good photometry over sufficiently long baselines but lacking spectral information, the required physical information on quasars could be obtained by indirect methods of estimating the coarse spectral information from broadband photometry~\citep{kozlowski2015}. This would benefit from better catalogs of existing spectroscopically confirmed quasars (SDSS DR14) to improve the calibration, as well as better methods of estimating the redshift based on photometry alone (e.g., photo-$z$; \citealt{richards2015,yang2017,curran2019,jin2019}).  This will be possible in the short term with the ZTF \citep{bellm2019} and in the long term with the LSST~\citep{ivezic2019}. Occasional coverage adding a few epochs to some quasars may be possible with other surveys (e.g., TESS; \citealt{ricker2014}), but to improve the statistics of an entire sample of S82 quasars would require longer baselines. Combining SDSS and PS1 with LSST would provide an unprecedented 35 yr baseline, which, assuming timescales below 1000 days, is over 10 times longer, allowing unbiased DRW parameter retrieval. This, coupled with correlations with quasar properties, would provide an estimate of black hole masses and bolometric luminosities for millions of quasars~\citep{ivezic2019}.

\section{Acknowledgments}

The Pan-STARRS1 Surveys (PS1) and the PS1 public science archive have been made possible through contributions by the Institute for Astronomy, the University of Hawaii, the Pan-STARRS Project Office, the Max Planck Society and its participating institutes, the Max Planck Institute for Astronomy, Heidelberg, and the Max Planck Institute for Extraterrestrial Physics, Garching, The Johns Hopkins University, Durham University, the University of Edinburgh, the Queen's University Belfast, the Harvard-Smithsonian Center for Astrophysics, the Las Cumbres Observatory Global Telescope Network Incorporated, the National Central University of Taiwan, the Space Telescope Science Institute, the National Aeronautics and Space Administration under grant No. NNX08AR22G issued through the Planetary Science Division of the NASA Science Mission Directorate, the National Science Foundation grant No. AST-1238877, the University of Maryland, Eotvos Lorand University (ELTE), the Los Alamos National Laboratory, and the Gordon and Betty Moore Foundation.

Funding for the Sloan Digital Sky Survey IV has been provided by the Alfred P. Sloan Foundation, the U.S. Department of Energy Office of Science, and the Participating Institutions. The SDSS-IV acknowledges support and resources from the Center for High-Performance Computing at
the University of Utah. The SDSS website is www.sdss.org. The SDSS-IV is managed by the Astrophysical Research Consortium for the  Participating Institutions of the SDSS Collaboration, including the  Brazilian Participation Group, the Carnegie Institution for Science, the Carnegie Mellon University, the Chilean Participation Group, the French Participation Group, the Harvard-Smithsonian Center for Astrophysics, Instituto de Astrof\'isica de Canarias, The Johns Hopkins University, the Kavli Institute for the Physics and Mathematics of the Universe (IPMU) / 
University of Tokyo, the Korean Participation Group, Lawrence Berkeley National Laboratory, 
Leibniz Institut f\"ur Astrophysik Potsdam (AIP),  Max-Planck-Institut f\"ur Astronomie (MPIA Heidelberg), Max-Planck-Institut f\"ur Astrophysik (MPA Garching), Max-Planck-Institut f\"ur Extraterrestrische Physik (MPE), National Astronomical Observatories of China, New Mexico State University, New York University, the University of Notre Dame, Observat\'ario Nacional / MCTI, the Ohio State University, Pennsylvania State University, Shanghai Astronomical Observatory, 
United Kingdom Participation Group, Universidad Nacional Aut\'onoma de M\'exico, the University of Arizona, the University of Colorado Boulder, the University of Oxford, the University of Portsmouth, 
the University of Utah, the University of Virginia, the University of Washington, the University of Wisconsin, Vanderbilt University, and Yale University.

The CSS survey is funded by the National Aeronautics and Space Administration under grant No. NNG05GF22G issued through the Science Mission Directorate Near-Earth Objects Observations Program.  The CRTS survey is supported by the U.S.~National Science Foundation under grants AST-0909182.

Based on observations obtained with the Samuel Oschin 48 inch telescope at the Palomar Observatory as part of the Zwicky Transient Facility project. The ZTF is supported by the National Science Foundation under grant No. AST-1440341 and a collaboration including Caltech, IPAC, the Weizmann Institute for Science, the Oskar Klein Center at Stockholm University, the University of Maryland, the University of Washington, Deutsches Elektronen-Synchrotron and Humboldt University, Los Alamos National Laboratories, the TANGO Consortium of Taiwan, the University of Wisconsin at Milwaukee, and Lawrence Berkeley National Laboratories. Operations are conducted by COO, IPAC, and UW.

\appendix
\section{Measuring Quasar properties} 
\label{app:measureBHmass}
In this work, we employ black hole masses, bolometric luminosities, and $k$-corrections from the \cite{shen2011} catalog based on single-epoch SDSS spectra. Here we explain the choices made in the difficult art of estimating each of these quasar physical properties.

It is nontrivial to measure the mass of black holes living in the centers of active galaxies, even provided with a detailed spectrum. The most common  approach to estimate black hole masses in AGNs is to assume that the BLR is virialized,

\begin{equation}
M_{\mathrm{BH}} = f \frac{ R\Delta V^{2} }{G} = f M_{\mathrm{vir}}
\end{equation}

where $f$ is a constant of order unity, $R$ is the size of the BLR (estimated from  emission line lag $\Delta t$ as $R = c \Delta t$), $\Delta V$ is the virial velocity, and $G$ is the gravitational constant \citep{shen2008}.  From reverberation mapping (RM) studies (e.g., \citealt{shen2019}), we know that continuum luminosity $L$ is related to the size of the BLR region as $R \propto L^{\gamma}$ \citep{vestergaard2006}, with $\gamma$ very close to $1/2$ (e.g., \citealt{bentz2009} found from RM studies $\gamma = 0.519 \pm 0.06$). Thus, we find  that $R \Delta V^{2} \propto L^{\gamma} \Delta V^{2} \equiv \mu$. The virial velocity $\Delta V$ is usually estimated from the width of the broad emission lines (or line dispersion).  In the absence of a quasar spectrum, there are alternative methods using a conversion of the broadband photometry into monochromatic fluxes in the vicinity of reverberating lines (e.g., \citealt{kozlowski2015}, used in \citealt{kozlowski2017b} to estimate black hole masses for 280,000 AGNs). Depending on the redshift, different rest-frame calibrated emission lines shift into the observed passband: broad H$\alpha$ at  $6562\mbox{\AA}$, H$\beta$ at  $5100 \mbox{\AA}$, Mg\,{\sc ii} at $2800\mbox{\AA}$, and C\,{\sc iv} at $1350\mbox{\AA}$ (see Figure 7 in \citealt{shen2019}, and \citealt{vestergaard2002}). Some authors even separately consider  C\,{\sc iv}- and  Mg\,{\sc ii}-based  black hole mass estimates. We refer the reader to \citet{shen2008}, who described in detail the various biases and inherent assumptions of virial black hole mass measurements.

Another important quasar property, bolometric luminosity, is most often estimated from the absolute $i$-band magnitude, $M_{i}$ (see \citealt{shen2008}, Figure 2). Here $M_{i}$ is derived from the observed $i$-band magnitude by correcting for Galactic extinction and the fact that at different redshifts, different portions of the spectral energy distribution are observed by the telescope filter bandpass. The latter, known as $k$-correction $K(z)$ \citep{oke1968},  is defined as $m_{\mathrm{intrinsic}} = m_{\mathrm{observed}} - K(z)$. In the early 2000s, the common approach was to $k$-correct to a redshift of zero, but as \citep{richards2006a} pointed out, since the distribution of quasars peaks at a redshift of 2, for most quasars, correcting to a redshift of zero required shifting the observed spectrum into the far infrared. Moreover, the procedure was to correct separately for the continuum and emission line contributions, assuming a particular spectral shape (e.g., power law  $f_{\nu} \propto \nu^{\alpha}$, with $\alpha=-0.5$; see \citealt{vandenberk2001,richards2006a,schneider2010}).  This introduces a larger error for $K(z=0)$ than for $K(z=2)$ if the assumed spectral shape $\alpha=-0.5$ is far from the real spectral index. In early 2010s, after  \citealt{richards2006a, wisotzki2000, blanton2003},  the practice started shifting toward $k$-correcting to a redshift of 2,  and including custom quasar spectral shapes, as reflected by the content of the \cite{shen2011} quasar catalog. Thus, in this study, we use  the absolute $i$-band magnitude $k$-corrected to $z=2$: $M_{i}(z=2)$.

These methods were used to create catalogs of quasar properties derived from spectra. Since quasars are variable at the ${\sim}0.2$ mag level, the ideal is to use a single-epoch calibrated spectrum to estimate the continuum luminosity and find virial black hole masses using relationships based on the monochromatic fluxes and broad line widths described above. A glance at the available quasar catalogs reveals that, given any SDSS data release, there is indeed first a catalog of basic quasar properties (redshift and photometry; e.g., \citealt{schneider2007, schneider2010}), and more detailed catalogs containing black hole masses and bolometric luminosities  follow (e.g., \citealt{shen2008, shen2011}). More recently, once the SDSS DR12 Quasar Catalog \citep{paris2017} was released,  K17 followed using SDSS photometry as a proxy for monochromatic luminosities. \citet{chen2018} added a detailed analysis of continuum luminosities in the  H$\alpha$, H$\beta$ regions for low-redshift quasars. Using the spectra from the Chinese LAMOST survey, \citet{dong2018} also sought to estimate virial black hole masses, and the results, while consistent with \citet{shen2011}, suffered from the necessity of pegging the noncalibrated spectra to the SDSS photometry, which was taken at a different epoch. Thus, even though the SDSS DR12 Quasar Catalog of \citet{paris2018}  is the most recent, like \citet{paris2017}, it lacks black hole masses and bolometric luminosities, and there is no recent work that reanalyzed the spectral data. Therefore, we use black hole mass estimates and monochromatic luminosities from \citet{shen2011} based directly on single-epoch spectra.

\section{CLQSO candidates}
\label{app:clqso_cands}

\begin{figure*} 
	\plotone{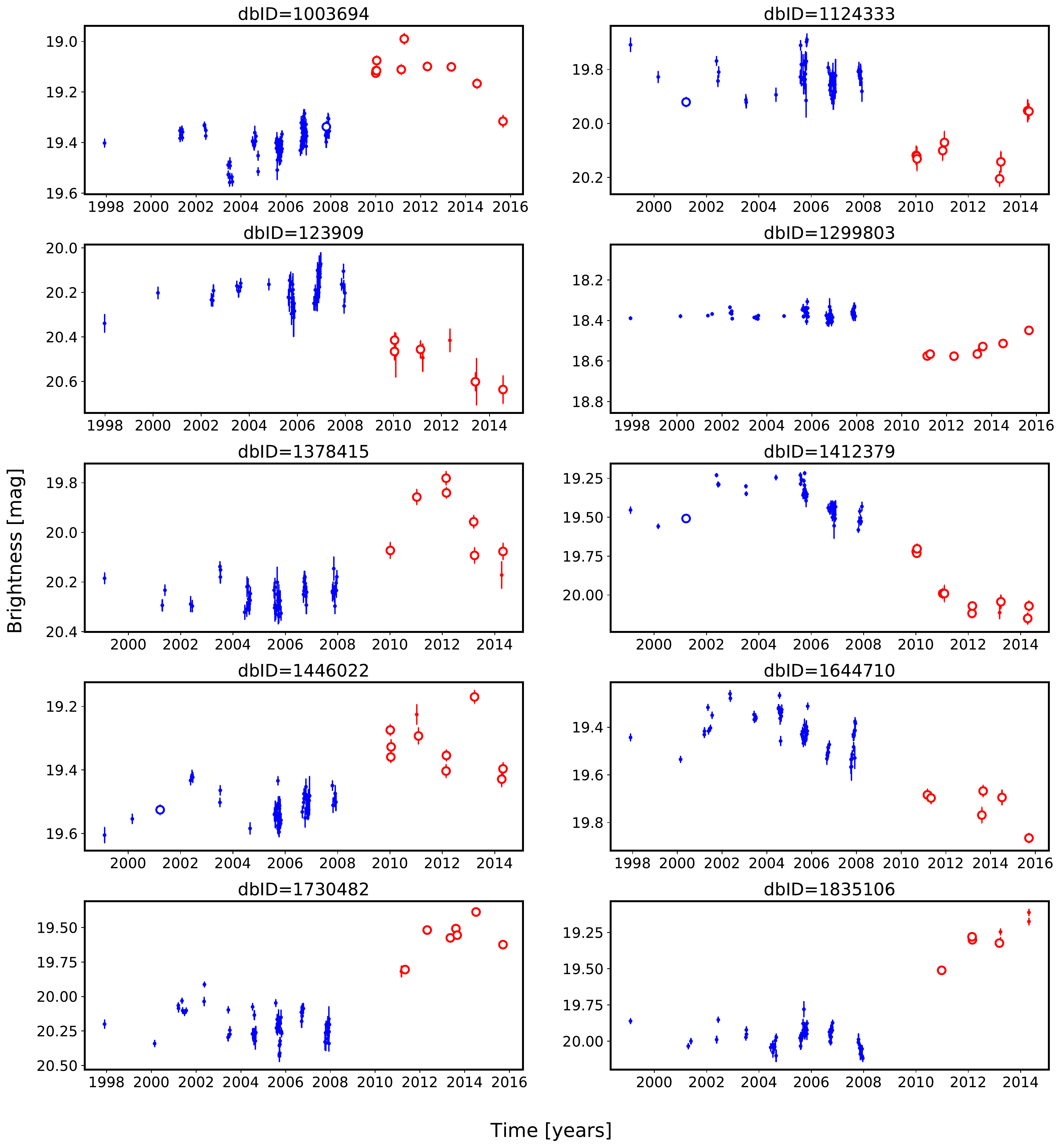}
	\caption{Outliers in the space of the recovered DRW parameters between SDSS and SDSS--PS1, as well as median offsets. Page 1/4 (continued on Figure~\ref{fig:clqso2}).}
	\label{fig:clqso1}
\end{figure*}

\begin{figure*}
\plotone{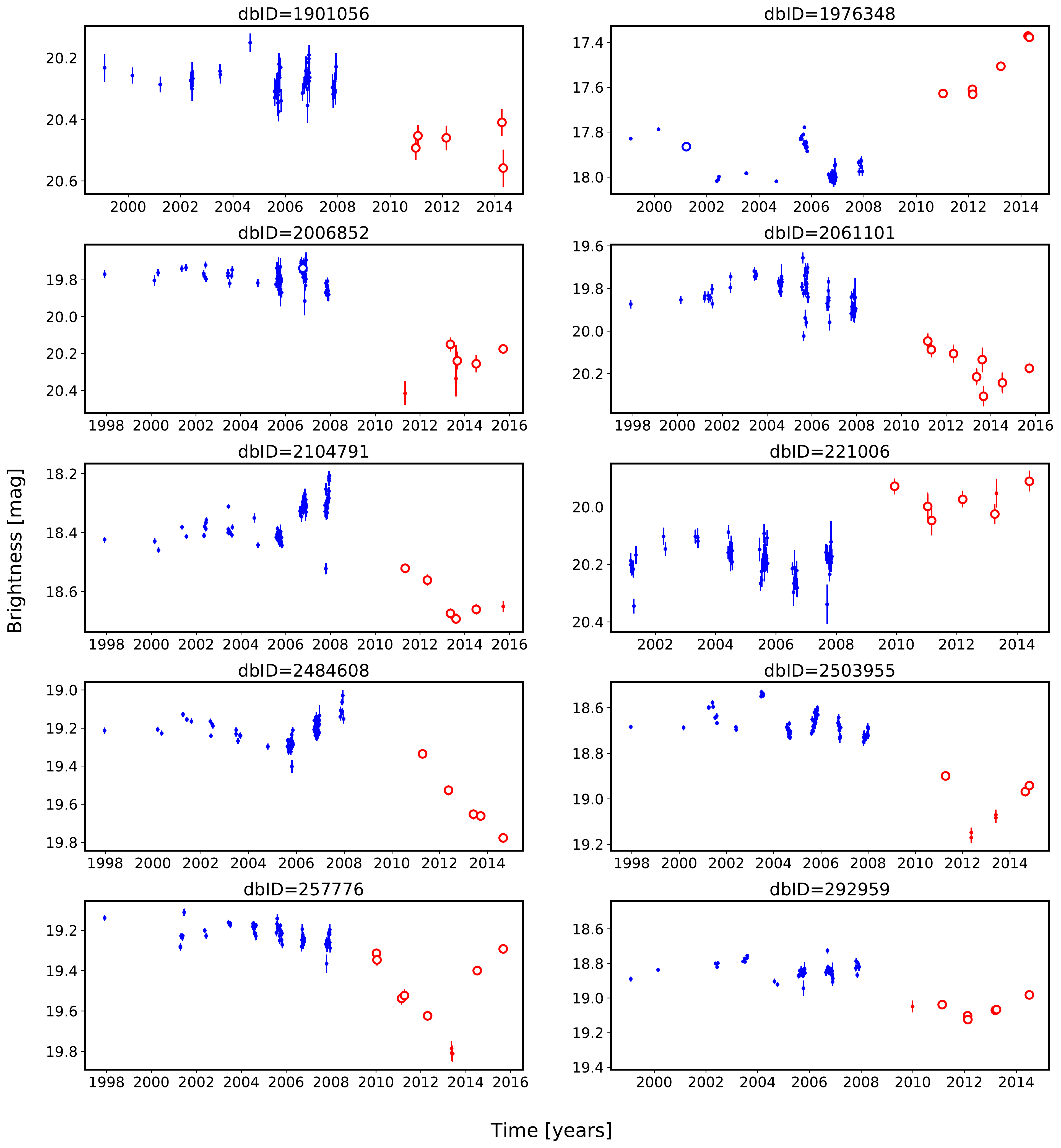}
\caption{As Figure~\ref{fig:clqso1}, page 2/4 (continued on Figure~\ref{fig:clqso3}). }
\label{fig:clqso2}
\end{figure*}

\begin{figure*}
\plotone{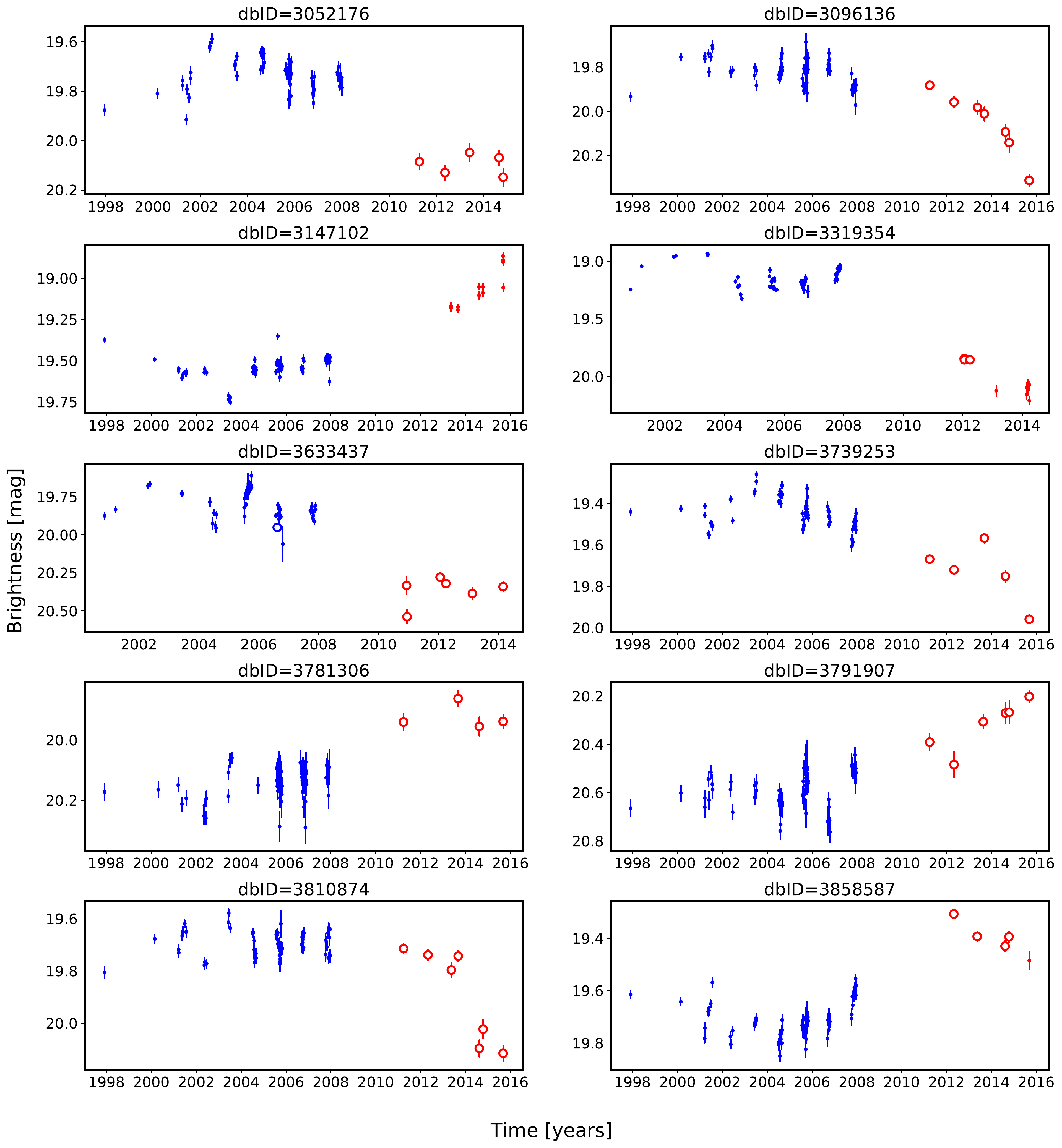}
\caption{As Figure~\ref{fig:clqso1}, page 3/4 (continued on Figure~\ref{fig:clqso4}). }
\label{fig:clqso3}
\end{figure*}

\begin{figure*}
\plotone{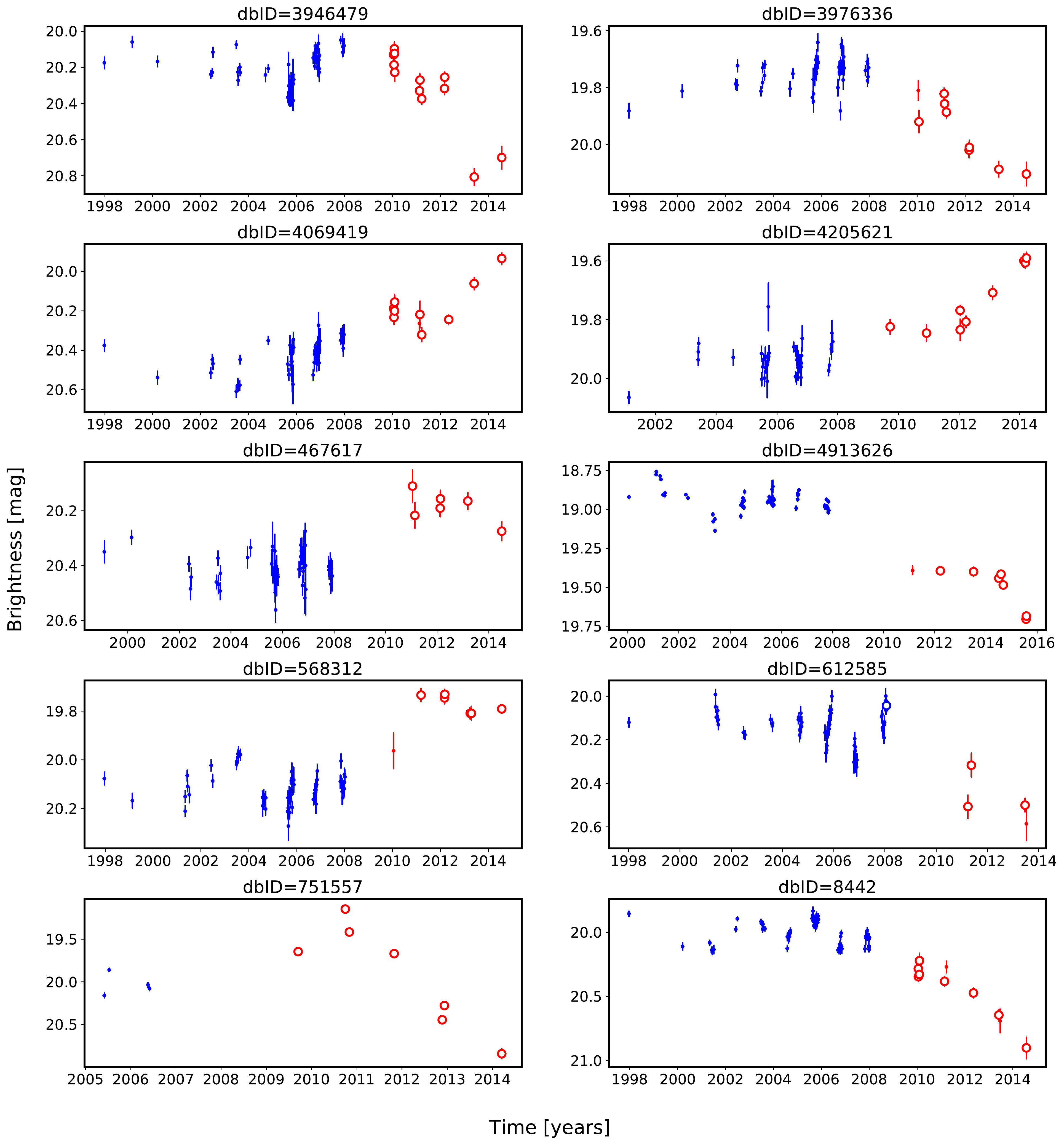}
\caption{As Figure~\ref{fig:clqso1}, page 4/4. }
\label{fig:clqso4}
\end{figure*}

Based on the DRW model  parameters $\tau$, $\sigma$ fitted with \project{celerite} using the SDSS and PS1 data, we find that there are quasars for which there is a pronounced difference between $\tau$, $\sigma$  inferred from combined SDSS--PS1 data vs just SDSS. Specifically, Figure~\ref{fig:sigma_tau_ratios} shows that there are objects where $f_{\sigma} = \log_{10}{\left( \sigma_{\mathrm{SDSS-PS1}} / \sigma_{\mathrm{SDSS}} \right)} > 0.4 $ and $f_{\tau} = \log_{10}{\left( \tau_{\mathrm{SDSS-PS1}} / \tau_{\mathrm{SDSS}} \right)} > 1$ (a tenfold increase in $\tau$ and over twofold increase in $\sigma$). Visual inspection of objects simultaneously satisfying $f_{\sigma}> 0.4$ and $f_{\tau}> 1 $ shows that these underwent a significant ($>0.5$ mag) change in brightness between the SDSS (baseline 1998 - 2008) and PS1 DR2 observations (2009-2014; see Figure~\ref{fig:lc_extent}). Thus, DRW fitting could also be a way of finding changing-look quasar (and AGN) candidates. Figures~\ref{fig:clqso1}--~\ref{fig:clqso4} show the SDSS--PS1 $r$-band light curves of 40 CLQSO candidates with median PS1 brightnesses larger  than 20.5 mag. The open circles indicate day-averaged epochs (see Sec.~\ref{sec:data}). Table~\ref{tab:clqso} contains the basic physical parameters for these quasars. Some quasars show a downward trend, like turn-off CLQSOs (e.g., 123909, 1412379, and 1644710), while some are seen in a brightening stage, like turn-on CLQSOs (e.g., 1976348, 221006, 4069419, and 4205621). Quasar 612585 has the largest amount of auxiliary multiwavelength coverage-- X-ray from XMM-Newton, UV from the Galaxy Evolution Explorer (GALEX), and IR from the UKIRT Infrared Deep Sky Survey (UKIDSS), VHS, the Wide-field Infrared Survey Explorer (WISE)--and has been analyzed as part of the X-ray-targeted sample of S82 quasars, S82X, by \cite{lamassa2016a}. Quasar 751557 was previously identified by \cite{macleod2019} as a CLQSO candidate, with detailed Magellan spectroscopy described therein.  Two quasars, 1003694 and 1299803, have WISE data in the S82X catalog \citep{lamassa2016a}. Quasars 612585 and 3633437 have X-ray detections in the 3XMM DR5 catalog \citep{rosen2016}, but they have no matches in the Chandra point-source catalog (second release; \citealt{evans2010, evans2018}). There are no matches against the unified radio catalog of \cite{kimball2008}, which includes FIRST and NVSS data. We especially recommend spectroscopic follow-up of the brightest targets: 1976348 (mean 17.8 mag, turn-on; top left panel in Figure~\ref{fig:clqso2}) and 2104791 (mean 18.4 mag, turn-off; left column, third row panel in Figure~\ref{fig:clqso2}).

\begin{deluxetable*}{c|CCCCCCCCCC}
\tablecaption{CLQSO Candidates.\label{tab:clqso}}
\tablehead{\colhead{dbID} & \colhead{SDSSJID} & \colhead{$\alpha$} & \colhead{$\delta$} & \colhead{Redshift} & \colhead{$\log_{10}{(L_{\mathrm{Bol}})}$} & \colhead{$\log_{10}{(M_{\mathrm{BH}}/ M_{\odot})}$} & \colhead{$f_{\mathrm{Edd}}$} & \colhead{$\Delta(\mathrm{mag})$} & \colhead{$\Delta(\sigma_{G})$} & \colhead{Median PS1} }
\startdata 
123909 & \object[SDSS J001626.54+003632.4]{001626.54{+}003632.4} & 4.111 & 0.609 & 3.24 & 46.57 & 9.47 & -1.0 & -0.29 & -0.054 & 20.48 \\ 
8442 & \object[SDSS J001731.70+004910.1]{001731.70{+}004910.1} & 4.382 & 0.819 & 2.43 & 46.61 & 9.09 & -0.58 & -0.36 & -0.107 & 20.36 \\ 
4069419 & \object[SDSS J003359.39+000230.0]{003359.39{+}000230.0} & 8.497 & 0.042 & 1.64 & 45.95 & 9.05 & -1.21 & 0.22 & 0.017 & 20.2 \\ 
221006 & \object[SDSS J005142.20+002129.0]{005142.20{+}002129.0} & 12.926 & 0.358 & 1.55 & 45.95 & 8.24 & -0.39 & 0.21 & -0.01 & 19.97 \\ 
257776 & \object[SDSS J005513.15-005621.2]{005513.15{-}005621.2} & 13.805 & -0.939 & 3.61 & 47.13 & 9.58 & -0.54 & -0.32 & -0.242 & 19.53 \\ 
612585\tablenotemark{a} & \object[SDSS J010812.00-000516.5]{010812.00{-}000516.5} & 17.05 & -0.088 & 1.0 & 45.52 & 9.06 & -1.64 & -0.37 & 0.006 & 20.5 \\ 
1003694\tablenotemark{b} & \object[SDSS J012114.19-010310.8]{012114.19{-}010310.8} & 20.309 & -1.053 & 1.89 & 46.59 & 8.83 & -0.34 & 0.28 & 0.032 & 19.11 \\ 
1299803\tablenotemark{b} & \object[SDSS J014303.23-004354.0]{014303.23{-}004354.0} & 25.763 & -0.732 & 0.53 & 45.78 & 8.68 & -1.0 & -0.18 & -0.008 & 18.56 \\ 
1644710 & \object[SDSS J021259.00-000550.1]{021259.00{-}000550.1} & 33.246 & -0.097 & 0.81 & 45.67 & 8.38 & -0.81 & -0.28 & 0.022 & 19.7 \\ 
1730482 & \object[SDSS J021529.02-005314.9]{021529.02{-}005314.9} & 33.871 & -0.887 & 1.37 & 45.98 & 8.8 & -0.92 & 0.66 & -0.009 & 19.57 \\ 
2104791\tablenotemark{*} & \object[SDSS J022239.83+000022.5]{022239.83{+}000022.5} & 35.666 & 0.006 & 0.99 & 46.28 & 9.33 & -1.16 & -0.31 & 0.013 & 18.66 \\ 
2061101 & \object[SDSS J022505.06+001733.2]{022505.06{+}001733.2} & 36.271 & 0.293 & 2.42 & 46.38 & 8.09 & 0.2 & -0.34 & -0.016 & 20.15 \\ 
2006852 & \object[SDSS J023917.86-001916.8]{023917.86{-}001916.8} & 39.824 & -0.321 & 1.41 & 46.07 & 8.73 & -0.76 & -0.45 & -0.013 & 20.24 \\ 
2503955 & \object[SDSS J025316.46+010759.7]{025316.46{+}010759.7} & 43.319 & 1.133 & 1.03 & 46.31 & 8.94 & -0.73 & -0.39 & -0.071 & 19.07 \\ 
2484608 & \object[SDSS J025654.42-011455.4]{025654.42{-}011455.4} & 44.227 & -1.249 & 0.54 & 45.57 & 8.48 & -1.01 & -0.43 & -0.023 & 19.65 \\ 
3052176 & \object[SDSS J030504.07+011324.5]{030504.07{+}011324.5} & 46.267 & 1.223 & 0.61 & 45.29 & 9.2 & -2.01 & -0.35 & 0.016 & 20.09 \\ 
3096136 & \object[SDSS J031401.11+011131.6]{031401.11{+}011131.6} & 48.505 & 1.192 & 1.31 & 46.06 & 9.21 & -1.25 & -0.27 & -0.181 & 20.09 \\ 
3147102 & \object[SDSS J031846.13-005622.8]{031846.13{-}005622.8} & 49.692 & -0.94 & 2.12 & 46.57 & 9.07 & -0.6 & 0.47 & -0.078 & 19.07 \\ 
3781306 & \object[SDSS J032745.74+005217.2]{032745.74{+}005217.2} & 51.941 & 0.871 & 1.16 & 45.77 & N/A & N/A & 0.19 & 0.032 & 19.94 \\ 
3858587 & \object[SDSS J032825.19-003252.3]{032825.19{-}003252.3} & 52.105 & -0.548 & 0.77 & 45.61 & 8.68 & -1.17 & 0.33 & 0.031 & 19.39 \\ 
3810874 & \object[SDSS J033047.73+004859.4]{033047.73{+}004859.4} & 52.699 & 0.816 & 0.86 & 45.8 & 8.35 & -0.65 & -0.1 & -0.177 & 19.8 \\ 
3739253 & \object[SDSS J033059.05+010952.0]{033059.05{+}010952.0} & 52.746 & 1.164 & 0.56 & 45.39 & 8.16 & -0.86 & -0.27 & 0.01 & 19.72 \\ 
3791907 & \object[SDSS J033431.17-000904.0]{033431.17{-}000904.0} & 53.63 & -0.151 & 1.64 & 46.01 & 9.12 & -1.21 & 0.28 & 0.02 & 20.29 \\ 
4913626 & \object[SDSS J034512.62+002245.7]{034512.62{+}002245.7} & 56.303 & 0.379 & 0.42 & 45.5 & 8.81 & -1.41 & -0.48 & -0.055 & 19.43 \\ 
4205621 & \object[SDSS J203932.41-001818.3]{203932.41{-}001818.3} & 309.885 & -0.305 & 1.58 & 46.21 & 8.66 & -0.55 & 0.17 & -0.126 & 19.77 \\ 
3319354 & \object[SDSS J204952.62+011306.6]{204952.62{+}011306.6} & 312.469 & 1.219 & 1.09 & 46.14 & 9.52 & -1.48 & -0.91 & -0.048 & 20.08 \\ 
3633437\tablenotemark{c} & \object[SDSS J205105.02-005847.5]{205105.02{-}005847.5} & 312.771 & -0.98 & 0.54 & 45.34 & 8.47 & -1.23 & -0.51 & 0.071 & 20.34 \\ 
1835106 & \object[SDSS J215015.05-005331.4]{215015.05{-}005331.4} & 327.563 & -0.892 & 1.9 & 46.33 & 9.21 & -0.98 & 0.69 & 0.035 & 19.28 \\ 
1901056 & \object[SDSS J215055.51-001739.4]{215055.51{-}001739.4} & 327.731 & -0.294 & 1.54 & 46.26 & 8.6 & -0.44 & -0.19 & 0.009 & 20.46 \\ 
1976348\tablenotemark{*} & \object[SDSS J215841.40-001507.7]{215841.40{-}001507.7} & 329.673 & -0.252 & 1.46 & 46.92 & 9.39 & -0.57 & 0.39 & -0.051 & 17.56 \\ 
1446022 & \object[SDSS J220535.23+000756.3]{220535.23{+}000756.3} & 331.397 & 0.132 & 1.69 & 46.45 & 9.25 & -0.9 & 0.17 & -0.039 & 19.34 \\ 
1378415 & \object[SDSS J221347.32+001928.4]{221347.32{+}001928.4} & 333.447 & 0.325 & 2.31 & 46.41 & 8.59 & -0.29 & 0.23 & -0.113 & 20.02 \\ 
1412379 & \object[SDSS J221831.58-004548.9]{221831.58{-}004548.9} & 334.632 & -0.764 & 1.23 & 46.15 & 9.48 & -1.43 & -0.61 & -0.057 & 20.04 \\ 
1124333 & \object[SDSS J222918.25-004003.6]{222918.25{-}004003.6} & 337.326 & -0.668 & 1.16 & 45.81 & 8.35 & -0.64 & -0.28 & -0.003 & 20.12 \\ 
751557\tablenotemark{d} & \object[SDSS J225240.37+010958.7]{225240.37{+}010958.7} & 343.168 & 1.166 & 0.53 & 45.32 & 8.88 & -1.66 & 0.39 & -0.536 & 19.67 \\ 
467617 & \object[SDSS J231032.17-011449.5]{231032.17{-}011449.5} & 347.634 & -1.247 & 1.82 & 46.03 & 7.86 & 0.06 & 0.23 & 0.024 & 20.18 \\ 
568312 & \object[SDSS J231953.07-010139.0]{231953.07{-}010139.0} & 349.971 & -1.028 & 1.15 & 45.6 & 8.29 & -0.79 & 0.33 & 0.005 & 19.79 \\ 
292959 & \object[SDSS J232030.97-004039.2]{232030.97{-}004039.2} & 350.129 & -0.678 & 1.72 & 46.69 & 9.39 & -0.8 & -0.22 & -0.009 & 19.07 \\ 
3976336 & \object[SDSS J235213.27-004326.3]{235213.27{-}004326.3} & 358.055 & -0.724 & 0.9 & 45.64 & 8.85 & -1.3 & -0.18 & -0.08 & 19.92 \\ 
3946479 & \object[SDSS J235248.71-001518.4]{235248.71{-}001518.4} & 358.203 & -0.255 & 1.34 & 45.8 & 9.01 & -1.3 & -0.07 & -0.025 & 20.26 \\ 
\enddata 
\tablecomments{Catalog information from \cite{shen2011} concerning DR7 name (dbID), SDSSJID, location $\alpha$ and $\delta$ (in degrees, J2000), distance (spectrum-based redshift),  physical parameters (bolometric luminosity  $L_{\mathrm{Bol}}$ erg s$^{-1}$, black hole mass $M_{\mathrm{BH}}/ M_{\odot} $, Eddington ratio $f_{\mathrm{Edd}} = L_{\mathrm{Bol}} / L_{\mathrm{Edd}}$),  and light-curve properties (the difference between median SDSS and PS1 magnitudes $\Delta(\mathrm{mag})$, the difference in scatter between SDSS and PS1 segments $\Delta(\sigma_{G})$, and the median PS1 magnitude; see Figure~\ref{fig:median_offsets}).}
\tablenotetext{\tiny *}{Recommended for follow-up}
\tablenotetext{\tiny a}{S82X \citep{lamassa2016a}, XMM-Newton, GALEX UV, UKIDSS, VHS, WISE}
\tablenotetext{\tiny b}{S82X \citep{lamassa2019}, WISE}
\tablenotetext{\tiny c}{In 3XMM DR5 X-ray catalog \citep{rosen2016}}
\tablenotetext{\tiny d}{M19, CLQSO candidate, Magellan follow-up}
\end{deluxetable*}

\section{Mg\,{\sc ii} variability}
\label{app:mgii_variability}
\begin{figure*}
	\plotone{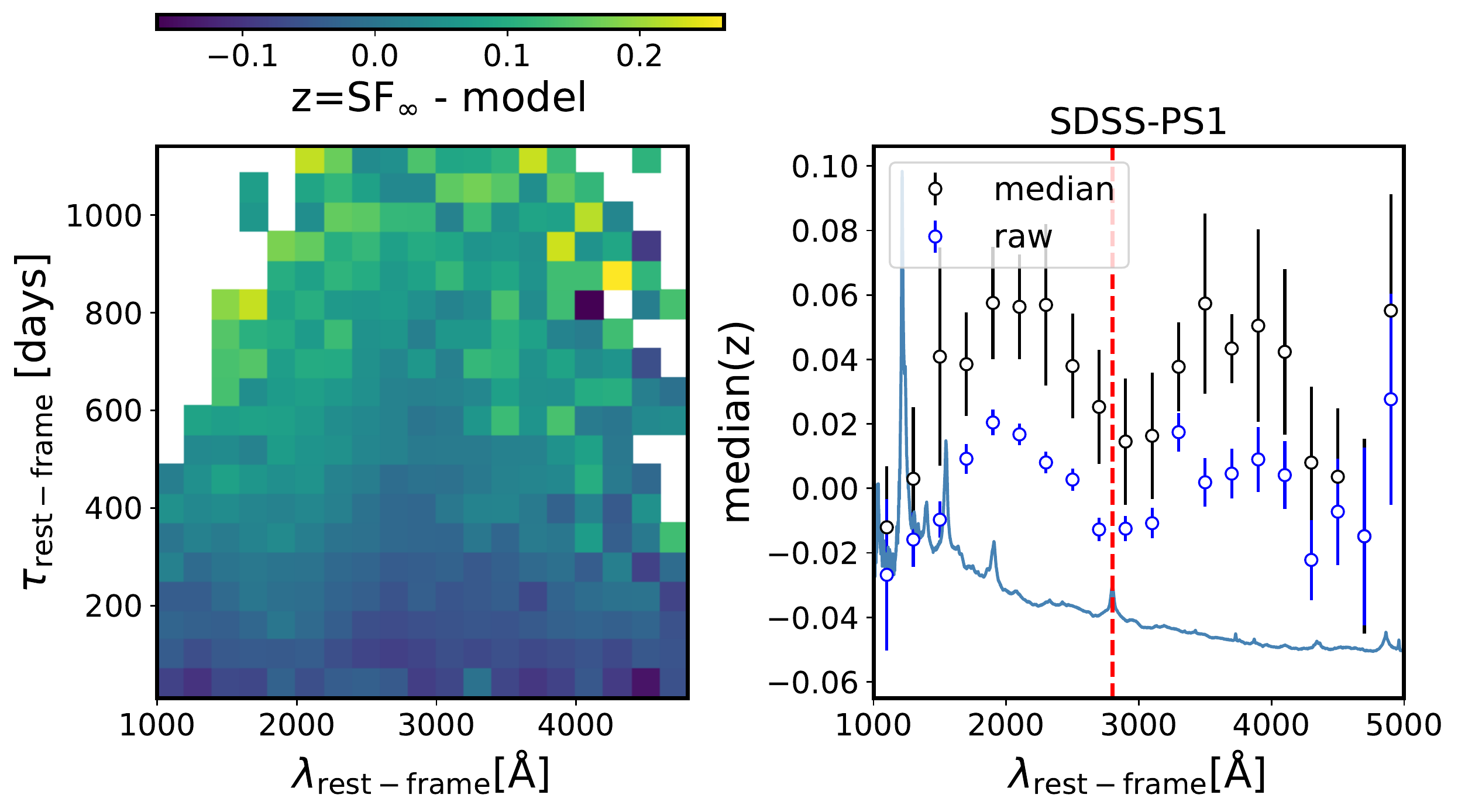}
	\caption{The residuals (z) after fitting Equation~\ref{eq:powlawmodel} to the quasar SF$_{\infty}$, using SDSS--PS1 quasar light curves. The left panel shows the residuals as a function of  $\lambda_{\mathrm{RF}}$ and $\tau_{\mathrm{RF}}$. The right panel shows the median of the residuals marginalized along $\tau_{\mathrm{RF}}$, with an error on the median based on the scatter in the $\lambda_{\mathrm{RF}}$ bins, $\sigma_{median} = 1.25 \mathrm{RMS}_{\mathrm{bin}} / \sqrt{N_{\mathrm{bin}}}$, where $\mathrm{RMS}_{\mathrm{bin}}$ is estimated from the robust Gaussian interquartile-based $\sigma_{G}$ \citep{ivezic2014}. Black dots are medians based on the averages plotted on the left ($N_{\mathrm{bin}}$ is between 7 and 20, depending on the number of nonempty bins in the left panel), and blue dots are medians based on the raw data ($N_{\mathrm{bin}}$ varies as a function of  $\lambda_{\mathrm{RF}}$, from a few to 800, peaking at about 2500 $\mbox{\AA}$, reflecting the fact that the quasar distribution peaks at redshift $z=2$).  We overplot the composite quasar spectrum from \citet{vandenberk2001}, and mark the location of the Mg\,{\sc ii} 2798 $\mbox{\AA}$ line with a vertical dashed red line.}
	\label{fig:mgii}
\end{figure*}

We searched for the dependency of variability parameters on other physical properties beyond the black hole mass and quasar luminosity employed in Equation~\ref{eq:powlawmodel}. The quasar optical spectrum has certain strong emission lines, depending on the redshift. In particular, the Mg\,{\sc ii} emission line lags the continuum and is less variable \citep{reichert1994}. The Mg\,{\sc ii} line  is also an important virial black hole mass estimator for quasars; \cite{mclure2002} noticed that the FWHM of the Mg\,{\sc ii} doublet trails that of the  H$\beta$ line (also see ~\citealt{shen2013}). Following  \cite{ivezic2004} and \cite{macleod2012},  we investigated the residuals after fitting the model ($f$=SF$_{\infty}$ in Equation ~\ref{eq:powlawmodel}) to the SDSS--PS1 data as a function of wavelength. The left panel of Figure~\ref{fig:mgii} shows the SF$_{\infty}$model residuals in the $\lambda_{\mathrm{RF}}$ vs $\tau_{\mathrm{RF}}$ space. The right hand side panel shows the residuals marginalized along $\tau_{\mathrm{RF}}$. The decrement around 2800 $\mbox{\AA}$ in the right panel is more pronounced when using combined SDSS--PS1 light curves. We see that both aggregates based on raw data (blue dots) or on medians (black dots) agree; the two statistical methods show a small (${\sim}5\%$), but statistically significant detection of a depression in the 2800 $\mbox{\AA}$  region, as expected from theory.

\bibliographystyle{aasjournal} 
\bibliography{references}

\end{document}